\documentclass[pre,twocolumn
,aps,superscriptaddress]{revtex4-2}
\usepackage [dvips]{graphicx}
\usepackage{amsmath}
\usepackage{amssymb}
\usepackage{mathrsfs}
\usepackage{color}
\usepackage{dsfont}

\usepackage{latexsym}
\usepackage{graphicx}
\usepackage{ucs}
\usepackage[utf8]{inputenc}
\usepackage{eurosym}
\usepackage[english]{babel}
\usepackage{graphicx}
\usepackage{siunitx}
\usepackage{hyperref}
\hypersetup{colorlinks=true,breaklinks=true,urlcolor=blue}
\usepackage{wrapfig}
\usepackage{amsmath,amsfonts,amssymb}
\usepackage[mathscr]{euscript}
\usepackage{url}
\usepackage{gensymb} 
\usepackage{natbib}
\usepackage{multirow}
\usepackage{dcolumn}
\usepackage{bm}
\usepackage{xcolor}
\usepackage[normalem]{ulem}
\usepackage[left=2.6cm,right=2.6cm,top=2.5cm,bottom=2.5cm]{geometry}
\usepackage{tikz}
\usepackage[compat=1.1.0]{tikz-feynman}

\usepackage{circuitikz}
\usetikzlibrary{arrows,shapes,positioning}
\usetikzlibrary{decorations.pathmorphing, decorations.pathreplacing, decorations.shapes}

\usetikzlibrary{arrows,shapes,positioning}
\usetikzlibrary{decorations.markings}
\tikzstyle arrowstyle=[scale=1]
\tikzstyle directed=[postaction={decorate,decoration={markings,
    mark=at position .65 with {\arrow[arrowstyle]{stealth}}}}]
\tikzstyle reverse directed=[postaction={decorate,decoration={markings,
    mark=at position .65 with {\arrowreversed[arrowstyle]{stealth};}}}]

\newcommand{\argc}[1]{\left[#1\right]} 
\newcommand{\arga}[1]{\left\lbrace #1\right\rbrace } 
\newcommand{\argp}[1]{\left(#1\right)} 
\newcommand{\argx}[1]{\left<#1\right>} 

\newcommand{\bt}[1]{\boldsymbol{#1}}

\newcommand{\funcder}[2]{\frac{\delta #1}{\delta #2}}

\newcommand{\phib}{\bar{\phi}}
\newcommand{\hb}{\bar{h}}

\newcommand{\chib}{\bar{\chi}}

\newcommand{\rhob}{\bar{\rho}}

\newcommand\arrow[2]{{\tikz[baseline=-.5ex] {\draw[ultra thick,#1,->] (0,0) -- (#2,0);} }}

\newcommand{\bxi}{\boldsymbol{\xi}}

\newcommand{\dd}{\text{d}}
\newcommand{\bz}{\text{\bf z}}

\newcommand{\ee}{\text{e}}

\newcommand{\p}{\partial}
\newcommand{\bx}{\text{\bf x}}
\newcommand{\br}{\text{\bf r}}

\newcommand{\bk}{\text{\bf k}}

\newcommand{\bq}{\text{\bf q}}
\newcommand{\bp}{\text{\bf p}}

\newcommand{\bv}{\text{\bf v}}

\newcommand{\bSigma}{\boldsymbol{\Sigma}}

\newcommand{\bnabla}{\boldsymbol{\nabla}}

\newcommand{\red}[1]{\color{red}#1\color{black}}

\newcommand*\circled[1]{\tikz[baseline=(char.base)]{
    \node[shape=circle,draw,inner sep=2pt] (char) {#1};}}

\def\wentzel{0}
\begin{document}

\title{From bulk to interface dynamics, in and out of equilibrium}

\author{Lila Sarfati}
\affiliation{Laboratoire Mati\`ere et Syst\`emes Complexes, Université Paris Cité  \& CNRS (UMR 7057), 75013 Paris, France}
\author{Julien Tailleur}
\affiliation{ Department of Physics, Massachussets Institute of Technology, Cambridge, USA}
\author{Frédéric van Wijland}
\affiliation{Laboratoire Mati\`ere et Syst\`emes Complexes, Université Paris Cité  \& CNRS (UMR 7057), 75013 Paris, France}
\affiliation{Yukawa Institute for Theoretical Physics, Kyoto University, Kyoto, 606-8502, Japan}

\begin{abstract}
We study the dynamics of weakly deformed interfaces separating two stable phases, starting from the fluctuating hydrodynamics of the phase-separating fields. 
Using a well-chosen definition for the interface and the dynamical-action formalism to represent path probabilities, we derive the linear relaxation of the interface and the fluctuations around it for a large class of models.
Our method applies to equilibrium dynamics, where it recovers and complements existing results, but also extends to their non-equilibrium counterparts. We explain how non-linear terms can be systematically computed and illustrate their derivations in the case of (active) model A.
We highlight the danger of a popular ansatz used to derive interface dynamics, which was rigorously established in equilibrium but is uncontrolled for active field theories. 
\end{abstract}

\maketitle



\section{Introduction}

Interfaces separating stable phases of matter are ubiquitous across a
wide variety of systems. From wetting to surface growth dynamics,
interfaces control a wealth of phenomena and their dynamics have
correspondingly attracted a long-standing interest~\cite{Bray1994_ReviewCoarsening,halpin1995kinetic,perlin2000capillary,hou2001boundary,casademunt2004viscous,lindner2009viscoelastic,gallaire2017fluid}. A
recent surge of studies in interface physics has been fueled by the
rise of biophysics and active matter, where interfacial phenomena
abound~\cite{hallatschek2007genetic,douezan2011spreading,joanny2012drop,sepulveda2013collective,maitra2014activating,nikola2016active,tjhung2018cluster,soni2019odd,perez2019active,cagnetta2019statistical,agudo2021wetting,Fausti2021_Capillary,adkins2022dynamics,Besse2023_Interface,pallares2023stiffness,langford2024theory,fins2024steer,zhao2024active,Caballero2025_Interface,sessa2026interfacial,laprade2026coarsening}.

For passive, equilibrium systems, much progress has been done to derive the relaxation dynamics of a weakly deformed interface starting from a coarse-grained description of the phase-separating fields~\cite{Bausch1981_Critical,Kawasaki1982_KineticDrumhead1,Kawasaki1982_KineticDrumhead2,kawasaki1983variational,Kawasaki1983_Kinetics,ohta1984scaling,Ohta1984_Dynamics,Jasnow1987_Crossover,Zia1988_Dynamics,Bausch1988_Dynamics,Bausch1991_Effects,Shinozaki1993_Dispersion,shinozaki1993dispersionH}.
The underlying field-theoretical approaches are rather demanding from an analytical standpoint and shortcuts have been designed, which provide equivalent results without resorting to heavy computations. For equilibrium systems, such shortcuts have been mathematically justified \cite{Ohta1984_Dynamics} and frequently used~\cite{Kawasaki1983_Kinetics,kawasaki1983variational,zia1985normal,Jasnow1987_Crossover,Zia1988_Dynamics,caballero2020bulk}.

Interface physics is even richer out of equilibrium, where it has
attracted a lot of interest over the past forty
years~\cite{Kuramoto1980_Instability,KPZ1986,kardar1998nonequilibrium,Bray2001_Interface,lucassen2009current,kardar2018,dean2020effect,romano2025dynamics}. Unfortunately,
controlled methods to derive interface dynamics starting from the
nonequilibrium stochastic dynamics of a phase-separating
nonequilibrium fluid remain in great demand.
The simplified route
proposed in equilibrium~\cite{Ohta1984_Dynamics} has been applied to
sheared~\cite{Bray2001_Interface}, driven~\cite{dean2020effect} and
active~\cite{Fausti2021_Capillary,Besse2023_Interface,langford2024theory,Caballero2025_Interface,langford2025phase,maire2025hyperuniform, sun2025interfacial,burekovic2026active}
systems, despite the absence of justification in this case.
A general approach out of equilibrium thus remains an outstanding challenge.

In this work, we fill this glaring methodological gap by introducing a field-theoretical framework that allows to systematically derive the dynamics of an interface between coexisting phases, in and out of equilibrium. 
In doing so, we critically discuss the application of the simplified route~\cite{Ohta1984_Dynamics} to nonequilibrium systems. We show that it is generally not predictive at the linear level and that predictions beyond the linear order in the deformation field are erroneous. 

This work is structured as follows. We briefly review in Sec.~\ref{sec:context} the various ways to define an interface starting from a bulk field theory. 
We introduce the simplified route in Sec.~\ref{sec:ansatz}, where we show its generic lack of consistency and predictability. We lay out our formalism in Sec.~\ref{sec:derivation}, which we first put to work in the equilibrium context throughout Sec.~\ref{sec:equilibrium}. 
We note that many of our equilibrium results can be partially found in the literature (e.g.~for models A~\cite{Bausch1981_Critical,Kawasaki1982_KineticDrumhead1} and B~\cite{Kawasaki1982_KineticDrumhead2,Ohta1984_Dynamics,Jasnow1987_Crossover,Shinozaki1993_Dispersion} of the Hohenberg and Halperin classification~\cite{HalperinHohenberg1977,Cates2019_LN} and, to some extent, for model H~\cite{Kawasaki1983_Kinetics,kawasaki1983variational,Ohta1984_Dynamics,shinozaki1993dispersionH}), sometimes lacking the noise term in the interface dynamics (e.g.~for model C~\cite{Ohta1984_Dynamics,Bausch1991_Effects}), yet some appear to be entirely new contributions 
(e.g.~for model D). At the very least, we offer here a unified framework to derive consistently all these interesting results. 
Beyond pointing out the pitfalls of simplified approaches, this section also helps us set the stage for nonequilibrium systems. 
Active versions of models A and B are then explored in Sec.~\ref{sec:noneq}. 
This is where we show how to deal technically with the breaking of time-reversal symmetry, which turns an important Hermitian operator of the equilibrium approach into a non-Hermitian operator~\cite{Kuramoto1980_Instability}. Throughout our work, we focus on the linear relaxation dynamics of the interface, at the exception of active model A, where we show how to recover the KPZ contribution. The derivation of non-linear terms in the presence of a conservation law out of equilibrium is much more involved and will be deferred to a future work. Our concluding Sec.~\ref{sec:prospects} addresses promising research directions.

\section{From bulk dynamics to the definition of an interface}\label{sec:context}

\subsection{Bulk dynamics}
For concreteness, we first consider systems described by a single
fluctuating scalar order-parameter field $\phi$. We assume that, at
least at the mean-field level, the evolution of $\phi$ leads to a
smooth planar interface located at $z=0$, separating two spatially
stable phases along some direction $z$. Translational invariance is
maintained along the remaining $d-1$ directions, hereafter denoted by
$\br$. We start from a stochastic partial differential equation for
$\phi(\br,z,t)$ in the form~\cite{HalperinHohenberg1977},
\begin{equation}\label{eq:motion}
    \p_t\phi= -\frac{\delta F}{\delta\phi}+w[\phi]+\sqrt{2T}\eta(\br,z,t)\;,
\end{equation}
where $\eta$ is a Gaussian white noise with correlations $\langle\eta(\br,z,t)\eta(\br',z',t')\rangle=\delta(t-t')\delta(z-z')\delta^{(d-1)}(\br-\br')$. The functional $F$ is a Landau free energy of the form 
\begin{equation}\label{eq:F}
    F[\phi]=\int\dd^{d-1} r\,\dd z\Big[\frac{1}{2}(\bnabla\phi)^2 + f(\phi)\Big]\;,
\end{equation}
where $f$ exhibits two degenerate minima and $w$ is the nonconservative contribution to the evolution. This functional Helmholtz decomposition has been discussed in a series of recent works~\cite{o2023nonequilibrium,o2024geometric,o2025geometric}. The goal is to start from dynamics~\eqref{eq:motion} and derive the fluctuating dynamics for a properly-defined interface between the coexisting phases. The extension of the formalism to other cases will be discussed further down, including the possibility of conservation laws and the coupling to auxiliary fields (that may themselves be conserved or not). \\

In Eq.~\eqref{eq:motion}, the parameter $T$ refers to the amplitude of the noise experienced by the field $\phi$. This is a temperature only if $w=0$ and the system is in thermal equilibrium. We denote by $m_c(z)$ the phase-separated steady-state mean-field profile, which solves
\begin{equation}\label{eq:MF}
    0= -\frac{\delta F}{\delta\phi}[m_c]+w[m_c]\;.
\end{equation}
Let us now discuss how to define the deformation field of the interface in the presence of fluctuations.

\definecolor{redUP}{RGB}{138,21,56}
\definecolor{blueUP}{RGB}{50,40,130}
\definecolor{orangeUP}{RGB}{220,110,50}
\begin{figure}
\begin{tikzpicture}
\node at (0,0) [text width=1\columnwidth]{\includegraphics[width=0.7\columnwidth]{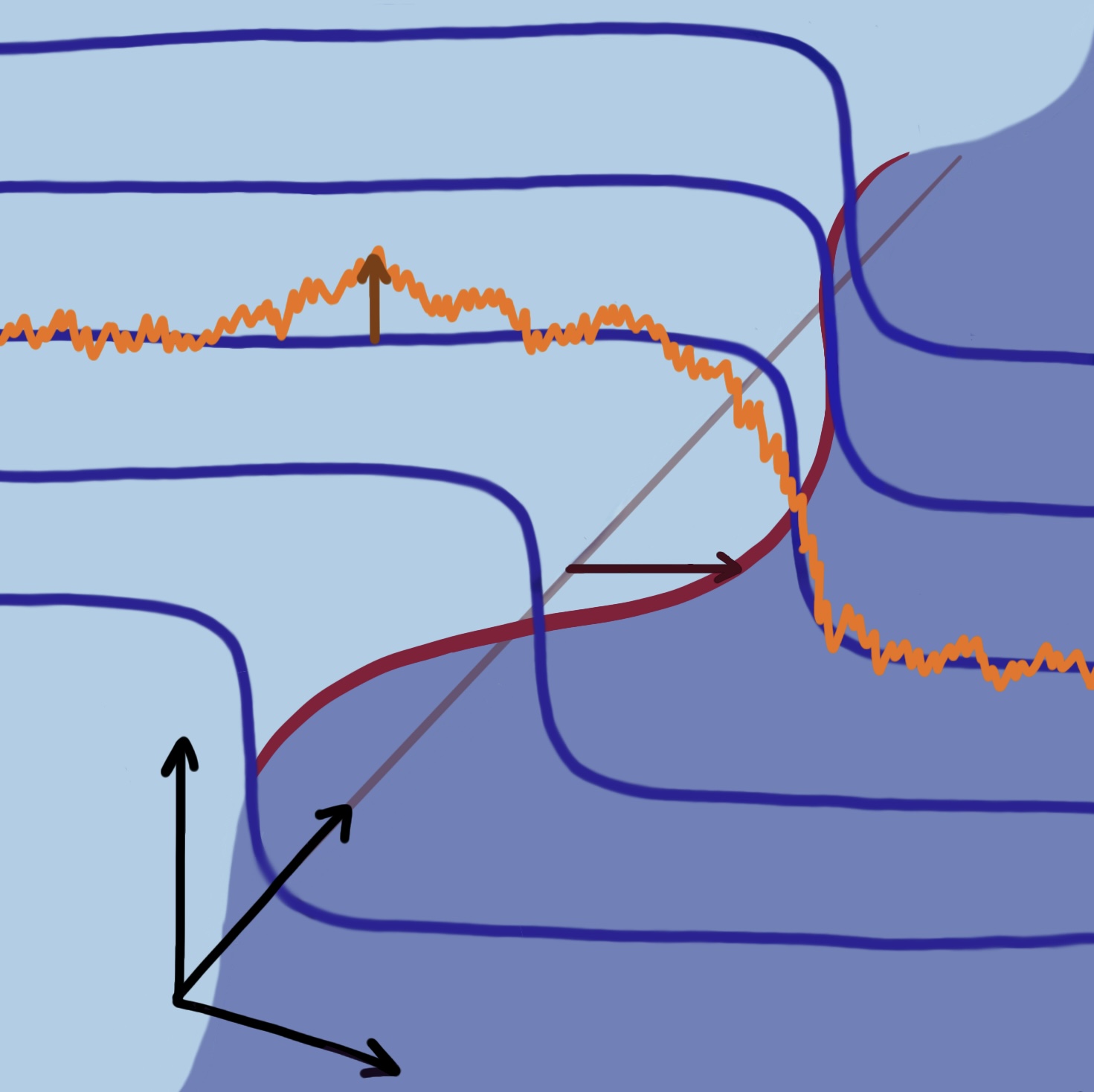}  };
\node at (-0.6,0.15) {$\color{redUP}\bt{h(\vec{\br},t)}$};
\node at (-2.5,-1.3) {$\vec{\br}$};
\node at (-3.0,-0.75){$\color{orangeUP}\bt{\phi(\vec{\br},z,t)}$};
\node at (-2.05,0.7) {$\color{orangeUP}\bt{\chi(\vec{\br},{z-h},t)}$};
\node at (-2.15,-2.35) {$\bt{z}$};
\node at (0.5,-1.68){$\color{blueUP}\bt{m_c({z-h})}$};
\node at (-4.8,-1.68) {};
\end{tikzpicture}
    \caption{Fluctuations of the interface between two stable phases around its flat average. The interface fluctuations result from both a shift of the mean-field profile $m_c$ (dark blue) and the fluctuations around it.
    }
    \label{fig:cartoon}
\end{figure}
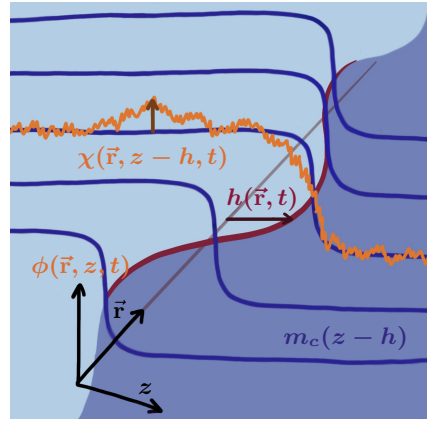
\subsection{Interface definition}\label{sec:definition}
Figure \ref{fig:cartoon} shows a cartoon of the field and of the
interface. We denote $h(\br,t)$ the amplitude of the deviation between
the fluctuating interface and its flat average. In what follows we
assume that the fluctuations of the interface occur over length scales
large with respect to its width $\xi$, that the amplitudes of
fluctuations remain small compared to the jump of the order parameter
across the interface, $\Delta m_c=m_c(+\infty)-m_c(-\infty)$, and that
the interface remains single valued (there are no overhangs). \\

To make progress, we need an operational definition of $h$ starting from the field $\phi$, beyond the cartoon. A natural definition is that the interface $h(\br ,t)$ is the set of $z$ coordinates where the field $\phi$ takes a prescribed value $\phi_0$ (be it half the step size, $m_c(0)$, {\it etc}): $\phi(\br,h(\br,t),t)=\phi_0$. For reasons that will become clear below, it is useful to recast this as the following equation for $h(\br,t)$:
\begin{equation}\label{eq:definterfacevalue}
    \int \dd z [\phi(\br,z,t)-\phi_0] \delta(z-h(\br,t))=0\;.
\end{equation}
Even though prescribing a value to $\phi(\br,h(\br,t),t)$ is an appealing definition for $h$, it is also very sensitive to initial condition, noise, and, for a given $\br$, there might exist several values of $z$ where the field takes the value $\phi_0$. This provides the incentive to search for a smoother definition of the interface, insensitive to such caveats.\\

Instead of using the fluctuating field, a possibility is to resort to its mean-field profile $m_c$ and to define $h$ as the solution of:
\begin{equation}\label{eq:rottenansatz}
    \phi(\br,z,t)=m_c(z-h(\br,t))\;.
\end{equation}
This however assumes that $\phi$ is simply a shift of the mean-field
profile along $\hat {\bf z}$, whose amplitude defines $h(\br,t)$. This
is indeed a smoother definition of $h$, but it does not account for
the possibility of fluctuations within the bulk of each phase. This is
the typical approach adopted in the theory of capillary
waves~\cite{davis1977capillary,turski1980dynamics}, and its use has
been justified when (local) equilibrium
holds~\cite{Ohta1984_Dynamics}. This is also the basis of recent works
on interfaces in nonequilibrium systems such as
sheared~\cite{Bray2001_Interface}, driven \cite{dean2020effect}, or
active
fluids~\cite{Fausti2021_Capillary,Besse2023_Interface,langford2024theory,Caballero2025_Interface,sun2025interfacial, maire2025hyperuniform,burekovic2026active}. We
shall see in Sec.~\ref{sec:ansatz} that this approach also conceals
its share of complications.\\

Inspired by the quantum-field-theory literature on solitons, instantons, and extended states~\cite{Dashen1974_Nonperturbative,gervais1975extended,Gervais1975_Perturbation}, it was proposed in \cite{Diehl1980,Kuramoto1980_Instability} that interfaces and fronts would be better defined using 
\begin{equation}
\label{eq:definterface2}
   \int\dd z \,\chi(\br,z-h(\br,t),t)   
   \p_z m_c(z-h(\br,t))=0
\end{equation}
where we have introduced the auxiliary field
\begin{equation}\label{eq:DefChi}
   \chi(\br,z-h(\br,t),t)=\phi(\br,z,t)-m_c(z-h(\br,t))\;.
\end{equation}
Comparing Eq.~\eqref{eq:DefChi} to Eq.~\eqref{eq:rottenansatz}, we see that $\chi$ vanishes when $\phi$ is  a shift of the mean-field profile. Otherwise, it captures the fluctuations of $\phi$ that are not a mere translation of $m_c$ and contains, for instance, bulk deformations away from the interface (See Fig~\ref{fig:cartoon}).

Comparing Eq.~\eqref{eq:definterface2} to
Eq.~\eqref{eq:definterfacevalue} shows that the sharp localization of
the interface using a $\delta$ function has been smoothed out on the
typical scale $\xi$ of the interface width. When $\xi$ is much smaller
than other length scales, including the wavelength of the interface
fluctuations, it is indeed reasonable to approximate
\begin{equation}
    \p_z m_c(z-h)\simeq\frac{\Delta m_c}{\xi}\delta\left(\frac{z-h}{\xi}\right)\;.
\end{equation}
In this limit, the definition of $h$ amounts to setting
$\phi(\br,h(\br,t),t)=m_c(0)$, recovering
Eq.~\eqref{eq:definterfacevalue} with $\phi_0=m_c(0)$. The
definitions~\eqref{eq:definterface2} and~\eqref{eq:DefChi} thus
circumvent the identified limitations of
Eqs.~\eqref{eq:definterfacevalue} and~\eqref{eq:rottenansatz}.  Note
that, in the integral entering Eq.~\eqref{eq:definterface2}, instead of
$\p_z m_c$, we could have used---and we will use---any function that
displays a similar physical behavior (namely a steep $\delta$ limiting
behavior as $\xi\to 0$), should it prove more convenient.

The interface definition through Eq.~\eqref{eq:definterface2} has been
used to study static fluctuations around the equilibrium profile of a
$\phi^4$ theory in~\cite{Diehl1980}. It is also the starting point to
study the dynamics of specific
passive~\cite{Bausch1981_Critical,Kawasaki1982_KineticDrumhead1,Kawasaki1982_KineticDrumhead2,zia1985normal,Zia1988_Dynamics,Bausch1988_Dynamics,Bausch1991_Effects}
and driven models~\cite{Kuramoto1980_Instability}. However, a general
framework, applicable in particular to active field theories, is still
missing.

\if{ Dynamics wise, the landmark works based on this starting point
  are those of Kuramoto \cite{Kuramoto1980_Instability}, Bausch {\it
    et
    al.}~\cite{Bausch1981_Critical,Zia1988_Dynamics,Bausch1988_Dynamics,Bausch1991_Effects}
  along with those of Kawasaki and
  Ohta~\cite{Kawasaki1982_KineticDrumhead1,Kawasaki1982_KineticDrumhead2}.}\fi

Before we embark in a description of the analytical route we have
followed, we briefly discuss the shortcomings of the ansatz given in
Eq.~\eqref{eq:rottenansatz}, which has recently gained popularity in
the active-matter and nonequilibrium
communities~\cite{Bray2001_Interface,dean2020effect,Fausti2021_Capillary,Besse2023_Interface,langford2024theory,langford2025mechanics,Caballero2025_Interface,sun2025interfacial,maire2025hyperuniform,burekovic2026active}.

\section{An appealing ansatz and its shortcomings}
\label{sec:ansatz}
Let us now show how, when the definition of the interface relies on
the mean-field ansatz of Eq.~\eqref{eq:rottenansatz}, inconsistencies
arise. Since the goal is to obtain an evolution equation for $h$, we
can substitute $\phi(\br,z,t)=m_c(z-h(\br,t))$ into
Eq.~\eqref{eq:motion}, ignoring at this stage, as commonly done in the literature, any pitfalls arising from its time-discretization. Space and time derivatives are then
computed as
\begin{align}
  \bnabla \phi(\br,z,t)&= \p_z m_c(z-h(\br,t))\hat{\bz}\notag\\&\;-\bnabla_\br h(\br,t)\,\p_z m_c(z-h(\br,t))\nonumber\\
      \p_t\phi(\br,z,t)&=-\p_t h(\br,t)\,\p_z m_c(z-h(\br,t))\;,
\end{align}
and similarly for higher-order spatial derivatives. In the simplest case of equilibrium model A~\cite{HalperinHohenberg1977}, the dynamics for $\phi$ reads
\begin{equation}
    \p_t\phi=\Delta\phi+\xi^{-2}(\phi-\phi^3)+\sqrt{2}\eta(\br,z,t)\;,
\end{equation}
which leads to
\begin{align}\label{eq:modelAansatz}
    -\p_t h \, \p_z m_c(z) =&-\bnabla_\br^2 h\, \p_z m_c(z)\nonumber 
 + (\bnabla_\br h)^2\,\partial_z^2 m_c(z)\\ &+  \sqrt{2}{\eta}(\br,z+h(\br,t),t)\;.
\end{align}
We then need to eliminate the $z$-dependence to get closed dynamics for $h(\br,t)$. This can, however, be achieved in many ways. 

For instance we can follow a first protocol:
\begin{itemize}
\item[$\circled{1}$] Multiply Eq.~\eqref{eq:modelAansatz} by $\p_zm_c(z+1)$ and integrate over $z$, which yields
  \begin{equation}\label{eq:hfakeKPZ}
    \p_t h= \bnabla_\br^2 h - \frac{\alpha}{\sigma_1}(\bnabla_\br h)^2 + \sqrt{2\frac{\sigma}{\sigma_1^2}} \tilde\eta(\br,t)\;,
  \end{equation}
  where $\tilde\eta$ is a Gaussian white noise with  correlations $\delta^{(d-1)}(\br-\br')\delta(t-t')$, and we have introduced
  \begin{equation}\label{eq:hfakeKPZsigma}
    \begin{split}
      \sigma =& \int \dd z\, \p_z m_c(z)^2\\
      \sigma_1 =& \int \dd z\, \p_z m_c(z) \, \p_z m_c(z+1)\\
      \alpha =& \int \dd z \, \p_z^2 m_c(z)\, \p_z m_c(z+1) \;.
    \end{split}
  \end{equation}
\end{itemize}
The mean-field ansatz Eq.~\eqref{eq:rottenansatz} can thus be used to predict that the interface of an equilibrium Ising model evolves according to the Kardar-Parisi-Zhang (KPZ) equation~\cite{KPZ1986}.

Of course, this cannot be correct. In fact, in Eq.~\eqref{eq:modelAansatz}, the unknown function $h$ depends solely on $\br$ and $t$, yet the resulting equation is valid for all $z$, which over constrains the problem. Out of the many manipulations of Eq.~\eqref{eq:modelAansatz} one can implement, one actually leads to the correct result. It consists in the following protocol:
\begin{itemize}
    \item[$\circled{2}$] Multiply Eq.~\eqref{eq:modelAansatz} by $\p_z m_c(z)$ and integrate over $z$ to get:
\begin{equation}\label{eq:EWansatz}
\p_t h=\bnabla_\br^2 h+\sqrt{\frac{2}{\sigma}}\tilde{\eta}(\br,t)\;.
\end{equation}
\end{itemize}
We know that Eq.~\eqref{eq:EWansatz} is correct simply because we have separate mathematically controlled procedures to derive it (incidentally, this also shows that the surface tension in Eq.~\eqref{eq:hfakeKPZsigma} was not correct). The ansatz of Eq.~\eqref{eq:rottenansatz} is thus useful only once one has established the correct accompanying procedure, provided it exists. At this stage, several remarks are in order.

\if{One may wonder whether Eq.~\eqref{eq:rottenansatz} could be improved by working in the Frenet frame, replacing the argument $z-h$ by $\frac{z-h}{\sqrt{1+(\bnabla_\br h)^2}}$ in Eq.~\eqref{eq:rottenansatz}. However, this does not alter the reasoning above, nor its accompanying mathematical inconsistency.}\fi 

First, if the dependence of $h$ on $\br$ could somehow be forgotten, and if the noise in Eq.~\eqref{eq:modelAansatz} played no role, Eq.~\eqref{eq:rottenansatz} would constitute a mathematically sound starting point. Although this does not apply to the problem at hand, it allows characterizing the deterministic coarsening of droplets~\footnote{See sections 2.3 and 2.4 of~\cite{Bray1994_ReviewCoarsening} for models A and B, respectively.}.

Then, on physical grounds, one might expect that the difference
$\chi=\phi-m_c(z-h)$ quickly relaxes to $0$ (which suggests, for
instance, the possibility of adiabatic approximations). However, this
is not what is needed to get a closed equation for $h$. Indeed, what
one has to do is the conditional average of $\chi(\br,z,t)$ \textit{at
  fixed realization of $h(\br,t)$}. As we show below this average
cannot be neglected.

Finally, as we have discussed above, there are cases where a precise
protocol to handle Eq.~\eqref{eq:rottenansatz} leads to the correct
interface equation. This prompts the following question: given the
ansatz in Eq.~\eqref{eq:rottenansatz}, is there always a protocol that
leads to the correct result?  For conservative equilibrium dynamics,
this was sorted out by Ohta, Ohta and Kawasaki~\cite{Ohta1984_Dynamics,ohta2025}, who used considerations of
energy balance and virtual displacement to derive such a
protocol. Such a derivation exists neither beyond linear order nor for
nonconservative dynamics.

All in all---and beyond the sole question of the ansatz of Eq.~\eqref{eq:rottenansatz}---some progress has been made in equilibrium~\cite{Bausch1981_Critical,Jasnow1981_Unstable,Kawasaki1982_KineticDrumhead1,Kawasaki1982_KineticDrumhead2,kawasaki1983variational,Kawasaki1983_Kinetics,Ohta1984_Dynamics,ohta1984scaling,jug1984renormalization,zia1985normal,jug1985dynamic,Jasnow1987_Crossover,Zia1988_Dynamics,Bausch1988_Dynamics,Bausch1991_Effects,Shinozaki1993_Dispersion,shinozaki1993dispersionH}, but a general approach, applying to equilibrium and nonequilibrium systems alike, is still missing. Our goal below is to fill this gap by deriving a framework that is simple enough to be adapted to modern problems, especially those emanating from the field of active matter.

\if{
It turns out that existing mathematically controlled approaches are limited, so much so that for models as simple as equilibrium model B~\cite{HalperinHohenberg1977}, available results are heavy and tedious to implement~\cite{Kawasaki1982_KineticDrumhead2,Shinozaki1993_Dispersion}. Our goal is to achieve a general presentation, applying to equilibrium and nonequilibrium systems alike, that is simple enough to be adapted to modern problems, especially those emanating from the field of active matter.}\fi

\section{Deriving the interface dynamics}\label{sec:derivation}
For the sake of concreteness and simplicity, we first present our approach in the case of Eq.~\eqref{eq:motion}.

\subsection{Path-integral formalism}
As discussed in Section~\ref{sec:context}, the definition we adopt for the interface is Eq.~\eqref{eq:definterface2}. 
Beyond its physical appeal, this  definition also has a geometric interpretation that will serve as the basis of our approach. 
\if{This suggests that we should introduce the auxiliary field
\begin{equation}\label{eq:DefChi}
    \chi(\br,z-h(\br,t),t)= \phi(\br,z,t) - m_c\argp{z-h(\br,t)}
\end{equation}}\fi
We introduce the scalar product
\begin{equation}\label{eq:SP}
\langle f,g\rangle=\int\dd z f(z) g(z)
\end{equation}
in the space of $z$-dependent functions, such that the condition
Eq.~\eqref{eq:definterface2} translates into
\begin{equation}\label{eq:definterface}
  \langle\chi(\br,z,t),u_0(z)\rangle=0\;,
\end{equation}
with $u_0(z)=\p_z m_c(z)$. The interface $h$ is thus defined in such a
way that $\chi$ is orthogonal to the function $u_0$.
Equation~\eqref{eq:definterface} serves several purposes.  First, it
removes an ambiguity in Eq.~\eqref{eq:DefChi}, where a shift of $h$
can be absorbed into a redefinition of $\chi$.  Then, this geometric
formulation as a scalar product will prove useful to integrate out
the $\chi$ degrees of freedom and obtain an effective dynamics for the
field $h$ only.  The precise choice of $u_0$ among all the possible
peaked functions with support localized near the interface will then
be made model by model to facilitate the algebra.

To proceed, we start from
\begin{equation}
    \mathbb{P}[h]=\langle  \delta[h-h[\phi]]\rangle_{\phi}\equiv \int{\mathcal D}\phi \,\delta[h-h[\phi]] \mathbb{P}[\phi]\;,
\end{equation}
where $\mathbb{P}[h]$ and $\mathbb{P}[\phi]$ are the path probabilities of the fields $h(\br,t)$ and $\phi(\br,z,t)$, respectively, and $h[\phi]$ is defined by Eqs.~\eqref{eq:definterface} and~\eqref{eq:DefChi}. 
To make progress, we then introduce the Janssen-De Dominicis action functional~\cite{janssen1976lagrangean,dominicis1976techniques},
\begin{equation}\label{eq:Z}
    Z=\int{\mathscr D}\bar{\phi}\,{\mathscr D}\phi\,\ee^{-\mathrm S[\bar{\phi},\phi]}
\end{equation}
with the It\=o-discretized action
\begin{equation}\label{eq:actionstart}
   \mathrm S[\bar{\phi},\phi]=\int\dd t\,\dd z\,\dd^{d-1}r\left[\phib\left(\p_t\phi+\frac{\delta F}{\delta \phi}-w\right)-T\phib^2\right]\;,
\end{equation}
so that:
\begin{equation}
 \mathbb{P}[h]=\frac 1 Z \int{\mathscr D}\phib\, {\mathscr D}\phi\, \delta[h-h[\phi]] \,\ee^{-\mathrm S[\bar\phi,\phi]}\;.
\end{equation}
We then use the definition Eq.~\eqref{eq:definterface} to rewrite
\begin{equation}
    \delta[h-h[\phi]]=\delta\argc{\langle \phi(\br,z,t)-m_c(z-h),u_0(z-h)\rangle}|J|\;,
\end{equation}
where the functional Jacobian $J$ is given by
\begin{equation}
J=-\argx{\phi(\br,z,t),\p_z u_0(z-h)}\;. 
\end{equation}
To enforce the $\delta$ constraint explicitly, we introduce the projector
\begin{equation}\label{eq:defPi}
    \Pi= \mathds 1 - \frac{\left|\p_z m_c\right>\left<u_0\right|}{\argx{ u_0,\p_z m_c}}\;.
\end{equation}
The projector $\Pi$ is well defined as long as $u_0$ and $\p_z m_c$ have a non-zero scalar product, 
which will prove a necessary condition for the choice of $u_0$. It verifies
\begin{equation}\label{eq:Piprop}
    \begin{aligned}
       & \Pi \p_z m_c = 0\;,\\
       & \Pi^\dagger u_0 = 0\;,
    \end{aligned}
\end{equation}
where $\dagger$ stands for the adjoint operator with respect to the scalar product defined in Eq.~\eqref{eq:SP}. Note that $\Pi$ is not an orthogonal projector in general, except if $u_0\propto\p_z m_c$.  We can then split the fluctuation $\chi$, defined in Eq.~\eqref{eq:DefChi}, between its longitudinal and transverse components, using
\begin{equation}\label{eq:DecompChi}
    \chi = \frac{\argx{u_0,\chi}}{\argx{u_0,\p_z m_c}}\p_z m_c + \chi^\perp,\;\text{where}\;\chi^\perp=\Pi\chi\;.
\end{equation}
Our choice of interface definition, which translates into the functional constraint in Eq.~\eqref{eq:definterface}, thus enforces the vanishing of the longitudinal component of $\chi$. 

This leads to an expression for $\mathbb{P}[h]$ in terms of a path integral
\begin{align}
\mathbb{P}[h]=&\frac 1  Z \int {\mathscr D}\bar{\phi}\,{\mathscr D}\phi\, \big\{\delta\argc{\langle \phi-m_c(z-h),u_0(z-h)\rangle}\notag\\
&\qquad\times \ee^{-S[h,\bar{\phi},\phi]}\big\}\;,
\end{align}
where $S[h,\bar{\phi},\phi]=\mathrm S[\bar{\phi},\phi]-\int_{t,\br} \ln |J|$. Using Eq.~\eqref{eq:DefChi} and defining $T \phib(\br,z,t) = \chib(\br,z-h,t)$ then leads to
\begin{align}
\mathbb{P}[h] =&\frac 1  Z \int{\mathscr D}\bar{\chi}\,{\mathscr D}\chi\,\delta\argc{\langle \chi,u_0\rangle}\ee^{-S[h,\bar{\chi},m_c+\chi]}\label{eq:Zforh}\\
    =&\frac 1     Z \int{\mathscr D}\bar{\chi}\,{\mathscr D}\chi^\perp\,\ee^{-S[h,\bar{\chi},m_c+\chi^\perp]}\;,
\end{align}
where the $\delta$ constraint of the first line restricts the path integration over $\chi$ to functions $\chi^\perp$ orthogonal to $u_0$. Note that all the actions written above imply a continuous time limit whose construction is delicate~\cite{de2022path}. These issues will prove crucial to go beyond the low-noise limit; initially, for clarity, we however omit the corresponding terms which arise whenever the chain rule is used in the action. After some algebra, the action then takes the form
\begin{align}
    S=\frac{1}{T}&\int_{t,\br,z}\Big\{\chib\Big[-\p_th(\p_z m_c + \p_z\chi^\perp)+ \p_t\chi^\perp\notag\\
    &\; +\left.\argp{\frac{\delta F}{\delta \phi}- w}\right|_{m_c+\chi^\perp}\Big] - \chib^2 \Big\}\notag\\
    &\;- \int_{t,\br}\ln\left|\argx{\p_z u_0,m_c} + \argx{\p_z u_0,\chi^\perp}\right|\;,\label{eq:Sfor4}
\end{align}
where the notation $(\frac{\delta F}{\delta \phi}- w)|_{m_c+\chi^\perp}$ means that we first evaluate the functional derivative in $\phi(\br,z,t)=m_c(z-h)+\chi^\perp(\br,z-h,t)$ and then do the shift $z\to z-h$. In order to find the explicit expression of this action, we use that, for $i = t,\br$,
\begin{align}
     \p_i\phi = -\p_i h[\p_z& m_c(z-h) + \p_z\chi^\perp(\br,z-h ,t)]\notag\\
     &+ \p_i\chi^\perp(\br,z-h ,t)\;.
\end{align}

Mirroring Eq.~\eqref{eq:DecompChi} for $\chi$, we decompose $\chib$ as:
\begin{equation}\label{eq:DecompChib}
    \chib = - \hb u_0 + \chib^\perp,\;\text{with}\; \chib^\perp=\Pi^\dagger\chib  \;,
\end{equation}
where
\begin{equation}
\hb(\br,t)=-\frac{\argx{\p_z m_c(z),\chib(\br,z,t)}}{\argx{\p_z m_c(z),u_0(z)}}
\end{equation}
captures the component of $\chib$ that we anticipate 
will drive the fluctuations of $h$. Note that $\chi^\perp$ is orthogonal to $u_0$ while $\chib^\perp$ is orthogonal to $\p_z m_c$.

Our goal will now be to integrate out the fields $\chib^\perp$ and $\chi^\perp$ to obtain an action for $\hb$ and $h$ only:
\begin{equation}\label{eq:Sforh}
    S_h[\hb,h]=-\ln \int{\mathscr D}\chib^\perp{\mathscr D}\chi^\perp\ee^{-S[h,-\hb u_0 +\chib^\perp,m_c+\chi^\perp]}\;,
\end{equation}
such that
\begin{equation}
\mathbb{P}[h] = \frac 1 Z \int{\mathscr D}\hb \,\ee^{-S_h[\hb,h]}\;.
\end{equation}
In practice, it will often prove more convenient to integrate over  $(\Gamma^\dagger \chib)^\perp$ instead of over $\chib^\perp$, where $\Gamma$ is model-dependent linear operator. 
We note that computing $S_h[\hb,h]$ is a formidable task, unless one accepts to narrow down the physical regime of interest. In this work we choose to consider the regime of weak noise in which fluctuations and deviations from the mean-field profile remain small \cite{Bausch1988_Dynamics,Bausch1991_Effects}.

\subsection{The low noise limit}\label{sec:LowNoise}

In the limit of low temperature, the path integral is dominated by the trajectories that minimize the first integral in Eq.~\eqref{eq:Sfor4}. By construction, owing to the definition of $m_c$ in Eq.~\eqref{eq:MF}, this corresponds to $\bar{\chi}=\chi^\perp=h=0$~\footnote{Strictly speaking, the saddle-point trajectory for $h(\br,t)$ is $h(\br,t) = \text{cst}$. The resulting dynamics on $h$ is independent of this constant, which we thus set to zero.}. This tells us that the three fields  scale as $\sqrt{T}$. We now carry out a perturbation expansion in $\sqrt{T}$ around the saddle point of the action.  Working to leading order in $\sqrt{T}$ amounts to truncating the action to quadratic order in the fields. This leads to an effective linear equation for $h$, while higher orders in $\sqrt{T}$ control the nonlinear behavior. Once the weak noise limit is taken, we shall also work in the large wavelength limit (with respect to the width of the interface). Note that these limits may not commute: working in the large wavelength limit at fixed field amplitudes requires to implement a renormalization group (RG) treatment if nonlinearities are relevant. In the low noise limit, the Jacobian
$J$, which is responsible for the third line in Eq.~\eqref{eq:Sfor4},
does not contribute to the leading order, but it will
have to be considered for higher-order corrections. 
The same holds not only for the terms coming from the application of the generalization of It\=o's lemma to integration paths~\cite{de2022path} but also for the term $\chib \p_t h \p_z\chi^\perp$ (these are all of $O(T^{3/2})$). The expansion of $\frac{\delta F}{\delta \phi}- w$, which is required to obtain the Gaussian action explicitly, is model dependent and will be detailed below on a case by case basis. In practice, it will often prove easier to work with Fourier transforms in time $t$ 
and in the space coordinates $\br$ along the interface, using: 
\begin{equation}\label{eq:FT}
    \begin{aligned}
        &\chi_{\bq,\omega} = \int_{t,\br} \ee^{-i\omega t+i\bt{q}\cdot\bt{r}} \chi(t,\br)\\
        &\bar{\chi}_{\bq,\omega} = \int_{t,\br}\ee^{i\omega t-i\bq\cdot\br}  \bar{\chi}(t,\br)\;,
    \end{aligned}
\end{equation}
and similarly for all the other physical and response fields. 

For simplicity, we use the next section to illustrate our methods on
equilibrium models, for which most of the physics is known with
varying degrees of accuracy. Our approach will not only fill some gaps
in existing derivations, but will also lead to new results for some
equilibrium dynamics.

\section{Equilibrium}\label{sec:equilibrium}
We adopt the terminology of the Hohenberg-Halperin classification \cite{HalperinHohenberg1977} for equilibrium dynamics in the presence of a Landau free energy, in which models are attributed letters based on their conservation laws and coupling with external fields, conserved or not. We start at the beginning of the complexity ladder by considering model A, where the field has a purely relaxational dynamics.

\subsection{Nonconserved relaxational dynamics: model A}\label{sec:A}
\subsubsection{Deriving the Edwards-Wilkinson equation}\label{sec:AL}
We consider the dynamics defined by Eq.~\eqref{eq:motion} with the nonequilibrium drive set to $w=0$:
\begin{equation}\label{eq:A}
    \p_t\phi= -\frac{\delta F}{\delta\phi}+\sqrt{2T}\eta(\br,z,t)\;,
\end{equation}
where $\eta$ is a $\delta$-correlated Gaussian white noise and $F[\phi]$ is given in Eq.~\eqref{eq:F}.  The steady--state mean-field profile satisfies Eq~\eqref{eq:MF}, which reads
\begin{equation}\label{eq:AMF}
    0 = \p_z^2 m_c -f'(m_c)\;.
\end{equation}
For 
\begin{equation}\label{eq:f_standard}
    f(\phi) = -\frac{\tau}{2}\phi^2 + \frac{g}{4!}\phi^4\;,
\end{equation}
the mean-field profile connecting the phases reads 
\begin{equation}\label{eq:profilmc}
    m_c(z) = \sqrt{\frac{6\tau}{g}} \text{tanh}\argp{\sqrt{\frac{\tau}{2}}z}\;.
\end{equation}

Introducing the linear Hermitian operator
\begin{equation}\label{eq:DefOmega0}
    \Omega_0\equiv -\p_z^2 + f''(m_c)\;,
\end{equation}
which stems from expanding $\delta F/\delta \phi$ around the mean-field profile $m_c$, the Janssen-De Dominicis
action can be split between quadratic and non-quadratic parts, as $S=S_0 + (S-S_0)$, with
 \begin{equation}\label{eq:AquadS}
    \begin{split}
       S_0 = \frac{1}{T}\int_{t,\br,z}&
       \Big\{\chib\big[-(\partial_t h-\bnabla_\br^2 h) \partial_z m_c 
       \\&
       + (\p_t-\bnabla_\br^2 +\Omega_0)\chi^\perp\big] - \chib^2 \Big\}\;.
    \end{split}
\end{equation}

\if\wentzel1{
The full action then reads:
\begin{align}
       S = \frac{1}{T}&\int_{t,\br,z}
       \Big\{\chib\big[-(\partial_t h-\bnabla_\br^2 h) \partial_z m_c 
       \notag\\
       &+(\p_t-\bnabla_\br^2 +\Omega_0)\chi^\perp\big] - \chib^2 \notag\\
       &+ \chib\Big[- (\partial_t h-\bnabla_\br^2 h) \partial_z \chi^\perp - (\bnabla_\br h)^2 \partial_z^2 m_c\notag\\
       &+ 2\bnabla_\br h\cdot\bnabla_\br \partial_z \chi^\perp 
       - (\bnabla_\br h)^2 \partial_z^2 \chi^\perp\notag\\
       &+f'(m_c+\chi^\perp)-f'(m_c)-f''(m_c)\chi^\perp\Big]\Big\}\notag\\
       &- \int_{t,\br}\ln\left|\argx{\p_z u_0,m_c} + \argx{\p_z u_0,\chi^\perp}\right|\;.\label{eq:actionAfull}
\end{align}
}\fi

In the low-$T$ regime, we truncate $S$ to quadratic order, 
$S\simeq S_0$. Next, we introduce the evolution operator
\begin{equation}\label{eq:GammaA}
    \Gamma=\p_t-\bnabla_r^2 +\Omega_0\;,
\end{equation}
using the definition of the scalar product~\eqref{eq:SP}, and decomposing $\chib$ as in Eq.~\eqref{eq:DecompChib} then leads to
\begin{align}
       S = \frac{1}{T}\int_{t,\br}&
       \Big\{\hb (\p_t h - \bnabla_\br^2h) \argx{\p_z m_c,u_0} - \hb^2 \argx{u_0,u_0}\notag\\
       &+\argx{\chib^\perp, \Gamma \chi^\perp} - \argx{\chib^\perp,\chib^\perp}\notag\\
        & -\hb\argx{u_0,\Omega_0\chi^\perp}+ 2\hb \argx{u_0, \chib^\perp}\Big\}\;.\label{eq:quadSA_halfdecoup}
\end{align}
where we have used $\langle \chib^\perp,\p_z m_c\rangle=\argx{\chi^\perp,u_0}=0$ and that $u_0(z)$ is independent of $t,\br$.
At this stage, we want to integrate over $\chi^\perp$ and $\chib^\perp$ to obtain an action involving solely $h$ and $\hb$. 
Inspection of Eq.~\eqref{eq:quadSA_halfdecoup} shows this to be unnecessary: 
the only couplings between the $\hb,h$ and $\chib^{\perp},\chi^\perp$ fields are the terms on the third line that can be made to vanish by a suitable choice of the function $u_0(z)$. 

To proceed, we state a few known facts about the  spectral properties of $\Omega_0$, which are well-known from the study of solitons in field theory for the explicit function $f$  of Eq.~\eqref{eq:f_standard}~\cite{Dashen1974_Nonperturbative,goldstone1975quantization,Gervais1975_Perturbation,rajaraman1975some,ohta1977renormalization}. Its spectrum comprises two discrete eigenvalues, $\lambda_0= 0$ and $\lambda_1 = 3\tau/2$, along with a continuum of eigenvalues $\tilde\lambda_k=\tau(2+k^2)$ indexed by $k\in\mathbb{R}$.  All the corresponding orthonormal eigenfunctions are known, but we only mention explicitly the first two, which will be used below:
\begin{equation}\label{eq:Om0_eigen}
    \begin{aligned}
       & e_0(z) = \frac{1}{\sqrt{\sigma}}\partial_z m_c\\
       & e_1(z) = \sqrt{\frac{3}{2}\sqrt{\frac{\tau}{2}}}\frac{\text{sinh}\argp{\sqrt{\frac{\tau}{2}} z}}{\text{cosh}\argp{\sqrt{\frac{\tau}{2}} z}^2}\;,
    \end{aligned}
\end{equation}
where $\sigma = \argx{\partial_z m_c, \partial_z m_c}$ is a normalization constant. Then, $\Omega_0$ reads
\begin{equation}
    \Omega_0 =\frac {3\tau}2\left|e_1\right>\left<e_1\right| +\int_k  \tilde \lambda_k\left|\tilde e_k\right>\left<\tilde e_k\right|\;,\label{eq:pabo}
\end{equation}
which, for clarity, we write as
\begin{equation}
    \Omega_0 = \sum\limits_i \lambda_i\left|e_i\right>\left<e_i\right|\;,
\end{equation}
where we have included $i=0$ since $\lambda_0=0$.
Note that the spectrum of $\Omega_0$ has not been determined for all choices of $f$, apart from $e_0=\p_z m_c /\sqrt{\sigma}$.
The latter is always associated to the zero eigenvalue,
since $\Omega_0$ is the linear operator obtained by
expanding $\delta F/\delta \phi$ around the mean-field
profile $m_c$.

Using the orthonormal basis of eigenvectors of $\Omega_0$ we expand $\chi^\perp$ and $\chib^\perp$ as
\begin{align}
    \chi^\perp&=\sum_i c_i(\br,t) e_i(z)\label{eq:chiDecompEigenv}\\
    \chib^\perp&=\sum_{i>0} \bar c_i(\br,t) e_i(z)\label{eq:chibDecompEigenv}\;,
\end{align}
where we have used that $\chib^\perp$ is orthogonal to $e_0=\p_z m_c/\sqrt{\sigma}$, leading to
\begin{align}\label{eq:AquadS_halfdecoup_decomp}
       S = \frac{1}{T}\int_{t,\br}&
       \Big\{\hb (\p_t h - \bnabla_\br^2h) \argx{\p_z m_c,u_0} - \hb^2 \argx{u_0,u_0}\notag\\
       &+\sum_{i>0}[\bar c_i(\p_t-\bnabla_\br^2 + \lambda_i)c_i - \bar c_i^2]\notag\\
        & -\sum_i\lambda_i\argx{u_0,e_i}\hb \,c_i + 2\sum_{i>0}\argx{u_0,e_i}\hb \,\bar c_i \Big\}\;
\end{align}
Choosing $u_0=e_0/\sqrt{\sigma}$~\footnote{Note that, for this choice of $u_0$, the projector $\Pi$ defined in Eq.~\eqref{eq:defPi} is the Hermitian projector orthogonal to $\p_z m_c$:
\begin{equation}
    \Pi = \mathds 1 - \frac{\left|\p_z m_c\right>\left<\p_z m_c\right|}{\sigma}\;.
\end{equation}} then makes the last line vanish and the actions of $\hb,h$ decouple from that of the modes $\bar c_i, c_i$, which describe bulk fluctuations. One then readily obtains
\begin{align}\label{eq:AquadS_final}
       S = \frac{1}{T}\int_{t,\br}&
       \Big\{\hb (\p_t h - \bnabla_\br^2h) - \frac{\hb^2}{\sigma} \notag\\
       &+\sum_{i>0}[\bar c_i(\p_t-\bnabla_\br^2 + \lambda_i)c_i - \bar c_i^2]\Big\}\;.
\end{align}
The integration over $\chib^\perp, \chi^\perp$ becomes trivial and yields
\begin{align}\label{eq:A_Sh}
       S_h[\hb,h] = \frac{1}{T}\int_{t,\br}
       \hb (\p_t h - \bnabla_\br^2h) - \frac{\hb^2}{\sigma}\;.
\end{align}

To determine the path probability
\begin{equation}
    \mathbb{P}[h] = \frac 1 Z \int \mathscr{D}\hb \, \ee^{-S_h[\hb,h]}\;,
\end{equation}
we use a Hubbard-Stratonovich transformation
\begin{equation}
        \mathbb{P}[h] =  \frac 1 Z \int \mathscr{D}\tilde\eta \, \mathscr{D}\hb \, \ee^{-S_\eta[\hb,h,\tilde\eta]}
\end{equation}
where
\begin{equation}
    \begin{split}
        S_\eta[\hb,h,\tilde\eta] = \int_{t,\br}&\Big[\frac{1}{2}\tilde\eta^2 - \sqrt{\frac{2}{\sigma T}}\hb\tilde\eta+ \frac{1}{T}\hb(\p_t h - \bnabla_\br^2h)\Big]\;.
    \end{split}
\end{equation}
Integrating over $\bar h$ then leads to
\begin{equation}
\begin{split}
        \mathbb{P}[h] =& \frac1Z \int \mathscr{D}\tilde\eta \; \delta\Big[\p_t h - \bnabla_\br^2h-\sqrt{\frac{2 T}{\sigma}}\tilde \eta\Big]\,\\& \times\exp\Big[- \frac 1 2 \int_{t,\br} {\tilde\eta}^2\Big] \;,           
\end{split}
\end{equation}
which is the probability density of a Gaussian process. One then finds the  Langevin equation for $h$:
\begin{equation}\label{eq:A_Inteq}
    \p_t h(\br,t) = \bnabla_\br^2 h(\br,t) + \sqrt{\frac{2T}{\sigma}}\tilde{\eta}(\br,t)\;,
\end{equation}
with $\argx{\tilde{\eta}(\br,t)\tilde{\eta}(\br,t)} = \delta(t-t')\delta^{(d-1)}(\br-\br')$. As discussed in Appendix~\ref{sec:choiceofu0}, choosing a different form for $u_0$ leads to a consistent equation in the small $\bq$ limit, at the cost of heavier algebra.   We emphasize that with an interface definition relying on the choice $u_0=\p_z m_c/\sigma$, Eq.~\eqref{eq:A_Inteq} is exact to all orders in gradients (but is still lacking nonlinearities).

Interestingly, Eq.~\eqref{eq:AquadS_final} not only characterizes the relaxation of the interface, but also the behavior of the bulk fluctuation $\chi^\perp$. 
Each of its components $c_i$ also follows a Langevin equation given by:
\begin{equation}
    \p_t c_i(\br,t) = (\bnabla_\br^2 -\lambda_i)c_i(\br,t) + \sqrt{2T}\tilde\eta_i(\br,t)\;.
\end{equation}
The finite relaxation rate $\lambda_i$ makes the bulk fluctuations decay much faster than the interface fluctuations. 
As we show in Sec.~\ref{sec:NLmodelA}, our framework can be used to derive the non-linear terms in the evolution for $h$. 
Let us first show how our approach allows understanding why the ansatz~\eqref{eq:rottenansatz} is inconsistent but how, nevertheless, it can be used to derive Eq.~\eqref{eq:A_Inteq} when paired with the proper protocol. 

\subsubsection{Fluctuations and simplified approach}\label{sec:A_ansatz}
Now that we have worked out a controlled procedure to derive the interface dynamics, let us come back to the discussion of the ansatz of Eq.~\eqref{eq:rottenansatz}. For simplicity, we stick to the choice $u_0=\p_z m_c/\sigma$, but our discussion holds beyond that case.

First, the ansatz of Eq.~\eqref{eq:rottenansatz} is quantitatively wrong 
because the fluctuation $\chi$ is in general not negligible.
It is the very presence of this bulk fluctuation that dictates how the dependency on the $z$ coordinate should be eliminated from (the linearized version of) Eq.~\eqref{eq:modelAansatz}, as detailed below.

We can now {\it a posteriori} understand why a specific manipulation using the ansatz Eq.~\eqref{eq:rottenansatz} as a starting point is able to lead to the correct result---at least to linear order---while other protocols give the wrong dynamics. 
Let us examine the protocols implemented to obtain Eq.~\eqref{eq:hfakeKPZ} and \eqref{eq:EWansatz} from the ansatz of Eq.~\eqref{eq:rottenansatz}. Both start from Eq.~\eqref{eq:modelAansatz}, which has set $\chi=0$.
Instead, we can inject in Eq.~\eqref{eq:A} the full decomposition $\phi(\br,z,t) = m_c(z-h) + \chi(\br,z-h,t)$, together with the interface definition of Eq.~\eqref{eq:definterface}, to get to lowest order in $\sqrt T$:
\begin{equation}\label{eq:A_EoM}
\begin{split}
    (\p_t -\bnabla_\br^2 + &\Omega_0)\chi(\br,z,t) =  \sqrt{2T}\eta(\br,z+h,t)\\
    &+[\p_t h(\br,t) -\bnabla_\br^2 h(\br,t)]\p_z m_c(z)
\end{split}
\end{equation}
with $\Omega_0$ the operator defined in Eq.~\eqref{eq:DefOmega0}. To get a closed equation for $h$, we need to eliminate the field $\chi$. Projecting Eq.~\eqref{eq:A_EoM} onto $\p_z m_c$ does precisely that. Since $\p_z m_c$ is independent from both $t$ and $\br$, and since $\Omega_0\p_z m_c = 0$, we have
\begin{align}
    \langle \p_z m_c,     (\p_t -\bnabla_\br^2)\chi \rangle &= (\p_t -\bnabla_\br^2) \langle \p_z m_c,\chi \rangle=0\\
    \langle \p_z m_c, \Omega_0 \chi \rangle &= \langle \Omega_0 \p_z m_c,\chi\rangle=0\;,
\end{align}
where we have used that $\Omega_0$ is Hermitian as well as the definition of Eq.~\eqref{eq:definterface}. The projection of the left-hand side of Eq.~\eqref{eq:A_EoM} onto $\p_z m_c$ thus vanishes.  Projecting the right-hand side of Eq.~\eqref{eq:A_EoM} onto $\p_z m_c$ is exactly the second protocol presented in section \ref{sec:ansatz} at linear order, which led to the right dynamics for $h$. Any other protocol that would not eliminate $\chi$ would not close the equation on $h$. Then, having neglected $\chi$ at the level of the ansatz of Eq.~\eqref{eq:rottenansatz} would lead to the wrong interface dynamics.

Note that, as discussed in Sec.~\ref{sec:AMANL}, the ansatz of Eq.~\eqref{eq:rottenansatz} also fails to predict the correct nonlinear terms.

\subsubsection{Beyond a linear equation}\label{sec:NLmodelA}
As alluded to before, nonlinear changes of the paths must be performed with care, since the chain rule does not apply in a straightforward manner~\cite{de2022path}. The corresponding extra contributions are similar to---though somewhat more complex than---those appearing in the famous It\=o's lemma, and we shall write them explicitly further down. Without these additional terms, the full action would read:
\begin{equation}\label{eq:actionAfull}
\begin{split}
       S = \frac{1}{T}&\int_{t,\br,z}
       \Big\{\chib\big[-(\partial_t h-\bnabla_\br^2 h) \partial_z m_c 
       \notag\\
       &+(\p_t-\bnabla_\br^2 +\Omega_0)\chi^\perp\big] - \chib^2 \notag\\
       &+ \chib\Big[- (\partial_t h-\bnabla_\br^2 h) \partial_z \chi^\perp - (\bnabla_\br h)^2 \partial_z^2 m_c\notag\\
       &+ 2\bnabla_\br h\cdot\bnabla_\br \partial_z \chi^\perp 
       - (\bnabla_\br h)^2 \partial_z^2 \chi^\perp\notag\\
       &+f'(m_c+\chi^\perp)-f'(m_c)-f''(m_c)\chi^\perp\Big]\Big\}\notag\\
       &- \int_{t,\br}\ln\left|\argx{\p_z u_0,m_c} + \argx{\p_z u_0,\chi^\perp}\right|\;.
\end{split}
\end{equation}
There is still a time derivative in the non-Gaussian part of the 
action, as well as the exponentiated Jacobian, in Eq.~\eqref{eq:actionAfull}.
Following \cite{Gervais1975_Extended,Kawasaki1982_KineticDrumhead1}, we eliminate them from $S_{\text{NG}}=S-S_0$  by 
changing variables from $\hb$ to $\hb'$, defined by 
\begin{equation}
    \hb = \frac{\hb' + \argx{\chib^\perp, \p_z \chi^\perp}}{1 + \argx{\p_z \chi^\perp, u_0}}\;.
\end{equation}
Several remarks are in order: first, this change of variables entirely eliminates $\chib(\p_t h -\bnabla_\br^2 h)\p_z\chi^\perp$ in $S_{\rm NG}$. It also induces a Jacobian that exactly cancels $J$, the one generated by the introduction of the definition \eqref{eq:definterface}.

However, to go beyond the linear equation for $h$, issues related to time-discretization of the action cannot be ignored anymore. A proper construction of the latter leads to~\cite{de2022path}:
\begin{equation}\label{eq:PIW_Sfull}
    \begin{split}
        S = &\frac{1}{T} \int_{t,\br} \Big\{\hb'(\p_t h-\bnabla_\br^2 h) + \argx{\chib^\perp, (\p_t -\bnabla_\br^2 + \Omega_0)\chi^\perp}\\
        &+  \Big<-\frac{\hb'+\argx{\chib^\perp,\p_z \chi^\perp}}{1 + \argx{\p_z \chi^\perp,u_0}} u_0+\chib^\perp,  \\
        &\;\frac{T}{\sigma(1+ \argx{\p_z \chi^\perp,u_0})^2}(\p_z^2 m_c + \p_z^2 \chi^\perp)\\
        &- \frac{2T}{\sigma(1+ \argx{\p_z \chi^\perp,u_0})^2}\p_z\Pi\p_z \chi^\perp\\
        &- (\bnabla_\br h)^2 (\p_z^2 m_c + \p_z^2 \chi^\perp) + 2\bnabla_\br h\cdot \p_z\bnabla_\br \chi^\perp\\
        &+f'(m_c+\chi^\perp)-f'(m_c) -f''(m_c) \chi^\perp\Big>\\
        &- \frac{1}{\sigma}\Big(\frac{\hb'+\argx{\chib^\perp,\p_z \chi^\perp}}{1 + \argx{\p_z \chi^\perp,u_0}}\Big)^2-\argx{\chib^\perp,\chib^\perp}\Big\}\;,
    \end{split}
\end{equation}
where $\Pi$ was defined in Eq.~\eqref{eq:defPi}.

As $T\to 0$, the full action in Eq.~\eqref{eq:PIW_Sfull} can be expanded beyond quadratic order, keeping in mind that each field contributes $\sim O(T^{1/2})$. With our choice of $u_0$, the quadratic action for the fields is
\begin{align}\label{eq:S0forNL}
       S_0 = \frac{1}{T}\int_{t,\br}&
       \Big\{\hb' (\p_t h - \bnabla_\br^2h) - \frac{\hb'^2}{\sigma} \notag\\
       &+\argx{\chib^\perp, \Gamma\chi^\perp} - \argx{\chib^\perp,\chib^\perp}\Big\}\;,
\end{align}
which is exactly the action we studied in the first part, such that all previous results on the Gaussian action $S_0$ and the linear interface dynamics still hold.
From Eq.~\eqref{eq:S0forNL}, we see that the fields $\chib^\perp,\chi^\perp$  have zero average~\footnote{
Strictly speaking, the averages verify
\begin{align}
    &(-\p_t -\bnabla_\br^2 + \Omega_0)\langle\chib^\perp\rangle=0\;,\\
    &(\p_t -\bnabla_\br^2 + \Omega_0)\langle{\chi^\perp}\rangle = 2\langle \chib^\perp\rangle\;,
\end{align}
which could in principle have nonzero components. For $\chib^\perp$, the nonzero solutions diverge so are excluded, but there could be exponential relaxations in $\chi^\perp$. They would be completely controlled by the initial conditions, which we set to zero, effectively making the averages of the fields vanish. 
}, which is convenient for the perturbation expansion we are about to set up. 
\if\wentzel1{However, before we implement it, note that there is 
still a time derivative in the non-Gaussian part of the 
action in Eq.~\eqref{eq:actionAfull}.
Therefore the final step is to eliminate the time 
derivative in $S_{\text{NG}}=S-S_0$. Following 
\cite{Kawasaki1982_KineticDrumhead1}, we do this by 
changing variables from $\hb$ to $\hb'$, defined by 
\begin{equation}
    \hb = \frac{\hb' + \argx{\chib^\perp, \p_z \chi^\perp}}{1 + \argx{\p_z \chi^\perp, u_0}}\;.
\end{equation}
Several remarks are in order: first, this change of 
variables entirely eliminates $\chib(\p_t h -\nabla_\br^2 
h)\p_z\chi^\perp$ in $S_{\rm NG}$. 
It also leaves $\hb=\hb' + O(T)$ unchanged to lowest order $\sqrt{T}$,
such that all previous results on the Gaussian action $S_0$ and the linear interface dynamics still hold.
Finally, this change of variables induces a Jacobian that exactly cancels $J$, the one generated by the introduction of the definition \eqref{eq:definterface}.

To leading order in $\sqrt T$, the non-Gaussian action then reads
\begin{equation}\label{eq:Aexpansion}
\begin{aligned}
    S_{\text{NG}} = &\frac 1 T \int_{t,\br,z}
       \Big\{\big(-\hb' u_0 + \chib^\perp \big)\times\\
       &\big[- (\bnabla_\br h)^2 \partial_z^2 m_c+ 2\bnabla_\br h\cdot\bnabla_\br \partial_z \chi^\perp \\
      &+ \frac{f'''(m_c)}{2}(\chi^\perp)^2\big]+ 2\hb'^2\argx{\p_z \chi^\perp, u_0}u_0^2 \\
      &-2\hb'\argx{\chib^\perp, \p_z \chi^\perp}u_0^2  + O(T^2)\Big\}\;.
\end{aligned}
\end{equation}}\fi

To leading order in $\sqrt T$, the non-Gaussian action then reads
\begin{equation}\label{eq:Aexpansion}
    \begin{split}
        S_{\text{NG}} = &\frac{1}{T} \int_{t,\br} \Big<-\hb' u_0+\chib^\perp,  \frac{T}{\sigma}\p_z^2 m_c \\
        &- (\bnabla_\br h)^2 \p_z^2 m_c + 2\bnabla_\br h\cdot \p_z\bnabla_\br \chi^\perp\\
        &+\frac{f'''(m_c)}{2}(\chi^\perp)^2\Big>\\
        &+\frac{2\hb'^2}{\sigma}\argx{\p_z \chi^\perp,u_0}-\frac{2\hb'}{\sigma}\argx{\chib^\perp, \p_z \chi^\perp}\;.
    \end{split}
\end{equation}

Let us denote by $S_0'$ the second line of Eq.~\eqref{eq:S0forNL}, which is the quadratic action for the fields $\chib^\perp$, $\chi^\perp$. We then implement a cumulant expansion with respect to $\ee^{-S_0'[\chib^\perp,\chi^\perp]}$ as:
\begin{equation}
    \int{\mathscr D}\chib^\perp{\mathscr D}\chi^\perp\ee^{-S_0'[\chib^\perp,\chi^\perp]-S_\text{NG}}=\ee^{-\langle S_\text{NG} \rangle_{0}'+\frac 12 \langle S_\text{NG}^2 \rangle_{c,0}'+\ldots}
\end{equation}
where $\langle\ldots\rangle_0'$ refers to an average over the $\chi^\perp$, $\chib^\perp$ fields with the weight $\ee^{-S_0'[\chib^\perp,\chi^\perp] }$,
keeping the fields $h$ and $\hb'$ fixed. 
Note that, due to the scaling of the fields as $\sqrt{T}$, all the terms in the cumulant expansion are small as $T\to 0$. 
We now determine $\argx{S_{\text{NG}}}_0'$ explicitly, to lowest order in $\sqrt T$. 
Any odd combination of the $\chi^\perp$ and $\chib^\perp$ fields lead to vanishing averages.
Furthermore, two-point correlations can be computed explicitly using the decompositions on the eigenvectors of $\Omega_0$ \eqref{eq:chiDecompEigenv},
\eqref{eq:chibDecompEigenv}:
\begin{align}
    \argx{\chib^\perp(\br,z,t)\chib^\perp(\br',z',t')}_0' &= 0\\
    \argx{\chib^\perp(\br,z,t)\chi^\perp(\br',z',t')}_0' & =T \sum_{i>0} e_i(z)\,e_i(z')\nonumber\\
     & \hspace{-1.7cm}\times\int_\bq \, \ee^{i\bq\cdot(\br-\br')} \, \ee^{-(\bq^2 + \lambda_i)(t'-t)}\Theta(t'-t) \label{eq:A_NGL_chibchi} \\
    \argx{\chi^\perp(\br,z,t)\chi^\perp(\br',z',t')}_0' &=T \sum_{i> 0}e_i(z)\,e_i(z')\nonumber\\
    &\hspace{-1.7cm}\times\int_\bq \, \ee^{i\bq\cdot(\br-\br')}\frac{1}{\bq^2+\lambda_i}\ee^{-(\bq^2 + \lambda_i)|t'-t|}\;.\label{eq:A_NGL_chi2}
\end{align}
Notice that $\argx{\chib^\perp\chi^\perp}_0'$ vanishes at equal time due to causality, as we are working in the It\=o convention. To leading order in $\sqrt{T}$, this leaves us with
\begin{equation}
    \langle S_\text{NG} \rangle_{0}'= -\frac{1}{2\sigma T}\int_{t,\br}\hb'\argx{\p_z f''(m_c),\argx{(\chi^\perp)^2}_0'}
\end{equation}
This correction to the Gaussian action $S_0$ is proportional to $\hb'$, which will thus contribute a  drift to the dynamics of $h$ with velocity
\begin{equation}\label{eq:expressioncKS}
    c=\frac{1}{2\sigma}\argx{\p_z f''(m_c),\argx{(\chi^\perp)^2}_0'}\;.
\end{equation} 
From Eq.~\eqref{eq:A_NGL_chi2}, we see that $\argx{(\chi^\perp)^2(\br,z,t)}_0'$---and thus $c$---is independent of $\br,t$ so that the velocity $c$  is constant. The evolution equation for the interface to the next order in temperature thus takes the form
\begin{equation}\label{eq:A_Inteq_NL}
    \p_t h(\br,t) = c + \bnabla_\br^2 h(\br,t) + \sqrt{\frac{2T}{\sigma}}\tilde{\eta}(\br,t)\;.
\end{equation}

The ballistic propagation of an interface in the presence of a degenerate but asymmetric double-well potential was first noted in~\cite{costantini2001asymmetric} and further analyzed in~\cite{Kado2024_MicroCutoff}. As in \cite{Kado2024_MicroCutoff}, we rewrite $c$ as
\begin{equation}\label{eq:newc}
c = \frac{\argx{\argc{f''(m_c)(\chi^\perp)^2 -(\p_z \chi^\perp)^2 + (\bnabla_\br\chi^\perp)^2}_{-\infty}^{+\infty}}_0'}{2\sigma}\;,
\end{equation}
where the limits refer to $z=\pm \infty$.
Deep in the bulk phases, the system is isotropic. In two space dimensions, the second and third contributions in Eq.~\eqref{eq:newc} cancel out.
Using Eq.~\eqref{eq:A_NGL_chi2}, we find
\begin{equation}\label{eq:Acfinal}
    c=\frac{T}{2\sigma}\int_\bq\sum_{i>0}\frac{\argc{f''(m_c)e_i^2}_{-\infty}^{+\infty}}{\bq^2+\lambda_i}\;.
\end{equation}

When $f$ is even, as in Eq.~\eqref{eq:f_standard}, the speed $c$ vanishes. Otherwise, the fluctuation-driven motion of the interface thus only occurs when the degenerate minima lie at the bottom of wells with distinct curvatures, in agreement with~\cite{costantini2001asymmetric,Kado2024_MicroCutoff}: fluctuations lift the degeneracy between the two minima. 
For an asymmetric $f$ with degenerate minima, assuming that the operator $\Omega_0$ has similar spectral properties as for the standard $\phi^4$ theory, the sum over eigenvalues $\lambda_i$ is dominated by the integral over the continuous part of the spectrum. 
At high values of $k$, we expect the eigenvectors $\tilde{e}_k$ to behave like plane waves and the eigenvalues to grow as $\tilde\lambda_k\sim k^2$. 
Integration over $q$ and $k$ will therefore lead to a logarithmic divergence which can be regularized by introducing a short-wavelength  cutoff~\cite{Kado2024_MicroCutoff}.

Note that, in three dimensions and more, the gradient terms in Eq.~\eqref{eq:newc} do not generically cancel out. 
One can then use the explicit expression of $\argx{(\chi^\perp)^2}_0'$ in terms of the eigenvectors of $\Omega_0$ to find an estimate for the interface velocity.

As a final remark, we would like to point out that a phase-separated system, whose potential is asymmetric and has degenerate minima, is out of equilibrium, since a drift velocity is generated. In this case, one expects to see the KPZ nonlinearity arise. Showing this in detail goes beyond the scope of this paper. It would require going to higher order in $\sqrt T$ in the expansion~\eqref{eq:Aexpansion} and studying the additional contributions that enter the cumulant expansion. Among these terms, we can already identify one that yields a KPZ term:
\begin{align}
    \frac{1}{T^2}\int_{\br,t}&-\hb (\bnabla_\br h)^2\int_{\br',z,z',t'}\p_z m_c(z)
    \frac{f'''(m_c)}{2\sigma}(z')\notag\\
    &\argx{\p_z^2 \chi^\perp(\br,z,t)[\chib^\perp(\chi^\perp)^2](\br',z',t')}'_{c,0}\;.
\end{align}
As expected, substituting the explicit expression of the average in terms of the eigenmodes of $\Omega_0$ shows that this term vanishes for a symmetric potential $f$, as detailed in Appendix~\ref{sec:passiveKPZ}.

Note that, if we were to continue the expansion leading to Eq.~\eqref{eq:A_Inteq_NL} to higher order in $\sqrt T$, we would also find the equilibrium nonlinearities corresponding to curvature corrections~\cite{Diehl1980,Kawasaki1982_KineticDrumhead1,ohta2025}, as shown in App.~\ref{sec:curvature} for the next order. 

This concludes the presentation of our framework for equilibrium model A. We now turn to more complex models, starting with model B.

\subsection{Conserved relaxational dynamics: Model B}

\subsubsection{Early results on interface dynamics in Model B}
We now study model B, whose dynamics is conserved and reads:
\begin{equation}\label{eq:B}
    \p_t\phi= \bnabla^2\frac{\delta F}{\delta\phi}+\sqrt{2T}\bnabla\cdot\boldsymbol{\eta}(\br,z,t)\;,
\end{equation}
where the components of $\boldsymbol{\eta}$ are independent white noises that obey $\argx{\eta_i(\br,z,t)\eta_j(\br',z',t')} = \delta_{ij}\delta(t-t') \delta(z-z')\delta^{(d-1)}(\br-\br')$. 

The Edwards-Wilkinson dispersion relation $\omega\propto k^2$ found in model A does not hold anymore in model B. Instead, a dispersion relation $\omega\propto k^3$ was predicted first using a phenomenological, hydrodynamic approach~\cite{Langer1977_Studies}. 
A statistical-mechanical derivation, directly based on model B, was developed soon after~\cite{Kawasaki1982_KineticDrumhead2}. 
This provided an expression for the various coefficients entering the interface evolution equation. 
Their better-known expressions in Fourier space where derived subsequently~\cite{Kawasaki1983_Kinetics,Ohta1984_Dynamics}.
Alternative derivations for the dispersion relation were proposed later, using variational~\cite{Jasnow1987_Crossover} or spectral~\cite{Shinozaki1993_Dispersion} approaches.

We now apply our field-theoretical framework to model B, mirroring the derivation employed for model A. 

\subsubsection{Dynamical action for interface dynamics}
\if{
For simplicity, we consider the case where the chemical potential $\mu=\langle f'(\phi)-\bnabla^2 \phi\rangle$ is set to zero. At mean-field level, this leads to an interface profile $m_c(z)$ identical to that of model A, which solves Eq.~\eqref{eq:AMF}. 

Even if a nonzero chemical potential favors one minimum of $f$ over the other, the conservation law warrants the stability of the interface (but the shape of the mean-field profile $m_c$ will be affected). 
For simplicity, we stick to the zero chemical potential case for which $m_c(z)$ is unchanged with respect to Eq.~\eqref{eq:AMF} (the derivation is essentially insensitive to a nonzero chemical potential).}\fi

For model B, the Janssen-De Dominicis action truncated to quadratic order takes the form
\begin{equation}\label{eq:actionB}
    \begin{split}
       S = \frac{1}{T}&\int_{t,\br,z}
       \Big\{\chib\big[-\partial_t h\, \partial_z m_c -\bnabla^2\argp{\bnabla_\br^2 h\, \p_z m_c}       \\
       &+\argp{\p_t-\bnabla(-\bnabla_\br^2 +\Omega_0)}\chi^\perp\big] - (\bnabla\chib)^2\Big\}\;,\\
    \end{split}
\end{equation}
where $m_c$ is the mean-field profile connecting the two bulk phases, which solves
\begin{equation}\label{eq:BMF}
    0 = \p_z^2[-\p_z^2 m_c + f'(m_c)]\;.
\end{equation}
Note that there is no steady-state mean-field current, so the integral of Eq.~\eqref{eq:BMF} also vanishes.
We remind that, as explained in Sec.~\ref{sec:LowNoise}, we have shifted the variables $(\br,z,t)$ as $(\br,z,t)\to (\br,z-h(\br,t),t)$ so that $m_c$ is solely a function of $z$ in Eq.~\eqref{eq:actionB}.

As for model A, the operator $\Omega_0$ is given by Eq.~\eqref{eq:DefOmega0}. To proceed, we work in Fourier space for both the time $t$ and the space coordinates $\br$ along the interface, using the conventions given in Eq.~\eqref{eq:FT}, which leads to:
\begin{align}
       S = \frac{1}{T}&\int_{\omega,\bq,z}
       \Big\{\chib_{\bq,\omega}\Big[-\big(i\omega + \bq^2(\bq^2 - \p_z^2)\big) h_{\bq,\omega}\, \partial_z m_c \notag \\
       &+\argp{i\omega+(\bq^2 - \p_z^2)(\bq^2 +\Omega_0)}\chi^\perp_{\bq,\omega}\Big] \notag\\
       &-\chib_{\bq,\omega}(\bq^2 - \p_z^2)\chib_{-\bq,-\omega} 
      \Big\}\;. 
\end{align}
We then introduce the evolution operator
\begin{equation}\label{eq:GammaB}
    \Gamma_{\bq,\omega} = i\omega + (\bq^2 - \p_z^2) (\bq^2 + \Omega_0)\;.
\end{equation}
Note that, due to the conservation law, $\Gamma$ includes an additional Laplacian compared to its definition in Eq.~\eqref{eq:GammaA} for model A. This leads to additional mathematical difficulties, which can be mitigated by decomposing $\Gamma_{\bq,\omega}^\dagger\chib_{\bq,\omega}$ instead of $\chib_{\bq,\omega}$, as~\footnote{Note that a derivation of the interface dynamics by decomposing $\chib$ instead of $\Gamma^\dagger \chib$ is doable and leads to the same result, at the cost of much heavier algebra, so that we  think that integrating over the fields $\chib$ or $\Gamma^\dagger \chib$ is equivalent.}
\begin{align}
    &\Gamma_{\bq,\omega}^\dagger\chib_{\bq,\omega} = -\tilde h_{\bq,\omega} u_0 + (\Gamma_{\bq,\omega}^\dagger\chib_{\bq,\omega})^\perp, \; \text{with}\notag\\ &(\Gamma_{\bq,\omega}^\dagger\chib_{\bq,\omega})^\perp = \Pi^\dagger \Gamma_{\bq,\omega}^\dagger\chib_{\bq,\omega}\;.\label{eq:Lila}
\end{align}
In this last equality, the $\dagger$ symbol refers to an 
adjoint solely with respect to the integration over $z$. To lighten notations, we introduce
\begin{align}
\mathrm{L}_\bq&= \bq^2 - \p_z^2\label{eq:defLq}\\
\mathrm{H}_\bq &= (\bq^2 + \Omega_0)^{-1}\mathrm{L}_\bq^{-1}\label{eq:defHq}\;.
\end{align}
The action for the fields $\tilde h,h,(\Gamma^\dagger \chib)^\perp, \chi^\perp$ then takes the form
\begin{equation}
    \begin{split}
       S = &\frac{1}{T}\int_{\omega,\bq,z}
       \Big\{\tilde h_{\bq,\omega}h_{\bq,\omega}\p_z m_c u_0 +( \Gamma_{\bq,\omega}^\dagger\chib_{\bq,\omega})^\perp \chi^\perp_{\bq,\omega} \\
       &-\mathrm{L}_\bq(\Gamma_{\bq,\omega}^\dagger)^{-1}\big[-\tilde h_{\bq,\omega} u_0 + (\Gamma_{\bq,\omega}^\dagger\chib_{\bq,\omega})^\perp\big]\times\\
       &\quad(\Gamma_{-\bq,-\omega}^\dagger)^{-1}\big[-\tilde h_{-\bq,-\omega} u_0 + (\Gamma_{-\bq,-\omega}^\dagger\chib_{-\bq,-\omega})^\perp\big]
      \Big\}\;.\\
    \end{split}
\end{equation}
where $(\Gamma_{-\bq,-\omega}^\dagger)^{-1}$ is an inverse with respect to the product over $\bq$ and $\omega$ and the convolution product over $z$,
and we have used 
\begin{equation}
\Gamma_{\bq,\omega}\p_z m_c = [i\omega + \bq^2(\bq^2 - \p_z^2)]\p_z m_c\;.
\end{equation}
The integration over $\chi^\perp$  leads to a $\delta$ constraint in the space of functions orthogonal to $u_0$, which enforces $(\Gamma^\dagger_{\bq,\omega}\chib_{\bq\omega})^\perp = a_{\bq,\omega} u_0(z)$, where $a_{\bq,\omega} $ is an unknown proportionality constant. Since by definition $\argx{(\Gamma^\dagger_{\bq,\omega}\chib_{\bq,\omega})^\perp,\p_z m_c} = 0$, we find $a_{\bq,\omega}=0$. This leads to the final form of the action
\begin{equation}\label{eq:BActionhugly}
\begin{split}
    S_h[&\tilde{h},h]=\frac{1}{T} \int_{\omega,\bq}\Big[\tilde{h}_{\bq,\omega}h_{\bq,\omega}\argx{\p_zm_c,u_0}\\
    &-\tilde{h}_{\bq,\omega}\tilde{h}_{-\bq,-\omega}\argx{(\Gamma^\dagger_{\bq,\omega})^{-1}u_0,\mathrm{L}_\bq(\Gamma^\dagger_{-\bq,-\omega})^{-1}u_0}\Big]\;,
\end{split}
\end{equation}
Let us now turn the action in Eq.~\eqref{eq:BActionhugly} into that of a fluctuating hydrodynamics for $h$ in the large wavelength limit. We note that, aside from requiring that $\argx{\p_z m_c,u_0} \neq 0$, $u_0$ is still arbitrary at this stage.

To simplify the $(\Gamma^\dagger_{\bq,\omega})^{-1}u_0$ contribution, we first note that
\begin{equation}
    \Gamma^\dagger_{\bq,\omega} = (\bq^2 + \Omega_0)\argp{1 + i\omega \mathrm{H}_\bq}\mathrm{L}_\bq\;,
\end{equation}
which leads to
\begin{equation}
    (\Gamma^\dagger_{\bq,\omega})^{-1} = \mathrm{L}_\bq^{-1}\argp{1 + i\omega \mathrm{H}_\bq}^{-1}(\bq^2 + \Omega_0)^{-1}\;.
\end{equation}
A convenient choice for $u_0$ is then, as in model A, $\p_z m_c/\sigma$. Indeed, with this choice we find
\begin{equation}\label{eq:Gu0}
    (\Gamma^\dagger_{\bq,\omega})^{-1}u_0 = \frac{1}{\sigma\bq^2}\mathrm{L}_\bq^{-1}\argp{1 + i\omega \mathrm{H}_\bq}^{-1}\p_z m_c\;.
\end{equation}
For concreteness, we now specialize to the case of the $\phi^4$ theory given in Eq.~\eqref{eq:f_standard}, for which $\Omega_0$ can be diagonalized explicitly, as explained in Sec.~\ref{sec:A}.
The Green's functions appearing in $(\Gamma^\dagger_{\bq,\omega})^{-1}$ and $\mathrm{H}_\bq$ then obey
\if{\begin{align}
        &\mathrm{L}_\bq^{-1}(z,z') = \frac{1}{2q} e^{-q|z-z'|} \;,\label{eq:L-1}\\
        &(\bq^2 + \Omega_0)^{-1}(z,z') = \sum\limits_i (\bq^2 + \lambda_i)^{-1}e_i(z)e_i(z')\;
\end{align}}\fi
\begin{align}
        &(\mathrm{L}_\bq^{-1}\varphi) (z) = \frac{1}{2q} \int_{z'} e^{-q|z-z'|}\varphi(z') \;,\label{eq:L-1}\\
        &[(\bq^2 + \Omega_0)^{-1}\varphi](z) = \int_{z'} \sum\limits_i \frac{e_i(z)e_i(z')}{\bq^2 + \lambda_i}\varphi(z')\;
\end{align}
where $q=|\bq|$.
We note that $\mathrm{L}_\bq^{-1}$ is not diagonal in the $e_i$ basis. Physically, the Laplacian enforces the conservation law, thereby coupling the different eigenfunctions $e_i$. 

In the limit of long wavelengths $q\xi\ll 1$, progress can be made. Upon splitting the $e_0=\sqrt{\sigma}u_0=\p_z m_c/\sqrt{\sigma}$ contribution from the rest, we see that
\begin{equation}\label{eq:B_Hq}
\begin{split}
    {\mathrm H}_\bq \p_z m_c &= \frac{A(q)}{2\sigma q^3} \p_z m_c \\
    &+ \sum\limits_{i>0} \frac{1}{q^2 + \lambda_i}\argx{e_i, \mathrm{L}_\bq^{-1} \p_z m_c} e_i\;,
\end{split}
\end{equation}
where $A(q) = 2q \argx{\p_z m_c, \mathrm{L}_\bq^{-1} \p_z m_c}$, which satisfies $A(0)=(\Delta m_c)^2$. Understanding the $q\to 0$ behavior of the various ingredients entering  Eq.~\eqref{eq:B_Hq} requires some care. A detailed derivation of the leading order and an estimate of the corrections in the $q\to 0$ limit are provided in Appendix~\ref{sec:LnighmareB}. Skipping the mathematical details that show the second line in Eq.~\eqref{eq:B_Hq} to be subleading, we write that
\begin{equation}
{\mathrm H}_\bq \p_z m_c = \frac{A(q)}{2\sigma q^3} \argc{\p_z m_c + O(q^2)}\;.
\end{equation}
For any $\omega$ that is $O(q^3)$ at most, we can deduce that
\begin{equation}\label{eq:Hqchi}
    \argp{1 + i\omega \mathrm{H}_\bq}^{-1}\!\p_z m_c = \argc{1 + \frac{i\omega A(q)}{2\sigma q^3}}^{-1}\!\!\!\!\!\!\argc{\p_z m_c + O(q)}
\end{equation}

We can now replace the scalar product in the second line of Eq.~\eqref{eq:BActionhugly} with its long wavelength approximation  
\begin{equation}
\begin{split}
    \langle(\Gamma^\dagger_{\bq,\omega})^{-1}&u_0,\mathrm L_\bq(\Gamma^\dagger_{-\bq,-\omega})^{-1}u_0\rangle = \\
    &\left| i\omega +\frac{2\sigma q^3}{ A(q)}\right|^{-2} \frac{2q}{A(q)}\argc{1 + O(q)}\;.
\end{split}
\end{equation}
To eliminate the frequency dependence from the contribution to the action $\propto \tilde h_{\bq,\omega}\tilde h_{-\bq,-\omega}$, it proves convenient to introduce the auxiliary field 
\begin{equation}\label{eq:hbarB}
    \hb_{\bq,\omega} = \argp{i\omega + \frac{2\sigma q^3}{A(q)}}^{-1} \tilde{h}_{\bq,\omega}\;.
\end{equation} 
Note that if, before Eq.~\eqref{eq:Gu0}, we had kept  $u_0$ arbitrary instead of choosing $u_0=\p_z m_c/\sigma$, we could have absorbed the extra bit into the change of fields Eq.~\eqref{eq:hbarB}. The action then takes the form
\begin{equation}\label{eq:B_Sforh}
\begin{split}
    S_h[\bar{h},h]=&\frac{1}{T} \int_{\omega,\bq}\Big[\hb_{\bq,\omega}\argp{i\omega + \frac{2\sigma q^3}{A(q)}}h_{\bq,\omega}\\
    &-\hb_{\bq,\omega}\hb_{-\bq,-\omega}\frac{2q}{A(q)}\argc{1+  O(q)}\Big]\;,
\end{split}
\end{equation}
which is equivalent to the Langevin equation
\begin{equation}\label{eq:Binterface}
\p_t h(\bq,t) = - \frac{2\sigma q^3}{A(q)} h(\bq,t) + \sqrt{2T\frac{2q}{A(q)}}\tilde{\eta}(\bq,t)\;,
\end{equation}
with $\argx{\tilde{\eta}(\bq,t)\tilde{\eta}(\bq',t')} = (2\pi)^{d-1}\delta(t-t')\delta^{(d-1)}(\bq+\bq')$. 

We recover the dispersion relation predicted in \cite{Kawasaki1982_KineticDrumhead2,Kawasaki1983_Kinetics,Ohta1984_Dynamics,Jasnow1987_Crossover,Shinozaki1993_Dispersion} as well as the noise derived in \cite{Kawasaki1982_KineticDrumhead2}. As noted in~\cite{Zia1988_Dynamics}, the conservation law correlates the relaxation of the interface to that of the bulk, leading to a slower dynamics than in model A that results in the $q^3$ scaling of the capillary relaxation. 

\begin{figure}
\centering
\includegraphics[width=1.0\columnwidth]{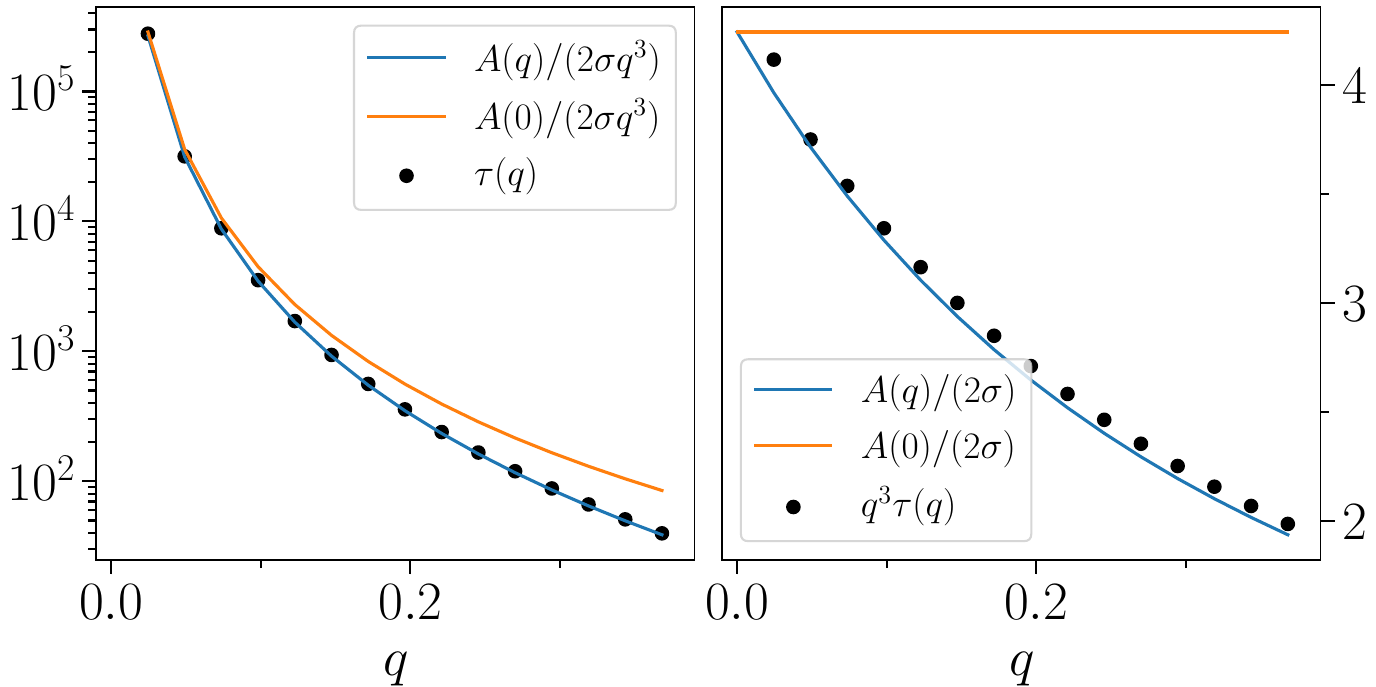}
    \caption{(Left) Relaxation time of a one-dimensional interface in a two-dimensional model B from an initial condition $\frac{0.1}{q} \cos\argp{qr}$, as a function of $q$. The dots are obtained from an exponential fit of the amplitude of the relaxing interface in model B simulations. The full lines are the theoretically predicted values for $\tau(q) = \frac{A(q)}{2\sigma q^3}[1+ O(q)]$, to lowest order in $q$ (orange), or using the full prediction (blue). (Right) Same as left, normalized by $q^3$. Numerical details are given in Appendix~\ref{sec:numerics}.}
    \label{fig:numerics model B}
\end{figure}

We have carried out our derivation of Eq.~\eqref{eq:B_Sforh} for the specific potential of Eq.~\eqref{eq:f_standard}, but the result Eq.~\eqref{eq:B_Sforh} holds for any potential that allows for stable phase separation. We also note that our dispersion relation $i\omega=-\frac{2\sigma q^3}{A(q)}[1+ O(q)]$ exhibits $O(q)$ corrections. In practice, this means that the $q$ dependence of $A(q)$ could be of the same order in $q$ as terms that have been neglected. As briefly conjectured in~\cite{Shinozaki1993_Dispersion} and further motivated in Appendix~\ref{sec:BShino}, it is likely that a more stringent mathematical derivation would show that $i\omega=-\frac{2\sigma q^3}{A(q)}[1+ O(q^2)]$. This suggests that the $A(q)$ is actually valid up to and including its order $q$ correction. This is consistent with our numerics shown in Fig.~\ref{fig:numerics model B}, which compare the numerically measured relaxation time and its theoretically predicted value, either to lowest order in $q$ or taking the full $A(q)$. Agreement beyond $A(0)$ is reported. 

%

\subsubsection{How to proceed using the mean-field ansatz}

As mentioned in Sec.~\ref{sec:ansatz}, there exists a sequence of operations that yields the correct interface dynamics starting from the mean-field ansatz Eq.~\eqref{eq:rottenansatz}, as used in~\cite{Kawasaki1983_Kinetics,Ohta1984_Dynamics,Bray2001_Interface,Fausti2021_Capillary,Besse2023_Interface}. 
In this paragraph, we explain why, in model B also, a flawed ansatz nevertheless yields the right interface equation~\eqref{eq:Binterface} upon using the appropriate recipe.

Starting from model B dynamics in Eq.~\eqref{eq:B}, using the decomposition~\eqref{eq:DefChi}, and linearizing in $h$ and $\chi$, which amounts to working to lowest order in $\sqrt{T}$ and to assuming $h,\chi\sim \sqrt{T}$, leads to
\begin{equation}\label{eq:B_EoM}
    \Gamma_{\bq,\omega}\chi_{\bq,\omega} = \Gamma_{\bq,\omega}\, h_{\bq,\omega}\,\p_z m_c + \sqrt{2T} (-i\bq + \p_z \hat{\bz})\cdot\boldsymbol{{\eta}'}_{\bq,\omega}
\end{equation}
 where $\bt{\eta}(\br,z,t) = \bt{{\eta}'}(\br,z-h,t)$.  
 To eliminate the field $\chi$ and get a closed equation on $h$, we  exploit $\argx{\p_z m_c, \chi} = 0$, which stems from Eq.~\eqref{eq:definterface} for $u_0=\p_z m_c/\sigma$. To this aim, we first invert $\Gamma$ and multiply the resulting equation by $\p_z m_c$, before integrating over $z$. In the long-wavelength limit, we  use
\begin{equation}\label{eq:B_gamma-1}
    (\Gamma_{\bq,\omega}^\dagger)^{-1}\frac{\p_z m_c}{\sigma} \simeq \frac{1}{\sigma q^2}\argp{1 + \frac{i\omega A(q)}{2\sigma q^3}}^{-1} \mathrm{L}_\bq^{-1}\p_z m_c\;.
\end{equation}
Multiplying Eq.~\eqref{eq:B_EoM} by~\eqref{eq:B_gamma-1} and integrating over $z$ then makes the contribution of $\chi$ vanish and leads to an equation for $h_{\bq,\omega}$, which is the Fourier transform in time of Eq.~\eqref{eq:Binterface}. All in all, this amounts to projecting Eq.~\eqref{eq:B_EoM} onto $(\Gamma_{\bq,\omega}^\dagger)^{-1}\frac{\p_z m_c}{\sigma} \propto \mathrm{L}_\bq^{-1}\p_z m_c$.

Thus, discarding $\chi$ from the outset in Eq.~\eqref{eq:DefChi} by using the ansatz~\eqref{eq:rottenansatz} and then projecting the resulting dynamics onto $\mathrm{L}_\bq^{-1}\p_z m_c$ results in the correct dynamics for $h$, since the projection eliminates $\chi$ properly. This protocol, however, is the only one that decouples the dynamics of $h$ and $\chi$, something hard to guess and justify \textit{a priori}. 
In equilibrium, this simplified route to the interface dynamics was introduced for model B in \cite{Kawasaki1982_KineticDrumhead2} 
(see below
Eq.~(2.55) and the last sentence of the appendix) and \cite{Ohta1984_Dynamics,ohta2025}.
It has since then been used out of equilibrium~\cite{Bray2001_Interface,Fausti2021_Capillary,Besse2023_Interface,langford2024theory,Caballero2025_Interface,maire2025hyperuniform,burekovic2026active}, despite the lack of any justification in this case. 

Much like for model A, we could at this stage discuss the emergence of non-linear terms in the interface dynamics for model B. We postpone this discussion to a later work as their computation is quite involved and goes beyond the scope of this paper.

\newcommand{\Lm}{\mathcal{L}}

\subsection{Coupling to a conserved field: Models C and D}
When the field $\phi$ is coupled to a conserved auxiliary field, whether $\phi$ itself is conserved (model D) or not (model C), the derivation of an interface equation follows the same principles. In this section, we make progress with respect to the existing literature~\cite{Ohta1984_Dynamics,Zia1988_Dynamics,Bausch1991_Effects} by deriving the expression of the noise for model C  and by deriving the full interface equation for model D. Models C and D dynamics read \cite{halperin1974renormalization,HalperinHohenberg1977}
\begin{align}
            &\p_t \phi = -D\Lm \funcder{F}{\phi} + \sqrt{2DT}\eta\;,\label{eq:C}\\
            &\p_t \rho = \Lambda \bnabla^2\funcder{F}{\rho} + \sqrt{2\Lambda T}\zeta\;,\label{eq:D}
\end{align}
where $\Lm = \mathrm{1}$ for model C and $\Lm = -\bnabla^2$ for model D \cite{Kawasaki1982_KineticDrumhead1,Kawasaki1982_KineticDrumhead2}. Using $\bx=(\br,z)$, the Gaussian white noises $\eta$ and $\zeta$ satisfy  
\begin{align}
    \argx{\eta(\bx,t)\eta(\bx',t')} &= \Lm \delta(t-t')\delta^{(d)}(\bx-\bx')\;,\\
    \argx{\zeta(\bx,t)\zeta(\bx',t')} &= -\bnabla^2 \delta(t-t')\delta^{(d)}(\bx-\bx')\;.
\end{align}
In this section, the free energy is a functional of both fields, given by \cite{Zia1988_Dynamics,Bausch1991_Effects}
\begin{equation}
    F[\phi,\rho] = \int_{\br,z}\left[\frac{1}{2}(\bnabla\phi)^2 +f(\phi) + \frac{1}{2}(\rho- V(\phi))^2\right]\;,
\end{equation}
where $V$ is a single-well potential. The Janssen-De Dominicis action reads \cite{Zia1988_Dynamics,Bausch1991_Effects}
\begin{equation}\label{eq:S_CD}
\begin{split}
    \mathrm S[\phib,\phi,\rhob,\rho] =&\int_{t,\br,z}\Big[ -DT\phib\mathcal{L}\phib + \Lambda T \rhob\bnabla^2\rhob\\
    &+ \phib\Big(\p_t \phi + D\mathcal{L}\funcder{F}{\phi}\Big)\\
    &+\rhob\Big(\p_t \rho - \Lambda\bnabla^2\funcder{F}{\rho}\Big)\Big]\;.
\end{split}
\end{equation}
The mean-field profiles $\rho_c$ and $m_c$ solve 
\begin{equation}\label{eq:CD_MF}
    \begin{split}
        &\rho_c = V(m_c) \\
        &\Lm_z(-\p_z^2 m_c + f'(m_c))=0\;.
    \end{split}
\end{equation}
In model D, the integral over $z$ of the second equation also vanishes, because the current vanishes in an equilibrium state. We choose to represent the coupled fields  in a vectorial form:
\newcommand{\Phib}{\bar\Phi}
\newcommand{\Psib}{\bar\Psi}
\begin{equation}
    \Phi = \begin{pmatrix}
        \phi\\
        \rho
    \end{pmatrix},\,\,\Phib = \begin{pmatrix}
        \phib\\
        \rhob
    \end{pmatrix}\;.
\end{equation}
As for models A and B, we introduce the interface height $h(\br,t)$ and the deformation $\chi$ of the mean-field profile:
\begin{equation}
    \Phi(\br,z,t) = \Phi_c(z-h) +\chi(\br,z-h,t)\;,
\end{equation}
where we have defined
\begin{equation}
    \Phi_c = \begin{pmatrix}
        m_c\\
        V(m_c)
    \end{pmatrix}\;,
\end{equation}
and we stress that $\chi$ is now a two-dimensional vector. We choose the same definition of the interface as in Eq.~\eqref{eq:definterface}, now written in the form
\begin{equation}
    \int \dd z \,\chi^T(\br,z,t) U_0 = 0\;,
\end{equation}
where
\begin{equation}
    U_0 = 
\begin{pmatrix}
        \p_z m_c/\sigma\\
        0
    \end{pmatrix}\;.
\end{equation}
We note that, as before, to lowest order in $q$ the choice of $U_0$ does not impact the derivation, as long as it is not orthogonal to $\partial_z \Phi_c$. The projector $\Pi$ that we use to split $\chi$ between longitudinal and transverse fluctuations is now a $2\times 2$ matrix with operator-valued elements
\begin{equation}
\begin{split}
    \Pi &=  \mathbb I - \left|\p_z \Phi_c\right> \left<U_0\right| \\
    &= \begin{pmatrix}
        \delta(z-z')-\frac{\p_z m_c(z)\p_z m_c(z')}{\sigma} & 0\\
        -\frac{\p_z V(m_c)(z)\p_z m_c(z')}{\sigma} & \delta(z-z')
    \end{pmatrix}\;.
\end{split}\end{equation}
In this section, the notation $\langle\ldots,\ldots\rangle$ includes both the scalar product of Eq.~\eqref{eq:SP} and the vector dot product. The corresponding  $\dagger$ symbol applies to both  products. We also introduce $\chib(\br,z,t) = \Phib(\br,z+h,t)$. 
Then, the action Eq.~\eqref{eq:S_CD} truncated to quadratic order takes the more compact form
\begin{align}
        S = \frac{1}{T}\int_{t,\br,z} \big\{&\chib^T\big[\p_t + \mathbb L\mathbb O\big](-h\, \p_z \Phi_c + \chi^\perp)\notag\\
        &- \chib^T \mathbb L\chib\big\}\;,
\end{align}
with 
\begin{equation}
    \mathbb L = \begin{pmatrix}
        D\Lm & 0\\
        0 &- \Lambda \nabla^2
    \end{pmatrix}
\end{equation}
and
\begin{equation}\label{eq:DefO}
    \mathbb O = \begin{pmatrix}
        -\nabla_\br^2+\Omega & -V_c'\\
        -V_c' & 1
    \end{pmatrix} = \mathbb O_0 + \begin{pmatrix}
        -\nabla_\br^2 & 0\\
        0 & 0
    \end{pmatrix}\;.
\end{equation}
Here, $V_c'$ is a short-hand notation for $V'(m_c)$,  $\Omega = -\p_z^2 + f''(m_c) + V_c'^2 $ and we stress that the Hermitian operator $\mathbb O_0$ satisfies $\mathbb O_0 \p_z \Phi_c = 0$. We Fourier transform in $t$ and $\br$, and introduce 
\begin{equation}\label{eq:defG}
    \mathbb {G}_{\bq,\omega} = i\omega\mathbb I + \mathbb L_\bq\mathbb O_\bq\;,
\end{equation}
which plays for models C and D the role of $\Gamma$ in Eq.~\eqref{eq:GammaB} for model B. This
leads to:
\begin{align}
        S = \frac{1}{T}\int_{\omega,\bq,z}& \Big[\chib_{\bq,\omega}^T\mathbb{G}_{\bq,\omega}(-h_{\bq,\omega}\, \p_z \Phi_c + \chi^\perp_{\bq,\omega})\notag\\
        &- \chib_{\bq,\omega}^T \mathbb L_\bq\chib_{-\bq,-\omega}\Big]\;.
\end{align}
In the same spirit as in Eq.~\eqref{eq:Lila}, we decompose
\begin{align}
    &\mathbb{G}_{\bq,\omega}^\dagger\chib_{\bq,\omega} = -\tilde h_{\bq,\omega} U_0 + (\mathbb{G}_{\bq,\omega}^\dagger\chib_{\bq,\omega})^\perp,\;\text{with}\notag \\
    &(\mathbb{G}_{\bq,\omega}^\dagger\chib_{\bq,\omega})^\perp=\Pi^\dagger \mathbb{G}_{\bq,\omega}^\dagger\chib_{\bq,\omega}\;,
\end{align}
which turns the action into
\begin{align}
        S =& \frac{1}{T}\int_{\omega,\bq} \Big\{\tilde h_{\bq,\omega}h_{\bq,\omega} + \argx{(\mathbb{G}_{\bq,\omega}^\dagger\chib_{\bq,\omega})^{\perp},\chi^\perp_{\bq,\omega}}\notag\\
        &- \Big<\mathbb L_\bq\argp{\mathbb{G}_{\bq,\omega}^\dagger}^{-1}( -\tilde h_{\bq,\omega} U_0 + (\mathbb{G}_{\bq,\omega}^\dagger\chib_{\bq,\omega})^\perp),\notag\\
        &\argp{\mathbb{G}_{-\bq,-\omega}^\dagger}^{-1}( -\tilde h_{-\bq,-\omega} U_0 + (\mathbb{G}_{-\bq,-\omega}^\dagger\chib_{-\bq,-\omega})^\perp)\Big>\Big\}\;,
\end{align}
where we have used that $\argx{U_0, \p_z \Phi_c} = 1$.
Performing the integration over $\chi^\perp$ yields a $\delta$ constraint in the space of functions orthogonal to $U_0$, which implies $(\mathbb{G}_{\bq,\omega}^\dagger\chib_{\bq,\omega})^{\perp} = a_{\bq,\omega}U_0$, with $a_{\bq,\omega}$ an unknown proportionality constant. 
Since by definition $\argx{(\mathbb{G}_{\bq,\omega}^\dagger\chib_{\bq,\omega})^{\perp},\p_z\Phi_c}=0$, we must have $a_{\bq,\omega}=0$. This leads to an action for $\tilde h$ and $h$ only:
\begin{equation}\label{eq:CDActionhugly}
\begin{split}
    S_h&[\tilde{h},h]=\frac{1}{T} \int_{\omega,\bq}\Big[\tilde{h}_{\bq,\omega}h_{\bq,\omega}\\
    &-\tilde{h}_{\bq,\omega}\tilde{h}_{-\bq,-\omega}\argx{\mathbb{L}_\bq(\mathbb G^\dagger_{\bq,\omega})^{-1}U_0,(\mathbb G^\dagger_{-\bq,-\omega})^{-1}U_0}\Big]\;.
\end{split}
\end{equation}
We respectively denote $L_k(q)$ and $R_k(q)$ the left and right eigenvectors of $\mathbb L_\bq\mathbb O_\bq$, and $\Lambda_k(q)$ the corresponding eigenvalue. 
From Eq.~\eqref{eq:defG}, the leading contribution to $(G^\dagger_{\bq,\omega})^{-1}$ (namely the slowest), once integrated over $\omega$, arises from the eigenvalue $\Lambda_k(q)$ of $\mathbb L_\bq\mathbb O_\bq$ whose real part is closest to zero, which we hereafter refer to as $\Lambda_0(q)$. The scalar product in Eq.~\eqref{eq:CDActionhugly} can then be rewritten as
\begin{equation}
    \begin{split}
       \langle\mathbb{L}_\bq(\mathbb G^\dagger_{\bq,\omega})^{-1}&U_0,(\mathbb G^\dagger_{-\bq,-\omega})^{-1}U_0\rangle\simeq \frac{1}{|i\omega + \Lambda_0(q)|^2}\\
        &\argx{L_0(q), \mathbb L_\bq L_0(q)}\argx{R_0(q), U_0}^2\;,
    \end{split}
\end{equation}
where $L_0(q)$ and $R_0(q)$ have been normalized such that $\argx{L_0(q),R_0(q)}=1$.
Using $\left<\partial_z \Phi_c\right| \mathbb O_0=0$ and the definition Eq.~\eqref{eq:DefO}, the eigenvalue $\Lambda_0$ satisfies
\begin{equation}\label{eq:lambda0}
    \Lambda_0(q) = \frac{q^2\argx{\p_z m_c, (R_0(q))_1}}{\argx{\mathbb{L}_\bq^{-1} \p_z \Phi_c, R_0(q)}}\;,
\end{equation}
where $(R_0(q))_1$ refers to the first component of the vector $R_0(q)$.

For long wavelengths, we assume that $\Lambda_0(q)$ is obtained as a perturbation of the ground state eigenvalue $\Lambda_0(0)=0$ of $\mathbb L_\bq\mathbb O_0$, which is the operator where $\bq$ has been set to zero in $\mathbb{O}_\bq$, but not in $\mathbb{L}_\bq$. Then, the lowest orders of the eigenvectors are  given by their unperturbed expressions: $R_0(0) = \p_z \Phi_c$ and $L_0(0)= \mathbb L_\bq^{-1} \p_z \Phi_c$, normalized by their scalar product. As for model B, we choose to work with $\hb_{\bq,\omega} = (i\omega + \Lambda_0(q))^{-1} \tilde{h}_{\bq,\omega}$, so that the action reads
\begin{equation}\label{eq:CDActionhugly2}
\begin{split}
    S_h[\hb&,h]=\frac{1}{T} \int_{\omega,\bq}\Big[\hb_{\bq,\omega}(i\omega + \Lambda_0(q))h_{\bq,\omega}\\
    &-\hb_{\bq,\omega}\hb_{-\bq,-\omega}\argx{\p_z \Phi_c ,\mathbb{L}_\bq^{-1}\p_z \Phi_c}^{-1}\Big]\;.
\end{split}
\end{equation}
At this point, we need to discuss models C and D separately.\\

For model C, $\Lm = 1$ and 
\begin{equation}
        \mathbb{L}_\bq^{-1} = \begin{pmatrix}
            D^{-1} & 0\\
            0 & \Lambda^{-1}\, \mathrm L_\bq^{-1}
        \end{pmatrix}\;,
\end{equation}
where the inverse Laplacian $\mathrm L_\bq^{-1}$ is already given in Eq.~\eqref{eq:L-1}.
Defining $B(q) = 2q \argx{\p_z V_c, \mathrm L_\bq^{-1}\p_z V_c}$ and using Eq.~\eqref{eq:lambda0}, we find
\begin{equation}
    \begin{split}\label{eq:CactionBefore}
    S_h[\hb,h]=\frac{1}{T} &\int_{\omega,\bq}\Bigg[\hb_{\bq,\omega}\argp{i\omega + \frac{D q^2}{1 + \frac{D}{2\Lambda \sigma q}B(q)}}h_{\bq,\omega}\\
    & - \frac D\sigma \frac{1}{1 + \frac{D}{2\Lambda\sigma q} B(q)}\hb_{\bq,\omega}\hb_{-\bq,-\omega}\Bigg]\;.
    \end{split}
\end{equation}
As pointed out in \cite{Zia1988_Dynamics,Bausch1991_Effects}, there are two different regimes depending on the symmetries of the auxiliary field $\rho_c$, which, we remind, satisfies $\rho_c  = V(m_c)$. The function $B$ in Eq.~\eqref{eq:CactionBefore} heavily depends on whether the difference of $\rho_c$ in the two phases is $\Delta \rho_c \neq 0$, or $\Delta \rho_c = 0$.

If $\Delta\rho_c=0$, to lowest order in $q$, $B(q)=-B_1q$ with $B_1 = \iint \dd z\, \dd z' |z-z'|\p_z V_c\,\p_z' V_c$ and the action takes the form
\begin{equation}\label{eq:CactionhSym}
    \begin{split}
    S_h[\hb,h]=\frac{1}{T} &\int_{\omega,\bq}\Bigg[\hb_{\bq,\omega}\argp{i\omega + \frac{D q^2}{1 -\frac{D B_1}{2\Lambda \sigma}}}h_{\bq,\omega}\\
    & - \frac D\sigma \frac{1}{1 - \frac{D B_1}{2\Lambda\sigma }}\hb_{\bq,\omega}\hb_{-\bq,-\omega}\Bigg]\;.
    \end{split}
\end{equation}
Thus at the special point $\Delta \rho_c = 0$, the interface follows Edwards-Wilkinson dynamics. \\

If instead $\Delta \rho_c \neq 0$, $B(q) \simeq (\Delta \rho_c)^2$ at long wavelengths, and we find
\begin{equation}\label{eq:CactionhAsym2}
    \begin{split}
    S_h[\hb,h]=\frac{1}{T} &\int_{\omega,\bq}\Bigg[\hb_{\bq,\omega}\argp{i\omega + \frac{2\Lambda \sigma q^3}{(\Delta \rho_c)^2}}h_{\bq,\omega}\\
    & - \frac {2\Lambda q}{(\Delta \rho_c)^2}\hb_{\bq,\omega}\hb_{-\bq,-\omega}\Bigg]\;.
    \end{split}
\end{equation}
Hence in this generic case, the interface exhibits conserved dynamics. A  discussion of this surprising behavior was presented in~\cite{Zia1988_Dynamics}. 

To our knowledge, our derivation of the noise in model C is new. We can compare our dispersion relations to those obtained previously. In the $\Delta \rho_c \neq 0$ case, our dispersion relation matches that of \cite{Ohta1984_Dynamics,Zia1988_Dynamics,Bausch1991_Effects}:
\begin{equation}
    i\omega = -\frac{2\Lambda\sigma }{(\Delta \rho_c)^2} q^3\;.
\end{equation}
In the symmetric case, the prefactor of $q^2$  differs from the one reported in \cite{Ohta1984_Dynamics}, but agrees with~\cite{Bausch1991_Effects}:
\begin{equation}
    i\omega = -\frac{D q^2}{1 -\frac{D B_1}{2\Lambda \sigma}}\;.
\end{equation}
We believe that there is a possibility that the $q\to 0$ behavior was improperly kept track of in~\cite{Ohta1984_Dynamics}. On the numerical side, results consistent with both the $q^2$ and $q^3$ scaling regimes have been observed~\cite{jorgenson1989roughening,Harris1992_MonteCarlo}.  


In the symmetric case, a more rigorous derivation for the $q^2$ prefactor would consist in establishing the absence of level crossing in the  spectrum of $\mathbb L_\bq \mathbb O_\bq$ as $q$ departs from $0$. This goes beyond the scope of this work.\\

Regarding model D in which both fields are conserved, we have instead $\Lm = -\bnabla^2$ and the action takes the form
\begin{equation}\label{eq:Dactionh}
    \begin{split}
    S_h[\hb,h]=\frac{1}{T} &\int_{\omega,\bq}\Bigg[\hb_{\bq,\omega}\argp{i\omega + \frac{2\sigma q^3}{\frac{(\Delta m_c)^2}{D} + \frac{(\Delta \rho_c)^2}{\Lambda}}}h_{\bq,\omega}\\
    & - \frac {2 q}{\frac{(\Delta m_c)^2}{D} + \frac{(\Delta \rho_c)^2}{\Lambda}}\hb_{\bq,\omega}\hb_{-\bq,-\omega}\Bigg]\;.
    \end{split}
\end{equation}
This corresponds to the dispersion relation
\begin{equation}
    i\omega = -\frac{2\sigma q^3}{\frac{(\Delta m_c)^2}{D} + \frac{(\Delta \rho_c)^2}{\Lambda}}\;.
\end{equation}
Given that both fields appear on equal footing, the $q^3$ behavior is insensitive to $V(m_c)$ being symmetric or not. This result, to the best of our knowledge, appears here for the first time in the literature. On the numerical side, the coarsening~\cite{laradji1992effect}
shows a $t^{1/3}$ growth law that we believe is consistent with the $q^3$ scaling of the relaxation spectrum. 

\subsection{Coupling to a momentum conserving fluid: Model H}
In a system of particles advected by a surrounding fluid in $d\geq 3$ dimensions, the order parameter $\phi$ couples to the velocity field ${\bf v}$ of the solvent. The resulting stochastic coupled dynamics for $\phi$ and $\bf v$ goes by the name of Model H~\cite{kawasaki1970kinetic,halperin1974renormalizationB,HalperinHohenberg1977, Cates2019_LN}. It reads
\begin{align}
        \p_t \phi + \bv\cdot\bnabla\phi =& \nabla^2 \mu +\sqrt{2T}\bnabla\cdot\bxi\\
        \rho\,(\p_t \bv + \bv\cdot\nabla\bv) =& \eta \nabla^2 \bv  -\phi\bnabla \mu\nonumber \\
         &- \bnabla P+ \sqrt{2T\eta}\,\bnabla\cdot \bSigma \label{eq:NS}\\
        \bnabla\cdot\bv =& 0 \;, \label{eq:incomp}
\end{align}
with $\mu = \delta F/\delta\phi$ the chemical potential, $\rho$ and $P$ the density and pressure of the fluid, and $\eta$ its viscosity (sending $\eta$ to infinity brings us back to model B). Denoting again $\bx = (\br,z)$, the Gaussian white noises have correlations
\begin{align}
    \argx{\xi_i(\bx,t)\xi_j(\bx,t')} &= \delta_{ij}\delta^{(d)}(\bx-\bx')\delta(t-t')\\
    \argx{\Sigma_{ij}(\bx,t)\Sigma_{kl}(\bx',t')} &= (\delta_{ik}\delta_{jl} + \delta_{jk}\delta_{il})\times\nonumber\\
    &\delta^{(d)}(\bx-\bx')\delta(t-t')\;.
\end{align}
In the limit of low Reynolds number, one neglects the left hand side of Eq.~\eqref{eq:NS}. Using the incompressibility condition~\eqref{eq:incomp}, one can solve the Stokes equation for $\bv$ as a convolution:
\begin{equation}
    v_i = -\mathrm T_{ij}*(\phi\,\p_j\mu - \sqrt{2T\eta}\,\p_k \Sigma_{jk})\;,
\end{equation}
where ${\mathrm T}_{ij}$ is the Oseen tensor, whose Fourier components read
\begin{equation}
    T_{ij}(\bk) = \frac{1}{\eta \bk^2} \argp{\delta_{ij} - \frac{k_ik_j}{\bk^2}}\;,
\end{equation}
where $\bk$ is the Fourier wavevector corresponding to the full space $\bx=(\br,z)$.
Then, upon eliminating the velocity field, the effective dynamics of the order parameter is given by
\begin{equation}
\begin{split}
    \p_t\phi = &\nabla^2 \mu + \sqrt{2T}\bnabla\cdot\bxi\\
    &+ \p_i\phi(\bx)\int_{\bx'}\mathrm T_{ij}(\bx-\bx')\phi(\bx')\p_j\mu(\bx')\\
    & - \sqrt{2T\eta}\,\p_i\phi(\bx)\int_{\bx'}\mathrm T_{ij}(\bx-\bx')\p_k\Sigma_{jk}(\bx')\;.
\end{split}\end{equation}
The interface dynamics in model H was previously studied in~\cite{Jasnow1981_Unstable,Kawasaki1982_KineticDrumhead2,Kawasaki1983_Kinetics,Ohta1984_Dynamics,ohta1984scaling,shinozaki1993dispersionH,Bray2001_Interface}, where the dispersion relation was derived using a variety of approaches. Here, we not only re-derive it consistently, but our approach also allows us to derive the noise term explicitly.

The dynamical action takes the form
\begin{equation}
    \begin{split}
        \mathrm S[\phib&,\phi] = \int_{t,\bx,\bx'}\Big\{\phib(\bx,t)\Big[\delta(\bx-\bx')(\p_t \phi - \nabla^2\mu)(\bx,t)\\
        & + \p_i\phi(\bx,t)\mathrm T_{ij}(\bx-\bx')\p_j\phi(\bx',t)\mu(\bx',t)\Big]\\
        &+ T\delta(\bx-\bx')\phib(\bx,t)\nabla^2\phib(\bx,t)\\
        &-T\phib(\bx,t)\p_i\phi(\bx,t) \mathrm T_{ij}(\bx-\bx')\p_j\phi(\bx',t)\phib(\bx',t)\Big\}\;,
    \end{split}
\end{equation}
where we have used two properties of the Oseen tensor, namely $\p_i \mathrm T_{ij}(\bx-\bx') = 0$ and $\mathrm T_{ij}(\bk)\mathrm T_{jl}(-\bk) = \frac{1}{\eta \bk^2} \mathrm T_{il}(\bk)$.

Defining $h(\br,t)$ and $\chi(\br,z,t)$ by Eqs.~\eqref{eq:definterface} and \eqref{eq:DefChi} and going to Fourier space in $\br$ and $t$, we obtain the following quadratic action:
\begin{align}
        S = &\frac{1}{T}\int_{\omega,\bq,z,z'}\Big\{\chib_{\bq,\omega}(z)\, \mathrm K_\bq(z,z')\chib_{-\bq,-\omega}(z') \notag\\
        &+ \chib_{\bq,\omega}(z)\Big[i\omega\delta(z-z')-\mathrm K_\bq(z,z')(\bq^2 + \Omega_0(z'))\Big]\notag\\
        &\quad \times(-h_{\bq,\omega}\p_z m_c(z') + \chi^\perp_{\bq,\omega}(z'))\Big\}\;,
\end{align}
with
\begin{equation}
\begin{split}
    \mathrm K_\bq(z,z') = &\delta(z-z')(\p_z'^2 - \bq^2)\\
    &- \p_z m_c(z) \mathrm T_{zz}(\bq,z-z')\p_z m_c(z')\;.
\end{split}\end{equation}
The operator that determines the behavior of the interface is $-\mathrm K_\bq (\bq^2 + \Omega_0)$: its smallest eigenvalue controls the dispersion relation.  Using the explicit form of $\mathrm T_{zz}(\bq,z-z')$~\cite{shinozaki1993dispersionH}
\begin{equation}
    \mathrm T_{zz}(\bq,z-z') = \frac 1{4\eta q}\argp{1 + q|z-z'|}\ee^{-q|z-z'|}\;,
\end{equation}
\if{the leading term in $\mathrm K_\bq(z,z')$ at large wavelengths is $-\frac{1}{4\eta q}\p_z m_c (z) \p_z m_c(z')$. This strong approximation applies only within the subspace of functions spanned by $\p_z m_c$. It is tempting to proceed in the same vein as before in Eq.~\eqref{eq:lambda0} and to consider the $\bq^2$ term in $-\mathrm K_\bq (\bq^2 + \Omega_0)$ as a perturbation of $-\mathrm K_\bq \Omega_0$. When following this route (which we expand on in App.~\ref{sec:appendixH}), we find the eigenvalue to be $\frac{\sigma q}{4\eta}$ to lowest order in $q$. And then, introducing $\hb_{\bq,\omega} = (i\omega + \frac{\sigma q}{4\eta})^{-1}\tilde h_{\bq,\omega}$, the action for the interface height reads
\begin{equation}
\begin{aligned}
    S_h[\hb,h] = \frac{1}{T}\int_{\omega,\bq}& \Big[\hb_{\bq,\omega}\Big(i\omega + \frac{\sigma q}{4\eta}\big)h_{\bq,\omega}\\
    &- \frac{1}{4\eta q}\hb_{\bq,\omega}\hb_{-\bq,-\omega}\Big]    \;,
\end{aligned}
\end{equation}
which leads to a dispersion relation $i\omega \sim q$.

We expect the next order to be $q^3$. Such a $q^3$ contribution is coming from the Cahn-Hilliard part of the field theory. Numerical studies confirm the two regimes of $i\omega \sim q$ and $i\omega \sim q^3$ for large viscosity and small-but-not-too-small $q$~\cite{shinozaki1993dispersionH,shinozaki1993spinodal}. At small viscosity and/or very small $q$, the lowest eigenvalue of $-\mathrm K_\bq (\bq^2 + \Omega_0)$ is instead contained in the continuous part of the spectrum,  so that the $q$ and $q^3$ scaling break down.
The lack of analytical predictions for the lowest part of the spectrum then prevents us from predicting the dispersion relation in this asymptotic regime.}\fi
the leading term in $\mathrm K_\bq(z,z')$ at large wavelengths is $-\frac{1}{4\eta q}\p_z m_c (z) \p_z m_c(z')$. 
It is tempting to proceed in the same vein as before in Eq.~\eqref{eq:lambda0} and to consider the $\bq^2$ term in $-\mathrm K_\bq (\bq^2 + \Omega_0)$ as a perturbation of $-\mathrm K_\bq \Omega_0$. 
In \cite{shinozaki1993dispersionH}, Shinozaki warns that this procedure, which works well for model B, seems to be numerically challenged in model H, as he observes that the branch continuing the ground state of $-\mathrm K_\bq \Omega_0$ is crossed by the continuous spectrum at some  wavevector that is a decreasing function of the viscosity.
For a large viscosity, the ground-state branch appears to be the lowest, and the ground-state eigenvalue controls the interface dynamics.
In this regime  we find the eigenvalue to be $\frac{\sigma q}{4\eta}$ to lowest order in $q$ (see App.~\ref{sec:appendixH} for details). Introducing $\hb_{\bq,\omega} = (i\omega + \frac{\sigma q}{4\eta})^{-1}\tilde h_{\bq,\omega}$, the action for the interface height reads
\begin{equation}
\begin{aligned}
    S_h[\hb,h] = \frac{1}{T}\int_{\omega,\bq}& \Big[\hb_{\bq,\omega}\Big(i\omega + \frac{\sigma q}{4\eta}\big)h_{\bq,\omega}\\
    &- \frac{1}{4\eta q}\hb_{\bq,\omega}\hb_{-\bq,-\omega}\Big]    \;,
\end{aligned}
\end{equation}
which leads to a dispersion relation $i\omega \sim q$. Shinozaki argues that the correction to the $q$ behavior goes as $q^3$~\cite{shinozaki1993dispersionH,shinozaki1993spinodal}. At small viscosity however, the continuous spectrum interferes with the ground state and no analytical prediction seems to be available.

\section{Out of equilibrium}\label{sec:noneq}
We now turn to studying how nonequilibrium driving impacts interface dynamics, and begin with one the simplest instances of a nonequilibrium field theory, namely active model A~\cite{caballero2020stealth,meissner2024introduction}.

\subsection{Active model A}\label{sec:AMA}

\subsubsection{Deriving the Edwards-Wilkinson equation}

The dynamics of active model A (AMA) is defined by Eq.~\eqref{eq:motion}  with a nonequilibrium drive $w[\phi] = -\lambda(\bnabla\phi)^2$, leading to:
\begin{equation}\label{eq:AMA}
    \p_t \phi = -\funcder{F}{\phi} - \lambda(\bnabla\phi)^2 + \sqrt{2T}\eta(\br,z,t)\;,
\end{equation}
where $F[\phi]$ remains given by Eq.~\eqref{eq:F}. Note that we keep the notation $T$ for the amplitude of the noise, even though $T$ is not, thermodynamically speaking, a temperature. The steady-state mean-field profile $m_c(z)$ solves the new equation
\begin{equation}\label{eq:AMAMF}
    0 = \p_z^2m_c - f'(m_c) -\lambda(\p_z m_c)^2\;.
\end{equation}
We introduce the linear operator $\Omega_\lambda$ defined by
\begin{equation}\label{eq:DefOmegaLambda}
    \Omega_\lambda = - \p_z^2  + f''(m_c) + 2\lambda(\p_z m_c) \, \p_z\;.
\end{equation}
The additional contribution $2\lambda(\p_z m_c) \, \p_z$ with respect to $\Omega_0$ in Eq.~\eqref{eq:DefOmega0} renders $\Omega_\lambda$ non Hermitian. Using $\Omega_\lambda$ and ommitting for now the additional terms coming from the It\=o discretization, we can write the Janssen-De Dominicis action as
\begin{widetext}
 \begin{align}\label{eq:AMAActionFull}
       S = &\frac 1 T  \int_{t,\br,z} \Big\{\chib (\p_t - \bnabla_\br^2 +\Omega_\lambda)(-h\p_z m_c + \chi^\perp)- \chib^2\notag\\
       & +\chib\Big[- (\p_t h-\bnabla_\br^2 h) \partial_z \chi^\perp- (\bnabla_\br h)^2 \argc{\partial_z^2 m_c - \lambda(\partial_z m_c)^2}+ 2\bnabla_\br h\cdot\bnabla_\br \p_z \chi^\perp \notag\\
       &  +\frac{f'''(m_c)}{2}(\chi^\perp)^2  +\lambda(\bnabla\chi^\perp)^2 - 2\lambda \bnabla_\br \chi^\perp\cdot \bnabla_\br h\, \partial_z m_c\notag\\
       &- (\bnabla_\br h)^2 \argc{\partial_z^2 - 2\lambda\p_z m_c\p_z}\chi^\perp- 2\lambda \bnabla_\br \chi^\perp\cdot \bnabla_\br h\,\p_z \chi^\perp\notag \\
       &+\lambda(\bnabla_\br h)^2 (\p_z\chi^\perp)^2+ f'(m_c+\chi^\perp)- f'(m_c) - f''(m_c)\chi^\perp -\frac{f'''(m_c)}{2}{\chi^\perp}^2\Big]\Big\}\notag\\
       &-\int_{t,\br}\ln\left|\argx{\p_z m_c,u_0} + {\argx{\p_z \chi^\perp,u_0}}\right|\;.
\end{align}   
\end{widetext}
The Gaussian part of this action, which controls the leading fluctuations, reads
\begin{equation}\label{eq:AMAquadS}
    \begin{split}
        S_0 = \frac{1}{T}\int_{t,\br,z}&\Big[\chib [-(\p_t h -\bnabla_\br^2 h)\p_z m_c\\
        &+ (\p_t - \bnabla_\br^2 + \Omega_\lambda)\chi^\perp] - \chib^2\Big]\;.
    \end{split}
\end{equation}
Comparing Eq.~\eqref{eq:AMAquadS} with Eq.~\eqref{eq:AquadS}, we realize that the actions for model A and active model A are the same up to the replacement of $\Omega_0$ with  $\Omega_\lambda$. Introducing $\Gamma = \p_t - \bnabla_\br^2 + \Omega_\lambda$, and decomposing $\chib^\perp$ as in Eq.~\eqref{eq:DecompChib} then leads to 
\begin{align}
       S_0 = \frac{1}{T}&\int_{t,\br}
       \Big\{\hb (\p_t h - \bnabla_\br^2h) \argx{\p_z m_c,u_0} - \hb^2 \argx{u_0,u_0}\notag\\
       &+\argx{\chib^\perp, \Gamma \chi^\perp} - \argx{\chib^\perp,\chib^\perp}\notag\\
        & -\hb\argx{u_0,\Omega_\lambda\chi^\perp}+ 2\hb \argx{u_0, \chib^\perp}\Big\}\;.\label{eq:quadSAMA_halfdecoup}
\end{align}
where we have used $\langle \chib^\perp,\p_z m_c\rangle=\argx{\chi^\perp,u_0}=0$ and that $u_0(z)$ is independent of $t,\br$.
As in model A, we have some freedom to choose $u_0$, which we shall use to cancel the terms coupling the bulk fluctuations to the interface deformation on the third line of Eq.~\eqref{eq:quadSAMA_halfdecoup}. Canceling the second one could be achieved by choosing $u_0\propto \p_z m_c$, but since $\Omega_\lambda$ is not Hermitian, $\p_z m_c$ is not an eigenstate of $\Omega_\lambda^\dagger$ and the first term does not vanish. Instead, we choose to cancel the first term (which drives a deterministic contribution of bulk fluctuations to the evolution of the interface deformation) using for $u_0$  an eigenvector of $\Omega_\lambda^\dagger$. We now explain which one to pick.

We assume that the property that $0$ is an isolated eigenvalue of $\Omega_\lambda$ still holds when $\lambda\neq 0$ (this is certainly true for $\lambda$ not too large). 
We introduce $\ell_i$ and $r_i$ the eigenvectors of $\Omega_\lambda^\dagger$ and  $\Omega_\lambda$ for the eigenvalue $\lambda_i$, respectively. The vector $r_0 = \p_z m_c$ is the ground state of $\Omega_\lambda$. In general, for $i\neq j$ the orthogonality of $\ell_i$ and $r_j$  is not guaranteed, but $\ell_0$ and $r_0$ are always orthogonal to $r_i$ and $\ell_i$ for $i>0$, respectively.
The simplest choice of $u_0$ is one that sets $\argx{u_0,\Omega_\lambda\chi^\perp}=0$,  and the sole choice that does not make the deterministic part of the action $\hb h$ vanish is $u_0 = \ell_0$. The eigenvector of $\Omega_\lambda^\dagger$ with eigenvalue $0$ is
\begin{equation}\label{eq:l0}
    \ell_0(z) =\frac 1 {\sigma_A} \ee^{-2\lambda m_c(z)}\p_z m_c(z)\;, 
\end{equation}
with 
\begin{equation}
    \sigma_A = \argx{\ee^{-2\lambda m_c}\p_z m_c,\p_z m_c}\;.
\end{equation}
We are thus left with the coupling term $2\hb \argx{\ell_0,\chib^\perp}$ which can be removed by shifting $\chi^\perp$ into
\begin{equation}\label{eq:AMAshift}
    \chi'^\perp = \chi^\perp + 2\Gamma^{-1}\Pi \hb\ell_0\;.
\end{equation}
Importantly, $\chi'^\perp$ still lives in the subspace perpendicular to $\ell_0$, since the operators $\Gamma$ and $\Pi$ commute. This leads to the following Gaussian action, which decouples the fields $\hb,h$ and $\chib^\perp,\chi'^\perp$,
\begin{equation}\label{eq:AMAS0forNL}
    \begin{split}
       S_0 = \frac{1}{T}&\int_{t,\br}
       \Big\{\hb(\p_t -\bnabla_\br^2)h- \frac{\Sigma}{\sigma_A} {\hb^2}\\
       & +\argx{\chib^\perp,\Gamma \chi'^\perp}- \argx{\chib^\perp,\chib^\perp} \Big\}\;,
    \end{split}
\end{equation}
with $\Sigma = \argx{\ee^{-2\lambda m_c}\p_z m_c,\ee^{-2\lambda m_c}\p_z m_c} $. 

At this stage, we note an important difference between the action $S_0$ in Eq.~\eqref{eq:AMAS0forNL} and that of model A for the transverse fluctuation $\chi$. Since the eigenvectors $r_i$ of $\Gamma$ coincide with those of $\Omega_\lambda$, we could decompose $\chi'^\perp(\br,z,t)=\sum_{i>0}c_i(\br,t)r_i(z)$ along those eigenvectors, and similarly decompose $\chib^\perp(\br,z,t)=\sum_{i>0}\bar c_i(\br,t)\ell_i(z)$ along the eigenvectors of $\Omega_\lambda^\dagger$. However, the resulting action is not diagonal in these modes, unlike in Eq.~\eqref{eq:AquadS_final}, and they do not decouple. 

The action for $h$ and $\hb$ can be directly read in Eq.~\eqref{eq:AMAS0forNL} as
\begin{equation}
    \begin{split}
        S_{0,h}[\hb,h] = \frac 1 T &\int_{t,\br}\Big[\hb(\p_t - \bnabla_\br^2)h- \frac{\Sigma}{\sigma_A^2}\hb^2\Big]\;.
    \end{split}
\end{equation}
The corresponding Langevin equation is
\begin{equation}\label{eq:AMA_Inteq}
    \p_t h(\br,t) = \bnabla_\br^2 h(\br,t) + \sqrt{\frac{2T\Sigma}{\sigma_A^2}}\tilde{\eta}(\br,t)\;.
\end{equation}
The linear interface dynamics in active model A still evolves according to an Edwards-Wilkinson equation, as for model A, albeit with a renormalized noise amplitude.

To interpret this renormalization, we first note that, in equilibrium, the interface dynamics must yield the same surface tension as that stemming from a purely static approach based, {\it e.g.}, on the free energy cost of small deformations. For active systems, it is known that, on the contrary, different definitions of surface tension that coincide in equilibrium take  different values~\cite{Bialke2015_Negative,tjhung2018cluster,Zakine2020_Surface,cates2023classical,zhao2024active,langford2025mechanics}. 
Equation~\eqref{eq:AMA_Inteq} can thus be used to infer a "capillary surface tension", which controls only the linearized dynamics of interfaces. There are, however, several ways to do so. For instance, one can read Eq.~\eqref{eq:AMA_Inteq} as stemming from a temperature $T$ and a surface tension given by $\sigma_A^2/\Sigma$. Alternatively, one can interpret this dynamics as stemming from a surface tension $\sigma_A$,  with a renormalized temperature $T\Sigma/\sigma_A$.

As in Sec.~\ref{sec:A_ansatz}, our derivation above shows that there is a specific protocol that allows deriving the linearized interface dynamics for active model A using the simplified ansatz~\eqref{eq:rottenansatz}.
Indeed, injecting the exact decomposition $\phi(\br,z,t) = m_c(z-h) + \chi(\br,z-h,t)$ into the equation of motion~\eqref{eq:AMA} for active model A leads, to order $\sqrt T$, to Eq.~\eqref{eq:A_EoM} with $\Omega_0$ replaced by $\Omega_\lambda$.
Projecting this equation onto $\ell_0$ would have eliminated $\chi$, thereby yielding the linear dynamics of the interface. 
This is exactly what would have been obtained starting from the ansatz of Eq.~\eqref{eq:rottenansatz} and specifically projecting onto $\ell_0$: As in model A, 
the simplified procedure works in active model A, except that one has to use $\ell_0$ to eliminate the $z$-dependence instead of $\p_z m_c$.

As a final remark, we note that the choice $u_0=\ell_0$ echoes a similar one made by Kuramoto in the study of propagating fronts~\cite{Kuramoto1980_Instability}---See also~\cite{birzu2018fluctuations} for a recent related work on fluctuations in pushed and pulled waves. In another context, the eigenvector $\ell_0$ turns out to be related to the generalized thermodynamics formalism introduced in \cite{solon2018generalized2,Solon2018_GeneralizedThermodynamics}. In these articles, the authors were able to find an equilibrium-like Maxwell construction to compute the binodals of an active phase-separated system, by working with a pseudodensity field $R$ instead of $\phi$. A possible derivation of the expression of $\ell_0$ found in Eq.~\eqref{eq:l0} goes as follows: if we search for $\ell_0$ in the form $\ell_0(z)=\p_zR(m_c(z))$, where $R$ is an arbitrary function, then
\begin{align}
    &\Omega_\lambda^\dagger \p_z R(m_c) = 0 \notag\\
    &\Leftrightarrow -(\p_z m_c)^3(R''(m_c) +2\lambda R'(m_c)) = 0\;.
\end{align}
The differential equation $R'' = -2\lambda R'$ is exactly the definition of the pseudo-density field in the case of active model A. 

We are now in a position to use $S_0$ in Eq.~\eqref{eq:AMAS0forNL} to construct a perturbative expansion in powers of $\sqrt{T}$  to determine the nonlinearities in the evolution of the field $h$. We shall implement this strategy to bring forth the presence of KPZ terms in the active model A dynamics.\\

\subsubsection{Deriving a KPZ equation}\label{sec:AMANL}

We shall now follow the same steps as for passive model A and implement a perturbation expansion in powers of $\sqrt{T}$. To proceed, we change variables from $\hb$ to $\hb'$ to eliminate time derivatives from the non-Gaussian part of the action:
\begin{equation}
    \hb = \frac{\hb' + \argx{\chib^\perp, \p_z \chi^\perp}}{1 + \argx{\p_z \chi^\perp, \ell_0}}\;.
\end{equation}
This redefinition is a technical step which cancels both $\chib(\p_t h -\bnabla_\br^2 h)\p_z\chi^\perp$ in the non-Gaussian action and the Jacobian.
Proper construction of the It\=o-discretized action leads to
\begin{widetext}
\begin{equation}\label{eq:PIWAMA_Sfull}
    \begin{split}
        S = &\frac{1}{T} \int_{t,\br} \Big\{\hb'(\p_t h-\bnabla_\br^2 h) + \argx{\chib^\perp, (\p_t -\bnabla_\br^2 + \Omega_\lambda)\chi^\perp}+  \Big<-\frac{\hb'+\argx{\chib^\perp,\p_z \chi^\perp}}{1 + \argx{\p_z \chi^\perp,\ell_0}} \ell_0+\chib^\perp,  \\
        &\;\frac{T\Sigma}{\sigma_A^2(1+ \argx{\p_z \chi^\perp,\ell_0})^2}(\p_z^2 m_c + \p_z^2 \chi^\perp)- 2T\p_z\Pi\argc{\frac{\p_z \chi^\perp \Sigma}{\sigma_A^2(1+ \argx{\p_z \chi^\perp,\ell_0})^2} - \frac{\ell_0}{1+ \argx{\p_z \chi^\perp,\ell_0}}}\\
        &- (\bnabla_\br h)^2 (\p_z^2 m_c + \p_z^2 \chi^\perp) + 2\bnabla_\br h\cdot \p_z\bnabla_\br \chi^\perp+f'(m_c+\chi^\perp)-f'(m_c) -f''(m_c) \chi^\perp\\
        &+\lambda\arga{(\bnabla_\br h)^2 \argc{(\p_z m_c)^2 + 2 \p_z m_c \p_z \chi^\perp + (\p_z \chi^\perp)^2} + (\bnabla \chi^\perp)^2 - 2\bnabla_\br \chi^\perp\cdot \bnabla_\br h(\p_z m_c + \p_z \chi^\perp)}\Big>\\
        &- \Big\langle-\frac{\hb'+\argx{\chib^\perp,\p_z \chi^\perp}}{1 + \argx{\p_z \chi^\perp,\ell_0}}\ell_0 + \chib^\perp,-\frac{\hb'+\argx{\chib^\perp,\p_z \chi^\perp}}{1 + \argx{\p_z \chi^\perp,\ell_0}}\ell_0 + \chib^\perp\Big\rangle\Big\}\;,
    \end{split}
\end{equation}
\end{widetext}
where we recall that $u_0=\ell_0$. 
The Gaussian action, after the shift \eqref{eq:AMAshift}, is given by Eq.~\eqref{eq:AMAS0forNL} with $\hb'$ instead of $\hb$. We hereafter relabel $\hb'$ as $\hb$, and the second line of Eq.~\eqref{eq:AMAS0forNL} is denoted by $S_0'[\chib^\perp, \chi'^\perp]$. 

The nonquadratic part of the action to order $5$ in the fields takes the appealing form
\newcommand{\hbr}{\color{red}\hb\color{black}}
\newcommand{\hr}{\color{red}h\color{black}}
\begin{widetext}
\begin{equation}\label{eq:AMAfullNGugly}
\begin{split}
       S_{\text{NG}} &= \frac 1 T  \int_{t,\br,z}\!\!\!\Big\{-\ell_0^2\big[\red{\hb^2} \rho(-2 + 3\rho) + \argx{\chib^\perp,\p_z\chi^\perp}^2+2 \hbr\argx{\chib^\perp,\p_z\chi^\perp}(1-2\rho)\big]+ 2\chib^\perp \ell_0\big[\hbr\rho(\rho-1)+ \argx{\chib^\perp,\p_z\chi^\perp}(1-\rho)\big] \\
        &+ \big[-\hbr \ell_0+\chib^\perp + \ell_0(\hbr\rho-\argx{\chib^\perp,\p_z\chi^\perp} - \hbr\rho^2 + \argx{\chib^\perp,\p_z \chi^\perp}\rho)\big]\\
        &\times \Big[\frac{T\Sigma}{\sigma_A^2}\p_z^2 m_c + 2T\p_z \Pi\ell_0-\red{(\bnabla_\br h)^2}[\p_z^2 m_c - \lambda(\p_z m_c)^2] + 2\red{\bnabla_\br h} \cdot\bnabla_\br(\p_z-\lambda\p_z m_c)\chi^\perp + \frac{f'''(m_c)}{2}(\chi^\perp)^2 + \lambda\big(\bnabla\chi^\perp\big)^2\\
        &\quad +\frac{T\Sigma}{\sigma_A^2}\big(-2\rho\p_z^2 m_c +\p_z^2 \chi^\perp\big)-2T\p_z\Pi\Big(\frac{\p_z\chi^\perp\Sigma}{\sigma_A^2} + \rho\ell_0\Big)-\red{(\bnabla_\br h)^2} [\p_z^2 - 2 \lambda\p_z m_c\p_z]\chi^\perp - 2\lambda\red{\bnabla_\br h}\cdot \bnabla_\br\chi^\perp\p_z\chi^\perp\\
        &\quad + \frac{f^{(4)}(m_c)}{3!}(\chi^\perp)^3 \\
        &\quad +\frac{T\Sigma}{\sigma_A^2}\big(3\rho^2\p_z^2 m_c -2\rho\p_z^2 \chi^\perp\big)-2T\p_z\Pi\Big(-2\rho\frac{\p_z\chi^\perp\Sigma}{\sigma_A^2} - \rho^2\ell_0\Big) + \lambda\red{(\bnabla_\br h)^2} (\p_z \chi^\perp)^2 + \frac{f^{(5)}(m_c)}{4!}(\chi^\perp)^4\Big]
       \Big\} \;,
    \end{split}
\end{equation}
\end{widetext}
where we introduced the short-hand notation 
\begin{equation}
    \rho = \argx{\p_z{\chi}^\perp,\ell_0}\;.
\end{equation}
Before computing averages, one final change of fields is in order, namely  $\chi^\perp = \chi'^\perp - 2\Gamma^{-1}\Pi \hbr\ell_0$ (which we have not written explicitly for legibility considerations). 
\if\wentzel1{
\begin{widetext}
\begin{equation}\label{eq:AMAfullNGugly}
\begin{split}
       S_{NG} &= \frac 1 T  \int_{t,\br,z}\Big\{-\ell_0^2\big[\red{\hb^2} \rho(-2 + 3\rho) + \argx{\chib^\perp,\p_z\chi'^\perp - 2\p_z \Gamma^{-1}\Pi\hbr\ell_0}^2+2 \hbr\argx{\chib^\perp,\p_z\chi'^\perp - 2\p_z \Gamma^{-1}\Pi\hbr\ell_0}(1-2\rho)\big]\\
        & + 2\chib^\perp \ell_0\big[-\hbr\rho(1-\rho)+ \argx{\chib^\perp,\p_z\chi'^\perp - 2\p_z \Gamma^{-1}\Pi\hbr\ell_0}(1-\rho)\big] \\
        &+ \big[-\hbr(1-\rho+ \rho^2)\ell_0- \argx{\chib^\perp,\p_z\chi'^\perp - 2\p_z \Gamma^{-1}\Pi\hbr\ell_0}(1-\rho)\ell_0 + \chib^\perp\big]\\
        &\times \big[-\red{(\bnabla_\br h)^2}[\p_z^2 m_c - \lambda(\p_z m_c)^2] + 2\red{\bnabla_\br h} \cdot(\bnabla_\br\p_z-\lambda\p_z m_c\bnabla_\br)(\chi'^\perp - 2 \Gamma^{-1}\Pi\hbr\ell_0)\\
        &\quad + \frac{f'''(m_c)}{2}(\chi'^\perp - 2 \Gamma^{-1}\Pi\hbr\ell_0)^2 + \lambda\big(\bnabla(\chi'^\perp - 2 \Gamma^{-1}\Pi\hbr\ell_0)\big)^2\\
        &\quad -\red{(\bnabla_\br h)^2} [\p_z^2 - 2 \lambda\p_z m_c\p_z](\chi'^\perp - 2 \Gamma^{-1}\Pi\hbr\ell_0) - 2\lambda\red{\bnabla_\br h}\cdot \bnabla_\br(\chi'^\perp - 2 \Gamma^{-1}\Pi\hbr\ell_0)\p_z(\chi'^\perp - 2 \Gamma^{-1}\Pi\hbr\ell_0)\\
        &\quad + \frac{f^{(4)}(m_c)}{3!}(\chi'^\perp - 2 \Gamma^{-1}\Pi\hbr\ell_0)^3+ \lambda\red{(\bnabla_\br h)^2} (\p_z \chi'^\perp - 2 \p_z\Gamma^{-1}\Pi\hbr\ell_0)^2 + \frac{f^{(5)}(m_c)}{4!}(\chi'^\perp - 2 \Gamma^{-1}\Pi\hbr\ell_0)^4\big]
       \Big\} \;,
    \end{split}
\end{equation}
\end{widetext}
where we introduced the short-hand notation 
\begin{equation}
    \rho = \argx{\p_z{\chi'}^\perp - 2\p_z \Gamma^{-1}\Pi \hbr \ell_0,\ell_0}\;.
\end{equation}}\fi
In the part of the action arising from $\bar{\phi}^2$ (namely the first line), we have only made explicit terms up to order $4$ in the fields, anticipating that higher orders will not contribute to the deterministic part of the evolution equation for $h$ to order $O(T^{2})$. For clarity, some terms with more than five fields in the product of the second and third-to-sixth lines have been kept, but they should be omitted since they do not contribute to the correponding Langenvin equation to $O(T^{2})$.

We implement a cumulant expansion with respect to the weight $\ee^{-S_0'[\chib^\perp,\chi'^\perp]}$ at fixed $\hb$ and $h$ (and denote by $\langle \dots \rangle_0'$ the corresponding average). Each field scales as $\sqrt T$, thus, the higher the order of the cumulant, the smaller its contribution is in powers of $\sqrt{T}$.

We begin by returning to the fluctuation-driven velocity acquired by the interface, for which interesting physics can already be extracted from the first order correction and first order cumulant. This amounts to searching, in $\argx{S_{\rm NG}}'_0$, for the terms proportional to  $\hb$ (but not to $h$ nor $\hb^2$). To lowest order in $\sqrt T$, we find the single contribution
\begin{equation}
\begin{split}
    -\frac{1}{T}\int_{\br,t}\hbr&\Big\langle\ell_0,\underbrace{\frac{T\Sigma}{\sigma_A^2} \p_z^2 m_c + 2T\p_z \Pi\ell_0}_{(a)}\\
    &+\underbrace{\frac{f'''(m_c)}{2}\argx{(\chi'^\perp)^2}'_0 + \lambda \argx{(\bnabla{\chi'}^\perp)^2}'_0 }_{(b)}\Big\rangle\;.
\end{split}\end{equation}

Introducing a Feynman diagram representation in which a black (arrowed) leg stands for $\chi'^\perp$ ($\bar{\chi}^\perp$), this corresponds to

\begin{tikzpicture}
\begin{scope}[local bounding box=diaga]
  \begin{feynman}
    \vertex (I2){\(\hbr\)};

    \vertex (V1) [left=1.27cm of I2];
    \vertex (V2) [left=0.9cm of V1];

    \diagram* {
        (V1) -- [red,thick,->] (I2),
      
      (V1) -- [white,half left, thick,looseness=1.75] (V2),
      (V1) -- [white,half right, thick, looseness=1.75] (V2),
    };
    
  \end{feynman}
\end{scope}
\begin{scope}[xshift=3.5cm,local bounding box=diagb]
      \begin{feynman}
    \vertex (i2){\(\hbr\)};

    \vertex (v1) [left=1.27cm of i2];
    \vertex (v2) [left=0.9cm of v1];

    \diagram* {
        (v1) -- [red,thick,->] (i2),
      
      (v1) -- [half left, thick,looseness=1.75] (v2),
      (v1) -- [half right, thick, looseness=1.75] (v2),
    };
  \end{feynman}
\end{scope}

\path (I2) -- (v2) node[midway] {$+$};
\node [below=-0.3cm of diagb] {($b$)};
\node [below=-0.3cm of diaga] {($a$)};
\end{tikzpicture}

\noindent where the $\chi'^\perp$ correlator is represented by an internal black line and non-contracted red legs are used for the interface height $\hr$ and the response field $\hbr$.

The expression Eq.~\eqref{eq:expressioncKS} found in model A by \cite{Kado2024_MicroCutoff} is changed into
\begin{equation}\label{eq:cAMA}
\begin{split}
        c=& \argx{\ell_0,\frac{f'''(m_c)}{2}\argx{(\chi'^\perp)^2}'_0 + \lambda \argx{(\bnabla{\chi'}^\perp)^2}'_0 }\\
        &- \lambda\frac{T\Sigma}{\sigma_A^2} \argx{\ell_0,(\p_z m_c)^2}
\end{split}
\end{equation}
in the Langevin equation for the interface dynamics. Here we have used that $\Pi = \mathds I - \left|\p_z m_c\rangle\langle\ell_0\right|$, $\ell_0$ vanishes at the boundaries, and $\argx{\ell_0,\p_z^2 m_c} = \lambda\argx{\ell_0,(\p_z m_c)^2}$. 
The first thing to note is that the velocity scales as $T$, similarly to that found in Eq.~\eqref{eq:newc}. Then, $c$ is now non-vanishing even if $f$ is symmetric. 
This is due to the second and third terms  in Eq.~\eqref{eq:cAMA}, which arise from the active addition to model A, $-\lambda(\bnabla\phi)^2$. Even though this is a nonlinear term, it affects the linear dynamics of the interface. It is also responsible for generating a KPZ nonlinearity, as we now explain. 

It requires a bit of book-keeping to understand which terms in Eq.~\eqref{eq:AMAfullNGugly} eventually give rise to such combinations. 
To first order in $\sqrt{T}$, it would require the presence of a  $\hb h^2$: there is such a term but it vanishes due to the $z$ integration. 
To second order, there are no such terms, and the KPZ terms we are after are thus of order $T^{3/2}$. Among those, some arise from the first cumulant (with the bulk fluctuation field appearing twice), some others from the second cumulant (with the bulk fluctuation field appearing four times) and finally some from the third cumulant (involving six bulk fluctuation fields). By visual inspection of Eq.~\eqref{eq:AMAfullNGugly} one sees that terms of the form $\hb h^2$ are actually already in a KPZ-like form $\hb (\bnabla_\br h)^2$,
where the fields are possibly evaluated at different times and locations. We now examine a few of those KPZ terms.

One such contribution from the first cumulant is
    \begin{equation}\label{eq:AMA_KPZ_1}
        -\frac{\lambda}{T}\int_{\br,t} \red{\hb(\bnabla_\br h)^2}\argx{\ell_0,\argx{(\p_z {\chi'}^\perp)^2}'_0}
    \end{equation}
This is a purely active contribution that corresponds to the Feynman diagram: 
\begin{center}
\begin{tikzpicture}
  \begin{feynman}
    \vertex (i1) {\(\hr\)};
    \vertex (i2) [below=.9cm of i1,xshift=.27cm] {\(\hbr\)};
    \vertex (i3) [below=1.8cm of i1] {\(\hr\)};

    \vertex (v1) [left=1.27cm of i2];
    \vertex (v2) [left=0.9cm of v1];

    \diagram* {
        (v1) -- [red, thick]  (i1),      
        (v1) -- [red,thick,->] (i2),
      (v1) -- [red,thick] (i3),
      
      (v1) -- [half left, thick,looseness=1.75] (v2),
      (v1) -- [half right, thick, looseness=1.75] (v2),
    };
  \end{feynman}
  \filldraw[red] ($(v1)!0.4!(i1)$) circle (2pt);
  \filldraw[red] ($(v1)!0.4!(i3)$) circle (2pt);
\end{tikzpicture}
\end{center}
where the dots indicate a spatial derivative $\bnabla_\br$. Note that in Eq.~\eqref{eq:AMA_KPZ_1}, $\hb(\bnabla_\br h)^2$ are already at equal time and location, so this is the exact form of a KPZ nonlinearity.

A contribution from the second cumulant is
\begin{align}
    -\frac{4\lambda}{T^2}&\int_{\br,t}\int_{\br',t'} \hbr(\br,t)\red{\bnabla_\br h}(\br,t)\cdot\, \red{\bnabla_\br h}(\br',t')\cdot\notag\\
    &\int_{z,z'}\left<\big[(\bnabla_\br\p_z {\chi'}^\perp - \lambda \p_z m_c\bnabla_\br{\chi'}^\perp)\chib^\perp\big](\br',z',t')\right.\notag\\
    &\qquad \left.\big[ \bnabla_\br{\chi'}^\perp \p_z {\chi'}^\perp\big](\br,z,t)\right>'_{c,0}\ell_0(z)\label{eq:AMA_KPZ_2}
\end{align}
\if{
A contribution from the second cumulant is
\begin{align}
    -\frac{4\lambda}{T^2}&\int_{\br,t}\int_{\br',t'} \hb(\br,t)\bnabla_\br h(\br,t)\cdot\, \bnabla_\br h(\br',t')\cdot\notag\\
    &\left<\int_{z,z'}\big[(\bnabla_\br\p_z {\chi'}^\perp - \lambda \p_z m_c\bnabla_\br{\chi'}^\perp)\ell_0\big](\br,z,t)\right.\notag\\
    &\qquad \left.\big[\chib^\perp \bnabla_\br{\chi'}^\perp \p_z {\chi'}^\perp\big](\br',z',t')\right>'_{c,0}\label{eq:AMA_KPZ_2}
\end{align}
This corresponds to the Feynman diagram
\begin{center}
\begin{tikzpicture}
  \begin{feynman}
    \vertex (i1) {\(h\)};
    \vertex (f1) [above=1cm of i1,xshift=3.9cm] {\(h\)};
    \vertex (f2) [below=2cm of f1] {\(\hb\)};

    \vertex (v1) [right=1.4cm of i1];
    \vertex (v2) [right=1.5cm of v1];
    \vertex (v3) [below=0.75cm of v1];

    \diagram* {
        (i1) -- [red, thick]  (v1),      
        (v2) -- [red,thick] (f1),
        (v2) -- [red,thick,->] (f2),
      
      (v1) -- [fermion, thick] (v2),
      (v1) -- [half left, thick,looseness=1.6] (v3),
      (v1) -- [half right, thick, looseness=1.6] (v3),

    };
  \end{feynman}
\filldraw[red] ($(v1)!0.4!(i1)$) circle (2pt);
\filldraw[red] ($(v2)!0.4!(f1)$) circle (2pt);
\filldraw[black] (2.65,0) circle (2pt);
\filldraw[black] (1.75,-0.375) circle (2pt);
\end{tikzpicture}
\end{center}}\fi
This corresponds to the Feynman diagrams:
\begin{tikzpicture}
  \begin{feynman}
    \vertex (i1) {\(\hr\)};
    \vertex (f1) [above=0.9cm of i1,xshift=3.27cm] {\(\hr\)};
    \vertex (f2) [below=1.8cm of f1] {\(\hbr\)};

    \vertex (v1) [right=1.27cm of i1];
    \vertex (v2) [right=1.1cm of v1];

    \diagram* {
        (i1) -- [red, thick]  (v1),      
        (v2) -- [red,thick] (f1),
        (v2) -- [red,thick,->] (f2),
      
      (v1) -- [fermion, thick,half left, looseness=1.5] (v2),
      (v1) -- [thick,half right, looseness=1.5] (v2),
      
      (v2) -- [red](f1),
    };
  \end{feynman}
\filldraw[red] ($(v1)!0.4!(i1)$) circle (2pt);
\filldraw[red] ($(v2)!0.4!(f1)$) circle (2pt);
\filldraw[black] (2.3,0.25) circle (2pt);
\filldraw[black] (1.34,-0.25) circle (2pt);
\end{tikzpicture}
\begin{tikzpicture}
  \begin{feynman}
    \vertex (i1) {\(\hr\)};
    \vertex (f1) [above=0.9cm of i1,xshift=3.27cm] {\(\hr\)};
    \vertex (f2) [below=1.8cm of f1] {\(\hbr\)};

    \vertex (v1) [right=1.27cm of i1];
    \vertex (v2) [right=1.1cm of v1];

    \diagram* {
        (i1) -- [red, thick]  (v1),      
        (v2) -- [red,thick] (f1),
        (v2) -- [red,thick,->] (f2),
      
      (v1) -- [ thick,half left, looseness=1.5] (v2),
      (v1) -- [fermion,thick,half right, looseness=1.5] (v2),
      
      (v2) -- [red](f1),
    };
  \end{feynman}
\filldraw[red] ($(v1)!0.4!(i1)$) circle (2pt);
\filldraw[red] ($(v2)!0.4!(f1)$) circle (2pt);
\filldraw[black] (2.3,0.25) circle (2pt);
\filldraw[black] (1.34,0.25) circle (2pt);
\end{tikzpicture}
where the oriented black line represents the response propagator. The fields $\chi'^\perp$ and $\chib^\perp$ have finite correlation lengths and times (these are given by the $\lambda_i$'s for $i>0$), hence at large time and spatial scales, these two diagrams generate effective local contributions of the KPZ form, along with contributions involving higher derivatives.

\if{This contribution is at first glance nonlocal in space and time, as opposed to the first cumulant contribution. In momentum space, it reads $\hb_{\bq_3}(t)h_{\bq_1}(t) h_{\bq_2}(t')i\bq_1\cdot i\bq_2\cdot$, contracted with correlators of the bulk fields. However, if we expand in $t-t'$, the most relevant term at low frequency is the first one, corresponding to equal time. As for the momenta, diagonality in momentum of the quadratic action $S'_0$ sets $\bq_1+\bq_2=\bq_3$. It is possible that the loop could contribute additional powers of $\bq_1,\bq_2$, but we can expand in the external momenta $\bq_1,\bq_2$ and keep only the $\bq_1^0\bq_2^0$ contribution from the loop. If it is nonvanishing, Eq.~\eqref{eq:AMA_KPZ_2} is exactly a KPZ term to lowest order in frequency and momentum; if it is zero, Eq.~\eqref{eq:AMA_KPZ_2} is of higher order and can be neglected with respect to \textit{e.g.} Eq.~\eqref{eq:AMA_KPZ_1}.
}\fi

As a final example, a contribution from the third cumulant is
\begin{align}
    -\frac{2}{T^3}&\int_{\br,t}\int_{\br',t'} \int_{\br'',t''}\hbr(\br'',t'')\red{\bnabla_\br h}(\br,t)\cdot\, \red{\bnabla_\br h}(\br',t')\cdot\notag\\
    &\int_{z,z',z''} \ell_0(z'')\left<\Big[\frac{f'''(m_c)}{2}({\chi'}^\perp)^2\Big](\br'',z'',t'')\right.\notag\\
    &\qquad\big[(\bnabla_\br\p_z {\chi'}^\perp - \lambda \p_z m_c\bnabla_\br{\chi'}^\perp)\chib^\perp\big](\br',z',t')\notag\\
    &\qquad \left.\big[(\bnabla_\br\p_z {\chi'}^\perp - \lambda \p_z m_c\bnabla_\br{\chi'}^\perp)\chib^\perp\big](\br,z,t)\right>'_{c,0}\label{eq:AMA_KPZ_3}
\end{align}
whose graphical representation is
\begin{tikzpicture}
  \begin{feynman}
    \vertex (i1) {\(\hr\)};
    \vertex (i2) [below=2.8cm of i1]{\(\hr\)};
    \vertex (f1) [below=1.4cm of i1,xshift=2.87cm] {\(\hbr\)};

    \vertex (v1) [right=0.9 of i1,yshift=-0.9cm];
    \vertex (v2) [right=0.9 of i1,yshift=-1.9cm];
    \vertex (v3) [left=1.27cm of f1];

    \diagram* {
        (i1) -- [red, thick]  (v1),      
        (i2) -- [red,thick] (v2),
        (v3) -- [red,thick,->] (f1),
      
      (v2) -- [fermion, thick] (v1),
      (v1) -- [fermion, thick] (v3),
      (v2) -- [thick] (v3),
    };
  \end{feynman}
\filldraw[red] ($(v1)!0.4!(i1)$) circle (2pt);
\filldraw[red] ($(v2)!0.4!(i2)$) circle (2pt);
\filldraw[black] (0.9,-1.15) circle (2pt);
\filldraw[black] (1.1,-1.75) circle (2pt);
\end{tikzpicture}
\begin{tikzpicture}
  \begin{feynman}
    \vertex (i1) {\(\hr\)};
    \vertex (i2) [below=2.8cm of i1]{\(\hr\)};
    \vertex (f1) [below=1.4cm of i1,xshift=2.87cm] {\(\hbr\)};

    \vertex (v1) [right=0.9 of i1,yshift=-0.9cm];
    \vertex (v2) [right=0.9 of i1,yshift=-1.9cm];
    \vertex (v3) [left=1.27cm of f1];

    \diagram* {
        (i1) -- [red, thick]  (v1),      
        (i2) -- [red,thick] (v2),
        (v3) -- [red,thick,->] (f1),
      
      (v1) -- [thick] (v2),
      (v1) -- [fermion,thick] (v3),
      (v2) -- [fermion,thick] (v3),
    };
  \end{feynman}
\filldraw[red] ($(v1)!0.4!(i1)$) circle (2pt);
\filldraw[red] ($(v2)!0.4!(i2)$) circle (2pt);
\filldraw[black] (0.9,-1.15) circle (2pt);
\filldraw[black] (0.9,-1.65) circle (2pt);
\end{tikzpicture}

Again, the time integrals appearing in these loops involve the finite characteristic times $\lambda_i^{-1}$ and thus lead to effective local vertices at large times (the same holds space-wise).
\if{The same reasoning as above applies: expanding in $t'-t$ and $t''-t'$, only the lowest order equal-time contribution remains; diagonality in momentum of $S'_0$ ensures that the momenta of $\hb_{\bq_3}$ and $h_{\bq_1}h_{\bq_2}$ are equal; expanding the loop contribution in $\bq_1,\bq_2$ and keeping the lowest order simplifies Eq.~\eqref{eq:AMA_KPZ_3} into a KPZ term.}\fi 
Unlike the previous contributions \eqref{eq:AMA_KPZ_1} and \eqref{eq:AMA_KPZ_2}, it is not immediately visible that the one of Eq.~\eqref{eq:AMA_KPZ_3} vanishes in equilibrium. Further comments are gathered in App.~\ref{sec:passiveKPZ}.

Note that all of the above contributions appear in the effective equation for $h$ as a $T(\bnabla_\br h)^2$ contribution, an order $T^{3/2}$ smaller than the linear part of the dynamics, and an order $\sqrt T$ smaller than curvature corrections~\cite{Kawasaki1982_KineticDrumhead1}.
There are several other KPZ contributions from the first three cumulants at this order in temperature which we haven't explicitly written. In order to obtain the microscopic (bare) coefficient of the KPZ contribution, all those diagrams, in their low frequency and low momentum limit, should be summed up. In a RG picture, the resulting coefficient is the initial value of the flow.

It is instructive to probe the extent to which a KPZ term could arise from the ansatz of Eq.~\eqref{eq:rottenansatz}. Had we used for $h$ the definition $\phi=m_c(z-h(\br,t))$, we would have found the following nonlinear equation in the Stratonovich discretization:
\begin{equation}\label{eq:ActiveModelAAnsatzInterm}
\begin{split}
    -\partial_t h \, \partial_z m_c(z) &=-\nabla_\br^2 h\, \partial_z m_c(z)+  \sqrt{2T}{\eta}(\bt{r},z+h,t) \\
    &+ (\nabla_\br h)^2\argc{\partial_z^2 m_c - \lambda(\partial_z m_c)^2}(z) \,.
\end{split}
\end{equation}
At the linear level, there is a definite series of steps that lead to the  correct equation: these consist in multiplying Eq.~\eqref{eq:ActiveModelAAnsatzInterm} by $\ell_0$ and then in integrating over $z$. This leads to
\begin{equation}
    \p_t h = \nabla_\br^2 h + \beta (\nabla_\br h)^2 + \sqrt{\frac{2T\Sigma}{\sigma_A^2}}\tilde \eta\;,
\end{equation}
with $\beta = \lambda\argx{\ell_0,(\p_z m_c)^2} - \argx{\ell_0,\p_z^2 m_c}$. But it turns out that using the explicit expression for $\ell_0$, the coefficient $\beta$ vanishes. (This term corresponds to the first order KPZ contribution that was said to vanish above Eq.~\eqref{eq:AMA_KPZ_1}.) Our systematic derivation above instead predicts the appearance of KPZ terms to an order $O(T)$ higher, which the simplified ansatz can never catch.

\subsection{Active model B+}

\subsubsection{Motivations for introducing active model B+}
Let us now turn to the simplest nonequilibrium generalization of model B.  
In active matter, the motility-induced phase separation~\cite{Tailleur2008_Statistical,fily2012athermal,redner2013structure,buttinoni2013dynamical,cates2015motility,liu2019self,van2019interrupted} undergone by pairwise repulsive self-propelled particles, presents features strikingly similar to equilibrium phase separation, in spite of being an intrinsically nonequilibrium phenomenon.

However, deriving exact (fluctuating) hydrodynamics for such systems is a challenging task~\cite{Tailleur2008_Statistical,bialke2013microscopic,Solon2018_GeneralizedThermodynamics,speck2021coexistence,omar2023mechanical,kafri2026active}. An alternative phenomenological route is to first complement Eq.~\eqref{eq:B} with an active contribution as in Eq.~\eqref{eq:AMA} affecting the chemical potential~\cite{Wittkowski2014_Scalar}. As noted in \cite{Nardini2017_Entropy,tjhung2018cluster}, model B can harbor another nonequilibrium term not present in active model A, thus leading to  active model B+:
\begin{align}
        \p_t \phi = &\bnabla^2\argc{\funcder{F}{\phi} + \lambda(\nabla\phi)^2} -\zeta\bnabla\cdot[(\bnabla^2\phi) \bnabla\phi]\notag\\
        &+ \sqrt{2T}\bnabla\cdot\bt{\eta}(\br,z,t)\;.\label{eq:AMB+}
\end{align}
where the components of $\boldsymbol{\eta}$ are independent Gaussian white noises with correlations $\argx{\eta_i(\br,z,t)\eta_j(\br',z',t')} = \delta_{ij}\delta(t-t') \delta(z-z')\delta^{(d-1)}(\br-\br')$.
Starting from this field theory, the interface dynamics was derived in \cite{Fausti2021_Capillary} using the ansatz of Eq.~\eqref{eq:rottenansatz}. 
A $q^3$ capillary relaxation spectrum was reported, similar to model B, albeit with a different capillary surface tension than in model B. 

Here we derive the linear interface dynamics using our path-integral formalism.

\subsubsection{Dynamical action for interface dynamics}

The mean-field steady-state profile of the active model B+ satisfies
\begin{equation}\label{eq:AMBMF}
    \p_z^2 \argc{-\p_z^2 m_c + f'(m_c) + \frac{2\lambda-\zeta}{2}(\p_z m_c)^2} = 0\;.
\end{equation}
The lack of steady-state current enforces the stronger constraint 
\begin{equation}
    j_c = -\p_z \argc{-\p_z^2 m_c + f'(m_c) + \frac{2\lambda-\zeta}{2}(\p_z m_c)^2} = 0\;.
\end{equation}

The corresponding dynamical action truncated to quadratic order reads
\begin{align}
    S_0 = \frac{1}{T}&\int_{t,\br,z}\big\{\chib[\p_t - \bnabla^2 (-\bnabla_\br^2 + \Omega)\notag\\
    &+2\zeta \p_z^2 m_c\bnabla_\br^2](-h\p_z m_c + \chi^\perp) - (\bnabla\chib)^2\big\}\;,\label{eq:AMBquadS}
\end{align}
where 
\begin{equation}\label{eq:omegaalpha}
    \Omega_\alpha = -\p_z^2 + f''(m_c) + 2\alpha\p_z m_c \p_z\;,
\end{equation}
with $\alpha ={\lambda-\zeta/2}$. We note that, up to $\lambda \to \lambda-\zeta/2$, the operator~\eqref{eq:omegaalpha} is the same as in Eq.~\eqref{eq:DefOmegaLambda} for active model A.
Going to Fourier space, we define the evolution operator
\begin{equation}\label{eq:GammaAMB+}
    \Gamma_{\bq,\omega} = i\omega + \mathrm L_\bq (\bq^2 + \Omega_\alpha) - 2\zeta \bq^2 \p_z^2m_c\;,
\end{equation}
where $\mathrm L_\bq$ was defined in Eq.~\eqref{eq:defLq}. We choose for $u_0$ the zero eigenvalue eigenvector of $\Omega_\alpha^\dagger$. It is given by Eq.~\eqref{eq:l0} upon replacing $\lambda$ with $\alpha=\lambda-\zeta/2$ (we keep the notation $\ell_0$ for this eigenvector).    We then decompose the field $\Gamma^\dagger_{\bq,\omega}\chib_{\bq,\omega}$ according to
\begin{align}
    &\Gamma^\dagger_{\bq,\omega}\chib_{\bq,\omega} = -\tilde h_{\bq,\omega}\ell_0 + (\Gamma^\dagger_{\bq,\omega}\chib_{\bq,\omega})^\perp,\;\text{with}\\
     &(\Gamma_{\bq,\omega}^\dagger\chib_{\bq,\omega})^\perp = \Pi^\dagger \Gamma_{\bq,\omega}^\dagger\chib_{\bq,\omega}\;.\end{align}
As for model B, the $\dagger$ solely refers to the adjoint with respect to the scalar product defined in Eq.~\eqref{eq:SP}. The action for the fields $\tilde h, h,(\Gamma^\dagger \chib)^\perp,\chi^\perp$ takes the form
\begin{equation}
\begin{split}
    S&= \frac{1}{T}\int_{\omega,\bq}\Big\{ \tilde h_{\bq,\omega} h_{\bq,\omega} + \argx{(\Gamma_{\bq,\omega}^\dagger\chib_{\bq,\omega})^\perp,\chi^\perp_{\bq,\omega}}\\
    &-\left<\mathrm{L}_\bq(\Gamma_{\bq,\omega}^\dagger)^{-1}\big[-\tilde h_{\bq,\omega} \ell_0 + (\Gamma_{\bq,\omega}^\dagger\chib_{\bq,\omega})^\perp\big],\right.\\
    &\left.(\Gamma_{-\bq,-\omega}^\dagger)^{-1}\!\big[\!-\tilde h_{-\bq,-\omega} \ell_0 + (\Gamma_{-\bq,-\omega}^\dagger\chib_{-\bq,-\omega})^\perp\big]\right>\!\Big\}\,.\\
      \end{split}
\end{equation}
Since the action is linear in $\chi^\perp$, integration over $\chi^\perp$ generates the constraint $(\Gamma^\dagger \chib)^\perp\propto\ell_0$, which, since $(\Gamma^\dagger \chib)^\perp$ has no component on $\ell_0$, sets $(\Gamma^\dagger \chib)^\perp$ to zero. The constraint is then enforced by the integration over $(\Gamma^\dagger \chib)^\perp$. This leads to the action for $\tilde h,h$
\begin{equation}\label{eq:AMBS1}
\begin{split}
    S_h&= \frac{1}{T}\int_{\omega,\bq}\Big\{ \tilde h_{\bq,\omega} h_{\bq,\omega}\\
    &-\tilde h_{\bq,\omega}\tilde h_{-\bq,-\omega}\left<\mathrm{L}_\bq(\Gamma_{\bq,\omega}^\dagger)^{-1} \ell_0 ,(\Gamma_{-\bq,-\omega}^\dagger)^{-1} \ell_0 \right>\Big\}\,.\\
      \end{split}
\end{equation}
We now need to simplify the scalar product $\argx{\mathrm L_\bq(\Gamma_{\bq,\omega}^\dagger)^{-1} \ell_0,(\Gamma_{-\bq,-\omega}^\dagger)^{-1} \ell_0}$. To this end, we look at the lowest eigenvalue $\lambda_0(q)$ of the operator $ \mathrm L_\bq (\bq^2 + \Omega_\alpha) - 2\zeta \bq^2 \p_z^2m_c$, which satisfies
\begin{equation}
    \lambda_0(q) = q^2\frac{\argx{\ell_0,R_0(q)}-2\zeta\argx{\mathrm L_\bq^{-1}\ell_0,\p_z^2 m_c R_0(q)}}{\argx{L_\bq^{-1}\ell_0, R_0(q)}}\;,
\end{equation}
where $R_0(q)$ is the corresponding right eigenvector. We denote $L_0(q)$ the corresponding left eigenvector.

As in passive model B (see App.~\ref{sec:BShino}), for large wavelengths, the eigenvalue $\lambda_0(q)$ 
can be obtained perturbatively from the eigenvalue $0$ of $\mathrm L_\bq \Omega_\alpha$. 
In this limit, we approximate $R_0(q)$ by its unperturbed expression $R_0(q)\simeq R_0(0)=\p_z m_c$.
Using  notations that mirror those of \cite{Fausti2021_Capillary,Besse2023_Interface}, we introduce $A_A(q) = 2\sigma_A q\argx{\p_z m_c, \mathrm L_\bq^{-1} \ell_0}$ and $B_A(q) = 2\sigma_A^2 q\argx{\ell_0, \mathrm L_\bq^{-1} \ell_0}$. Then, we find that:
\begin{equation}
\begin{split}
    \lambda_0(q) = \frac{2\sigma_A q^3}{A_A(q)}&\Big[1 + \frac{\zeta}{2}\iint_{z,z'}\text{sign}(z'-z)\ee^{-q|z-z'|}\\
    &\times\ell_0(z')(\p_z m_c)^2(z)\Big]\;,
\end{split}
\end{equation}
to lowest order in $q$. The scalar product in the action \eqref{eq:AMBS1} is dominated by the contribution from this eigenvalue, such that 
\begin{align}
    \langle\mathrm L_\bq(\Gamma_{\bq,\omega}^\dagger)^{-1} &\ell_0,(\Gamma_{-\bq,-\omega}^\dagger)^{-1} \ell_0\rangle \simeq \frac{1}{|i\omega + \lambda_0(q)|^2}\notag\\
    &\times \argx{\p_z m_c, \ell_0}^2 \frac{\argx{\ell_0,\mathrm L_\bq^{-1}\ell_0}}{\argx{\p_z m_c,\mathrm L_\bq^{-1}\ell_0}^2}\;.
\end{align}

Then, the effective action for $\tilde h,h$ simplifies into
\begin{equation}
\begin{split}
    S_h= \frac{1}{T}\int_{\omega,\bq}\Big\{ \tilde h_{\bq,\omega}h_{\bq,\omega} - \frac{\tilde h_{\bq,\omega}\tilde h_{-\bq,-\omega}}{|i\omega + \lambda_0(q)|^2}\frac{2 q B_A(q)}{A_A(q)^2}\Big\}\;.\end{split}
\end{equation}
Introducing $\hb_{\bq,\omega} = \tilde h_{\bq,\omega}(i\omega + \lambda_0(q))^{-1}$, the quadratic action for $\hb$ and $h$ reads to lowest order in $q$
\begin{equation}
\begin{split}
    S_{h}[\hb,h]= &\frac{1}{T}\int_{\omega,\bq}\Big\{\hb_{\bq,\omega}(i\omega + \lambda_0(q))h_{\bq,\omega}\\
    &- \frac{2 q B_A(q)}{A_A(q)^2}\hb_{\bq,\omega}\hb_{-\bq,-\omega}\Big\}\;.\end{split}
\end{equation}
The corresponding Langevin equation for $h$ is
\begin{equation}
\begin{split}
    \p_t h&(\bq,t) = \sqrt{2T\frac{2 q B_A(q)}{A_A(q)^2}}\tilde{\eta}(\bq,t) -\frac{2\sigma_A q^3}{A_A(q)}h(\bq,t)\times\\
    &\Big[1 + \frac{\zeta}{2}\iint_{z,z'}\!\!\!\!\!\!\text{sign}(z'-z)\ee^{-q|z-z'|}\ell_0(z')(\p_z m_c)^2(z)\Big]\;,
\end{split}
\end{equation}
where the Gaussian white noise $\tilde\eta$ has correlations $\argx{\tilde{\eta}(\bq,t)\tilde{\eta}(\bq',t')} = (2\pi)^{d-1}\delta(t-t')\delta^{(d-1)}(\bq+\bq')$.\\
This coincides with the linear dynamics  predicted in~\cite{Fausti2021_Capillary} starting from the mean-field ansatz of Eq.~\eqref{eq:rottenansatz}, upon noticing that the derivative of the auxiliary field $\psi'$ in~\cite{Fausti2021_Capillary} is exactly $\sigma_A \ell_0$. 

The fact that the simplified ansatz yields the same result as our controlled derivation can be understood in the following way.
Working at the level of the equation of motion for $h$ and $\chi$ to lowest order in $\sqrt T$, 
we would obtain Eq.~\eqref{eq:B_EoM} with $\Gamma_{\bq,\omega}$ now given by Eq.~\eqref{eq:GammaAMB+}. 
To decouple interfacial from bulk dynamics and eliminate $\chi$, one then projects onto $(\Gamma_{\bq,\omega}^\dagger)^{-1}\ell_0$, whose leading order in $q$ is proportional to $\mathrm L_\bq^{-1}\ell_0$. 
Therefore, as in model B, there exists---to lowest order in $\sqrt T$ and in $q$---a simplified route that in practice consists in neglecting bulk fluctuations \textit{a priori} and projecting the resulting linear equation for $h$ onto $\mathrm L_\bq^{-1}\ell_0$.

\begin{figure}
\centering
\includegraphics[width=1.0\columnwidth]{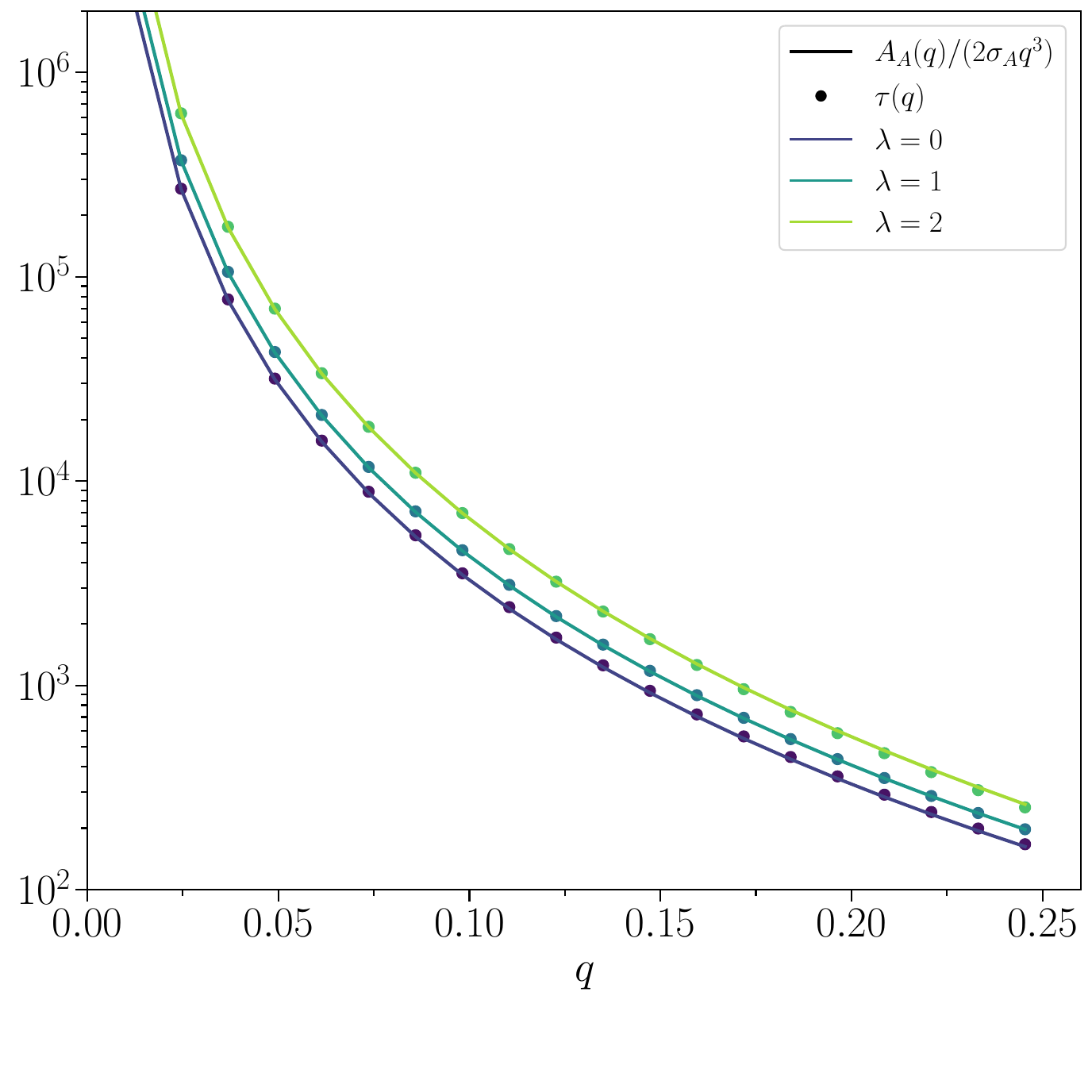}
    \caption{Relaxation time of a one-dimensional interface in a $512\times512$ system from an initial condition $\frac{0.1}{q} \cos\argp{qr}$, as a function of $q$, for different values of the activity $\lambda$. The dots are obtained from an exponential fit of the amplitude of the relaxing interface in active model B ($\zeta=0$) simulations. The full lines are the theoretically predicted values for $\tau(q) = \frac{A_A(q)}{2\sigma_A q^3}$.}
    \label{fig:multlambda}
\end{figure}

We numerically confirm the expression for the relaxation rate for $\zeta=0$ and different values of $\lambda$ in Fig.~\ref{fig:multlambda}. Note that we have stuck to $\zeta=0$ (active model B) which is easier to stabilize numerically than the full model Active B+. Much like for passive model B, it appears that the theoretical prediction for the relaxation rate holds beyond the lowest order in $q$.

\section{Prospects}\label{sec:prospects}

We have introduced a method, that is both well-controlled and versatile, 
to derive stochastic interface dynamics between two stable phases starting from a field theory. 
Instead of the popular shortcut assuming that the fluctuations of the order parameter can be described by a simple shift of the mean-field profile, we account for the coupling between interface deformations and bulk fluctuations.
We show that interface and bulk fluctuations are of the same order in $\sqrt T$, so that neglecting the latter is not justified \textit{a priori}. 
To obtain the effective dynamics of the interface, one then has to eliminate bulk fluctuations, which can be done in a dynamical-action formalism, by integrating them out, together with their associated response fields.
Our formalism first systematically recovers known equilibrium results, regardless of the presence of a conservation law or coupling to other fields. In addition, we derive by the same token both the relaxation rate and the noise amplitude. Providing a derivation for the noise in addition to the dispersion relation is a significant improvement, as both quantities are required to define the capillary surface tension, even in equilibrium. 

Bulk fluctuations generically relax faster than the interface. It may therefore come as a surprise that a naive adiabatic approximation may lead to inconsistent results. This is because, by working at fixed interface deformation, bulk fluctuations do relax faster, but not to zero. Instead they relax to a constrained interface-dependent expression. No guidelines had been established for the use of this shortcut in locally out-of-equilibrium systems, hence the necessity for a well-controlled approach. 

Upon adapting the definition of the interface to the new properties of the system, the formalism we introduced consistently applies out of equilibrium, such as to active versions of models A and B. For active model B+, we provide a consistent derivation of the results of \cite{Fausti2021_Capillary}, previously obtained with the simplified ansatz. The method we propose is to our knowledge the first well-controlled derivation of interface dynamics in an active system. 
We emphasize that the use of the shortcut is dangerous if no prior proof exists, which establishes the applicability of the ansatz in a given model together with the specific path to follow.

Another important contribution of the path-integral formulation is that it allows to systematically derive nonlinear orders in the low-temperature limit, which have attracted a lot of interest in the active community~\cite{Besse2023_Interface,toner2023roughening,Caballero2025_Interface,maire2025conservation}. 
In passive model A, we immediately recover the expression for the fluctuation-induced drift (and its divergence with the microscopic cutoff) in the case of an asymmetric potential with degenerate minima, as well as the curvature contribution.
For active model A, we also identify a fluctuation-induced velocity and KPZ nonlinearities, which are already at the bare level. All these predictions are out of reach from the simplified ansatz. 
An important next step will be to derive the bare KPZ-like terms of active model B+ and to discuss whether they renormalize to the proposed `$|\bq|$KPZ' nonlinearity.

Beyond the models that we have considered in this article, it would be interesting to investigate polar or nematic systems described by Landau-Lifschitz-Gilbert-like models~\cite{lucassen2009current}. We speculate that their analysis will bring us back to theories similar to models C and D.
Activity is one important way to drive a system out of equilibrium. 
However, interfacial dynamics arise in a variety of other nonequilibrium systems.
For instance, in sheared phase-separated systems, we expect to recover the predictions of \cite{Bray2001_Interface} to linear order, but nonlinear corrections might differ from the ones they predicted using the simplified ansatz. 
In a similar vein, it would be interesting to put our formalism to work in the presence of propagating pushed fronts into unstable phases, a topic that has attracted attention recently~\cite{birzu2018fluctuations}. Finally, equilibrium systems with quenched disorder have been studied using the simplified ansatz in~\cite{caballero2020bulk}. There, it is simply not known which predictions are affected by accounting for all fluctuations in the system, especially when the relaxation of bulk deformations are hindered by the disorder.

\acknowledgments We are indebted to T.~Arnoulx de Pirey, M.~Besse, H.~Hayakawa, K.~Mandadapu, C.~Nardini, S.-i.~Sasa, K.~Sekimoto, V.~Lecomte, A.~Solon, H.~Chaté and L.~Neville, and especially to T.~Ohta, for very useful discussions. This work was partly funded by the Agence Nationale de la Recherche grant THEMA No 20-CE30-0031-01. LS acknowledges financial support from the Fondation CFM pour la Recherche.

\appendix

\section{Choice of $u_0$ for model A}\label{sec:choiceofu0}
As discussed in Sec.~\ref{sec:A} for model A, a suitable choice of $u_0$ in the definition of the interface Eq.~\eqref{eq:definterface} is $u_0 = \sigma^{-1} \p_z m_c$. 
If, instead of choosing $u_0 \propto \p_z m_c$, we take $u_0$ with another nonzero overlap with any other $e_i$, then the resulting noise is non-Markovian, leading to a generalized Langevin equation with memory. However, to lowest order in the wavevector $\bq$, we recover the same linear dynamics as in Eq.~\eqref{eq:A_Inteq}. Indeed, if one writes $\alpha_i = \argx{e_i,u_0}$, with $\alpha_0 = 1/\sqrt{\sigma}$, Eq.~\eqref{eq:AquadS_halfdecoup_decomp} becomes
\begin{align}
       S = \frac{1}{T}&\int_{t,\br}
       \hb (\p_t h - \bnabla_\br^2h) - \frac{\hb^2}{\sigma}\notag\\
       &+\sum_{i>0}[\bar c_i(\p_t-\bnabla_\br^2 + \lambda_i)c_i - \bar c_i^2\notag\\
       &-\hb^2 \alpha_i^2 - \lambda_i \alpha_i \hb c_i + 2 \alpha_i \hb \bar c_i] \;.
\end{align}
In Fourier space in $t$ and $\br$, this reads
\begin{align}
       S &= \frac{1}{T}\int_{\omega,\bq}
       \hb_{\bq,\omega} (i\omega +\bq^2)h_{\bq,\omega} - \frac{\hb_{\bq,\omega}\hb_{-\bq,-\omega}}{\sigma}\notag\\
       &+\sum_{i>0}[\bar c_{i,\bq,\omega}(i\omega+\bq^2 + \lambda_i)c_{i,\bq,\omega} - \bar c_{i,\bq,\omega}\bar c_{i,-\bq,-\omega}\notag\\
       &-\hb_{\bq,\omega}\hb_{-\bq,-\omega} \alpha_i^2 + \alpha_i\hb_{\bq,\omega}(- \lambda_i c_{i,\bq,\omega} + 2\bar c_{i,-\bq,-\omega})] \;.\label{eq:AppSA}
\end{align}
With such a generic choice of $u_0$, there is no immediate decoupling between the $\hb,h$ and $\bar c_i,c_i$ fields. To decouple them, we need to shift the bulk fields $\bar c_i,c_i$ by their averages as follows:
\begin{align}
    \bar c'_{i,\bq,\omega} &= \bar c_{i,\bq,\omega}-\frac{\lambda_i \alpha_i \hb_{\bq,\omega}}{i\omega + \bq^2+\lambda_i}\\
    c'_{i,\bq,\omega} &= c_{i,\bq,\omega}-\frac{2 \alpha_i \hb_{-\bq,-\omega}}{i\omega + \bq^2+\lambda_i}\argp{\frac{\lambda_i}{-i\omega + \bq^2 + \lambda_i} - 1\!}\;.
\end{align}
Then the action \eqref{eq:AppSA} becomes
\begin{align}
       S &= \frac{1}{T}\int_{\omega,\bq}
       \hb_{\bq,\omega} (i\omega +\bq^2)h_{\bq,\omega} - \frac{\hb_{\bq,\omega}\hb_{-\bq,-\omega}}{\sigma}\notag\\
       & -\sum_{i>0}\alpha_i^2 \Big|\frac{i\omega+\bq^2}{i\omega + \bq^2 +\lambda_i}\Big|^2\hb_{\bq,\omega}\hb_{-\bq,-\omega}\notag\\
       &+\sum_{i>0}[\bar c'_{i,\bq,\omega}(i\omega+\bq^2 + \lambda_i)c'_{i,\bq,\omega} - \bar c'_{i,\bq,\omega}\bar c'_{i,-\bq,-\omega}]\;.\label{eq:SappA}
\end{align}
The fields $\hb,h$ and $\bar c_i',c_i'$ are now decoupled, so the action $S_h$ can directly be read off from the two first lines of Eq.~\eqref{eq:SappA}.
For $q \xi \ll 1$, anticipating that the relevant values of $\omega$ are of order $q^2$,  we can write the effective action for $\hb,h$ in the low-wavevector and low-frequency limits as
\begin{equation}
    \begin{split}
       S_h[\hb, h] = \frac{1}{T}&\int_{\omega,\bq}\Big\{\hb_{\bq,\omega}(i\omega + \bq^2)h_{\bq,\omega}\\
       &-\frac{\hb_{\bq,\omega}\hb_{-\bq,-\omega}}{\sigma}[1 + O(q^2)] \Big\}\;,
    \end{split}
\end{equation}
where we have used the fact that the $\lambda_i$'s are of order $\xi^{-2}$. To leading order in $q\xi$ this is the same as Eq.~\eqref{eq:A_Sh}. In other words, obtaining the linear interface dynamics to lowest order in $q$ can be done without specifying $u_0$ in the definition of the interface, as long as $\argx{u_0,\p_z m_c}\neq 0$.

There is a more physical reason for the choice  $u_0\propto \p_z m_c$, besides not requiring any large wavelength approximation, which goes as follows. At fixed bulk perturbation $\chi$, the dynamics should be invariant under a constant shift of $h$. 
Uniforms shifts of $h$ should therefore be excluded from the bulk fluctuations $\chi$. Since a shift $h\to h+\Delta h$ generates a terms $\propto\Delta h\p_z m_c$ in Eq.~\eqref{eq:DefChi}, $\chi$ should have no component along $\p_z m_c$~\cite{Kuramoto1980_Instability,Kawasaki1982_KineticDrumhead1,Kawasaki1982_KineticDrumhead2,ohta2025}.  
In equilibrium model A, the relevant operator is $\Omega_0$ and the relevant basis is the $e_i$'s, so writing that there is no $\p_z m_c$ component in $\chi$ amounts to enforcing its orthogonality to $e_0$, here proportional to $\p_z m_c$. For models with non-Hermitian operators $\Omega$, this reasoning has to be adapted accordingly: excluding the right eigenvector $\p_z m_c$ means that $\chi$ should be orthogonal to the corresponding left eigenvector~\cite{Kuramoto1980_Instability,Kawasaki1982_KineticDrumhead2,birzu2018fluctuations}.

\section{Higher order nonlinearities in model A}
\subsection{KPZ nonlinearity in model A and symmetry of the potential}\label{sec:passiveKPZ}
In section \ref{sec:NLmodelA}, we have shown that, in the presence of an asymmetric potential with degenerate minima, thermal fluctuations generate a nonvanishing drift velocity. This breaking of time-reversal symmetry is expected to generate a KPZ nonlinearity. Let us motivate this by looking at one of the contributions from the cumulant expansion:
\begin{align}
    \frac{1}{T^2}\int_{\br,t}&-\hb (\bnabla_\br h)^2\int_{\br',z,z',t'}\p_z m_c(z)
    \frac{f'''(m_c)}{2\sigma}(z')\notag\\
    &\argx{\p_z^2 \chi^\perp(\br,z,t)[\chib^\perp(\chi^\perp)^2](\br',z',t')}'_{c,0}\;.
\end{align}
Using the expression of the correlations as a function of the eigenmodes of $\Omega_0$ given in Eqs.~\eqref{eq:A_NGL_chibchi} and \eqref{eq:A_NGL_chi2}, the cumulant can be rewritten as
\begin{align}
    \langle\p_z^2 \chi^\perp(\br,z,t)&[\chib^\perp(\chi^\perp)^2](\br',z',t')\rangle'_{c,0}= T^2 \sum_{i,j> 0}\notag\\
    &\times \int_{\bq,\bp}\ee^{i\bq\cdot(\br'-\br)}\frac{\ee^{-(\bq^2 + \lambda_i)(t-t')}}{\bp^2 + \lambda_j}\Theta(t-t')\notag\\
    &\times e_i(z') e_j(z')^2 \p_z^2 e_i(z)\;.
\end{align}
Then, the integrals over $z$ and $z'$ become
\begin{align}
    \sum_{i>0}&\argx{\p_z m_c,\p_z^2 e_i}\argx{e_i,f'''(m_c)e_j^2}\ee^{-\lambda_i(t-t')}\notag\\
    &= \sum_{i>0}\argx{\p_z^3 m_c, e_i}\ee^{-\lambda_i(t-t')}\argx{e_i,f'''(m_c)e_j^2}\notag\\
    &=\left<\p_z^3 m_c\right|\big(\mathds \ee^{-\Omega_0(t-t')} - \left|e_0\right>\left<e_0\right|\big)\left|f'''(m_c)e_j^2\right>\notag\\
    &=\argx{\p_z^3 m_c,  \ee^{-\Omega_0(t-t')} f'''(m_c) e_j^2}  \notag\\
    &\qquad -\argx{\p_z^3 m_c,e_0}\argx{e_0, f'''(m_c)e_j^2}
\end{align}
If $f$ is a symmetric potential, the scalar products involving $f'''(m_c)$ vanish, whereas no such simplification occurs for a generic $f$. Generically, one thus expects the interface to follow KPZ dynamics in the case of an asymmetric potential with degenerate minima.

A similar reasoning can be applied to the KPZ term \eqref{eq:AMA_KPZ_3}, which appears in the third cumulant, even at $\lambda=0$, \textit{i.e.} in equilibrium.
Using the decompositions of $\chi^\perp$ and $\chib^\perp$ on the eigenvectors of $\Omega_0$ given by \eqref{eq:chiDecompEigenv} and \eqref{eq:chibDecompEigenv}, respectively, rewriting sums over $i>0$ as $G(\Omega_0)-G(0)\left|e_0\right>\left<e_0\right|$, and invoking parity arguments, it can be shown that this contribution also vanishes for a symmetric potential, as expected.\\

\subsection{Curvature correction in model A}\label{sec:curvature}
Diehl, Kroll and Wagner~\cite{Diehl1980} have shown at the level of the statics that the interfacial free energy for the Ginzburg-Landau $\phi^4$ field theory (the free energy of Eq.~\eqref{eq:F} with the potential of Eq.~\eqref{eq:f_standard}) is given, in the limit of low temperature, by
\begin{equation}\label{eq:Fdiehl}
    F[h] = \sigma\int_\br \sqrt{1+(\bnabla_\br h)^2}\;.
\end{equation}
It has been argued~\cite{Bausch1981_Critical} that the proper dynamical evolution consistent with the statics governed by $F[h]$ is the following multiplicative dynamics for $h$:
\begin{equation}\label{eq:Afullh}
    \p_t h = -\lambda\sqrt{1+(\bnabla_\br h)^2}\,\funcder{F}{h}+\sqrt{2T\lambda}\zeta\;,
\end{equation}
where the mobility $\sqrt{1+(\bnabla_\br h)^2}$ is required to have a covariant equation~\cite{Bausch1981_Critical,bitbol2012statistics}, and $\zeta$ is a Gaussian white noise with correlations 
\begin{equation}
    \argx{\zeta(\br,t)\zeta(\br',t')} = \sqrt{1+(\bnabla_\br h)^2}\,\delta^{(d-1)}(\br-\br')\delta(t-t')\;.
\end{equation}
The Edwards-Wilkinson dynamics that we derived in section \ref{sec:AL} is the lowest order of Eq.~\eqref{eq:Afullh}, with the free energy \eqref{eq:Fdiehl} and $\lambda=1/{\sigma}$. The full Eq.~\eqref{eq:Afullh} with the free energy \eqref{eq:Fdiehl} reads
\begin{equation}\label{eq:Afullhexpl}
    \p_t h =\bnabla_\br^2 h - \frac{\bnabla_\br h\cdot \bnabla_\br\bnabla_\br h\cdot\bnabla_\br h}{{1+(\bnabla_\br h)^2}}+\sqrt{\frac{2T}{\sigma}}\zeta
\end{equation}
Thus the next order of the deterministic part comes from the curvature and is exactly $-\bnabla_\br h\cdot \bnabla_\br\bnabla_\br h\cdot\bnabla_\br h$. We set out to derive this term in our formalism.

It would appear in the form of an $\hb h^3$ term in our effective action. Since its prefactor is of order $T^0$, we shall limit ourselves to order $T^{3/2}$ in the equation of motion.
We therefore start from the non-Gaussian action to order 4 in the fields
\begin{widetext}
\begin{equation}\label{eq:AfullNGugly}
\begin{split}
       S_{\text{NG}} &= \frac 1 T  \int_{t,\br,z}\!\!\!\Big\{-\Big(\frac{\p_z m_c}{\sigma}\Big)^2\big[\red{\hb^2} \rho(-2 + 3\rho) + \argx{\chib^\perp,\p_z\chi^\perp}^2+2 \hbr\argx{\chib^\perp,\p_z\chi^\perp}(1-2\rho)\big]\\
        &+ \Big[-\hbr \frac{\p_z m_c}{\sigma}+\chib^\perp +  \frac{\p_z m_c}{\sigma}(\hbr\rho-\argx{\chib^\perp,\p_z\chi^\perp})\Big]\\
        &\times \Big[\frac{T}{\sigma}\p_z^2 m_c-\red{(\bnabla_\br h)^2}\p_z^2 m_c + 2\red{\bnabla_\br h} \cdot\bnabla_\br\p_z\chi^\perp + \frac{f'''(m_c)}{2}(\chi^\perp)^2\\
        &\quad +\frac{T}{\sigma}\big(-2\rho\p_z^2 m_c +\p_z^2 \chi^\perp\big)-2T\p_z\Pi\Big(\frac{\p_z\chi^\perp}{\sigma}\Big)-\red{(\bnabla_\br h)^2} \p_z^2\chi^\perp + \frac{f^{(4)}(m_c)}{3!}(\chi^\perp)^3\big]
       \Big\} \;,
    \end{split}
\end{equation}
\end{widetext}

\if\wentzel1{\begin{widetext}
\begin{equation}\label{eq:AfullNGugly}
\begin{split}
       S_{NG} &= \frac 1 T  \int_{t,\br,z}\Big\{-\Big(\frac{\p_z m_c}{\sigma}\Big)^2\big[\red{\hb^2} \rho(-2 + 3\rho) + \argx{\chib^\perp,\p_z\chi^\perp}^2+2 \hbr\argx{\chib^\perp,\p_z\chi^\perp}(1-2\rho)\big]\\
        & + 2\chib^\perp \frac{\p_z m_c}{\sigma}\big[-\hbr\rho(1-\rho)+ \argx{\chib^\perp,\p_z\chi^\perp}(1-\rho)\big] \\
        &+ \big[-\hbr(1-\rho)\frac{\p_z m_c}{\sigma}- \argx{\chib^\perp,\p_z\chi^\perp}\frac{\p_z m_c}{\sigma} + \chib^\perp\big]\\
        &\times \big[-\red{(\bnabla_\br h)^2}\p_z^2 m_c + 2\red{\bnabla_\br h} \cdot\bnabla_\br\p_z\chi^\perp + \frac{f'''(m_c)}{2}(\chi^\perp)^2\\
        &\quad -\red{(\bnabla_\br h)^2} \p_z^2\chi^\perp + \frac{f^{(4)}(m_c)}{3!}(\chi^\perp)^3\big]
       \Big\} \;,
    \end{split}
\end{equation}
\end{widetext}\fi
where $\rho=\argx{\p_z m_c,\p_z\chi^\perp}/\sigma$. To keep the action legible, some terms with more than four fields have been kept in the product of the second and third-to-fourth lines, but they should be omitted since they do not contribute to order $T^{3/2}$ in the dynamics.

By visual inspection of Eq.~\eqref{eq:AfullNGugly}, no term of the form $\hb h^3$ is contributed by the first order cumulant $\argx{S_{\rm NG}}_0'$. Let us turn to the second cumulant: such a contribution would involve two bulk fluctuation fields. The only possible combination at order $T^{3/2}$ in the Langevin equation is
\begin{equation}
    \begin{split}
        -\frac{2}{T^2}&\int_{\br,t}\int_{\br',t'} \red{(\bnabla_\br h)^2}(\br,t) \hbr(\br',t') \red{\bnabla_\br h}(\br',t')\cdot\\
        &\int_{z,z'}\p_z^2 m_c(z) \frac{\p_z m_c(z')}{\sigma} \times\\
        &\argx{\chib^\perp(\br,z,t) \bnabla_\br\p_z \chi^\perp(\br',z',t')}_{c,0}'\;.
    \end{split}
\end{equation}
This corresponds to the Feynman diagram
\begin{tikzpicture}
\useasboundingbox (-2,-2) rectangle (2,0.3);
\begin{feynman}
    \vertex (i1) {\(\hr\)};
    \vertex (i2) [below=1.8cm of i1]{\(\hr\)};
    \vertex (f1) [right=2.9cm of i1] {\(\hbr\)};
    \vertex (f2) [below=1.8cm of f1] {\(\hr\)};

    \vertex (v1) [below=0.9cm of i1,xshift=0.9cm];
    \vertex (v2) [right=1.1cm of v1];

    \diagram* {
        (i1) -- [red, thick]  (v1),      
        (i2) -- [red, thick]  (v1),      
      
      (v1) -- [fermion, thick] (v2),
      
        (v2) -- [red,thick] (f2),
        (v2) -- [red,thick,->] (f1),
    };
  \end{feynman}
\filldraw[red] ($(v1)!0.4!(i1)$) circle (2pt);
\filldraw[red] ($(v1)!0.4!(i2)$) circle (2pt);
\filldraw[red] ($(v2)!0.4!(f2)$) circle (2pt);
\filldraw[black] (1.75,-0.9) circle (2pt);
\end{tikzpicture}
\\
where in the notations of Sec.~\ref{sec:AMANL}, non-contracted red legs represent the interface height $h$ and the response field $\hb$, a black (arrowed) leg stands for $\chi^\perp$ ($\chib^\perp$), the oriented black line represents the response propagator, and the dots indicate a spatial derivative $\bnabla_\br$.

Using the expression of the response propagator as a function of the eigenmodes of $\Omega_0$ given in Eq.~\eqref{eq:A_NGL_chibchi}, this contribution can be rewritten as
\begin{equation}\label{eq:Acurv_1}
    \begin{split}
        -\frac{2}{\sigma T}&\int_{\br,t}\int_{\br',t'}(\bnabla_\br h)^2(\br,t) \hb(\br',t') \bnabla_\br h(\br',t')\cdot \bnabla_\br \\
        &\!\!\!\!\!\!\!\sum_{i>0}\int_\bq \ee^{i\bq\cdot(\br-\br')}\ee^{-(\bq^2 + \lambda_i)(t'-t)}\Theta(t'-t) \argx{\p_z^2 m_c, e_i}^2,
    \end{split}
\end{equation}
where the last $\bnabla_\br$ of the first line applies to $\br$ and not $\br'$.
To make progress, we  follow \cite{Diehl1980} and notice that 
\begin{equation}\label{eq:commutator}
    [z,\Omega_0] = 2\p_z\;.
\end{equation}
Then, the sum over $i>0$ can be rewritten as
\begin{equation}
    \begin{split}
        \sum_{i>0}&\ee^{-\lambda_i(t'-t)} \argx{\p_z^2 m_c, e_i}^2 \\
        \if{&= \sum_{i>0}\argx{\p_z^2 m_c, e_i}\ee^{-\lambda_i(t'-t)} \argx{e_i,\p_z^2 m_c}\\}\fi
        &=\argx{\p_z^2 m_c, \ee^{-\Omega_0(t'-t)}\p_z^2 m_c}\\
        &=-\frac{1}{2}\argx{\p_z^2 m_c, \ee^{-\Omega_0(t'-t)}\Omega_0\, z\, \p_z m_c}\\
        &=\frac{1}{2}\p_{t'}\argx{\p_z^2 m_c, \ee^{-\Omega_0(t'-t)}z\, \p_z m_c}\;.
    \end{split}
\end{equation}
We have used that $\argx{e_0,\p_z^2m_c}=0$ when going from the first to the second line, and Eq.~\eqref{eq:commutator} when going from the second to the third line. Integrating by parts the resulting $\p_{t'}$, we identify the leading contribution in the low-frequency and low-momentum limit as the one where the derivative applies to the $\Theta(t'-t)$. Eq.~\eqref{eq:Acurv_1} then becomes 
\begin{equation}
    -\frac{2}{\sigma T} \int_{\br,t} \hb \bnabla_\br h\cdot \bnabla_\br\bnabla_\br h\cdot\bnabla_\br h\argx{\p_z^2 m_c, z \,\p_z m_c}\;.
\end{equation}
Using that $\argx{\p_z^2 m_c, z \,\p_z m_c} = -\sigma/2$, we arrive at
\begin{equation}
     \frac{1}{ T} \int_{\br,t} \hb \bnabla_\br h\cdot \bnabla_\br\bnabla_\br h\cdot\bnabla_\br h\;.
\end{equation}
In the effective dynamics of the interface, this is exactly the first deterministic correction coming from the curvature in Eq.~\eqref{eq:Afullhexpl}. Further adapting the static reasoning of~\cite{Diehl1980} to the dynamical case, we would find that diagrams with closed loops can be neglected, leading upon summation of the remaining contributions to the full curvature term in Eq.~\eqref{eq:Afullhexpl}.

\section{Long wavelength behavior in model B}
\subsection{Detailed derivation of the first order}\label{sec:LnighmareB}
Let us start from Eq.~\eqref{eq:B_Hq}. Because the first eigenvalue of $\Omega_0$ is zero, the first term dominates in the long wavelength limit $q \to 0$. However, to control this expansion, we need to account for the dependence of $\mathrm L_\bq^{-1}$  on $q$, and properly estimate the orders in $q$ of the remaining terms. To make the argument fully explicit, we exploit the knowledge of all the eigenfunctions and eigenvalues of $\Omega_0$ when using the potential $f$ from Eq.~\eqref{eq:f_standard}. Then, explicitly:
\begin{equation}
\begin{split}
    \mathrm H_\bq \p_z m_c = &\frac{A(q)}{2\sigma q^3}\p_z m_c + \frac{1}{q^2 + \lambda_1}\argx{e_1, \mathrm L_\bq^{-1} \p_z m_c}e_1 \\
    &+ \int_k \frac{1}{q^2 + \tilde\lambda_k} \argx{\tilde{e}_k, \mathrm L_\bq^{-1}\p_z m_c}\tilde e_k\;.    
\end{split}
\end{equation}
For the first two eigenvalues $\lambda_0 = 0$ and $\lambda_1 = 3\tau/2$ the eigenfunctions are given by Eq.~\eqref{eq:Om0_eigen}, which we recall here for convenience:
\begin{equation}
    \begin{aligned}
       & e_0(z) = \frac{1}{\sqrt{\sigma}}\partial_z m_c\\
       & e_1(z) = \sqrt{\frac{3}{2}\sqrt{\frac{\tau}{2}}}\frac{\text{sinh}\argp{\sqrt{\frac{\tau}{2}} z}}{\text{cosh}\argp{\sqrt{\frac{\tau}{2}} z}^2}\;.
    \end{aligned}
\end{equation}
In the continuum part of the spectrum we have $\tilde\lambda_k = \tau(2 + k^2)$, and the corresponding eigenfunctions are given by \cite{Gervais1975_Perturbation,Ohta1977_Renormalization}:
\begin{equation}\label{eq:Om0_eigensuite}
\begin{split}
    \tilde e_k(z) = \frac{e^{i \, k\sqrt{\tau}z}}{N_k}\Big[&2k^2 +1 - 3\tanh\Big(\sqrt{\frac{\tau}{2}}z\Big)^2\\
    &+ 3\sqrt{2} i \, k\tanh\Big(\sqrt{\frac{\tau}{2}}z\Big)\Big]\;,
\end{split}
\end{equation}
where $N_k$ is the normalization such that $\argx{\tilde e_k,\tilde e_k} = 1$. It is immediately visible that these eigenfunctions do not vanish at infinity and therefore do not have a well-defined integral for a strictly infinite system. We find it useful to work in a system of finite size $2L$ along the $\hat z$ direction. Then, taking the $L\to\infty$ limit whenever possible (in $\text{tanh}$ terms), the normalization reads
\begin{equation}
    N_k^2 = 4L(2k^2 +1 ) (k^2 +2) - 12 \sqrt{\frac{2}{\tau}}(k^2 +1)\;.
\end{equation}
With periodic boundary conditions between $-L$ and $L$, the Green's function of the Laplacian takes the form:
\begin{equation}\label{eq:LqAppC}
    \mathrm{L}_\bq^{-1} = \frac{1}{2q}\frac{1}{1-e^{-2qL}} \argp{ e^{-2qL} e^{q|z-z'|} + e^{-q|z-z'|}}\;.
\end{equation}
We are interested in the $q\to 0$ and $L\to \infty$ limits; we work at constant $qL$ and take $1/L$ as an expansion parameter. We need to estimate the orders of scalar products of the form $\argx{f,\mathrm{L}_\bq^{-1}g}$.
In the main text, we introduced $2q\argx{\p_z m_c, \mathrm L_\bq^{-1}\p_z m_c} = A(q)$, which has a finite limit $A(0)$ as $L\to\infty$, as can be check by direct substitution using Eq.~\eqref{eq:LqAppC}.
We look at all combinations of eigenfunctions of $\Omega_0$ that appear. The first two eigenfunctions $e_0$ and $e_1$ are exponentially decaying so, with the same reasoning, 
$2q\argx{e_1,\mathrm{L}_\bq^{-1}e_{1}} = O(1)$ and $2q\argx{e_0,\mathrm{L}_\bq^{-1}e_{1}} = O(1)$. In fact, for the choice of the free energy $f$ in Eq.~\eqref{eq:f_standard}, $e_1$ is an antisymmetric function, and these scalar products are of order $1/L$. For generalizability to various forms of double-well potentials $f$, we will not use this specific property.\\

For scalar products involving the $\tilde e_k$, more care is required. We first split $\tilde e_k$ into a `well-behaved' part $\tilde e_k^{wb}$ and a part that does not vanish at infinity $\tilde e_k^{c}$:
\begin{equation}\label{eq:Om0_eigensuite2}
\begin{split}
    \tilde e_k(z) &= \tilde e_k^c+ \tilde e_k^{wb}\\
    \tilde e_k^{wb}&=\frac{e^{i \, k\sqrt{\tau}z}}{N_k}\Big\{3\Big[1 - \tanh\Big(\sqrt{\frac{\tau}{2}}z\Big)^2\Big]\\
    &+ 3\sqrt{2} i \, k\Big[\tanh\Big(\sqrt{\frac{\tau}{2}}z\Big)-\text{sign}(z)\Big]\Big\}\\
    \tilde e_k^{c}&=\frac{e^{i \, k\sqrt{\tau}z}}{N_k}\Big[2k^2 -2+ 3\sqrt{2} i \, k\, \text{sign}(z)\Big]\;,
\end{split}
\end{equation}
Scalar products involving the well-behaved part and another exponentially decreasing function can  be upper bounded using:
\begin{align}
    \Big|2q \langle\tilde e_k^{wb}&,\mathrm L_\bq^{-1}e_0\rangle\Big|\leq\frac{1}{1-e^{-2qL}}\times\notag\\
    &\int_{z,z'} \argp{ e^{-2qL} e^{q|z-z'|} + e^{-q|z-z'|}}|e_0(z')|\notag\\
    &\times\frac{1}{N_k}\Big\{3\Big[1 - \tanh\Big(\sqrt{\frac{\tau}{2}}z\Big)^2\Big]\notag\\
    & + 3\sqrt{2}\, |k|\Big|\tanh\Big(\sqrt{\frac{\tau}{2}}z\Big)-\text{sign}(z)\Big|\Big\}\notag\\
    &\leq\frac{cst + cst\, |k|}{N_k} = O\Big(\frac{1}{\sqrt{L}}\Big) \frac 1 {|k|}\;,
\end{align}
where $1/|k|$ is the scaling in $k$ for large $k$. Likewise, $2q\argx{\tilde e_k^{wb},\mathrm L_\bq^{-1}e_1} = O\Big(\frac{1}{\sqrt{L}}\Big) \frac 1 {|k|}$ and  $2q\argx{\tilde e_k^{wb},\mathrm L_\bq^{-1}\tilde e_l^{wb}} = O\Big(\frac{1}{L}\Big) \frac 1 {|kl|}$.
$\tilde e_k^c$ can be approximated by a plane wave at large $k$. From this we can deduce that, for large $k$ and $l$
\begin{align}
    2q\argx{\tilde e_k^c,\mathrm L_\bq^{-1}e_0} &= O\Big(\frac{1}{\sqrt{L}}\Big) \frac 1 {|k|}\\
    2q\argx{\tilde e_k^c,\mathrm L_\bq^{-1}e_1} &= O\Big(\frac{1}{\sqrt{L}}\Big) \frac 1 {|k|}\\
    2q\argx{\tilde e_k^c,\mathrm L_\bq^{-1}\tilde e_l^{wb}} &= O\Big(\frac{1}{L}\Big) \frac 1 {|kl|}\\
    2q\argx{\tilde e_k^c,\mathrm L_\bq^{-1}\tilde e_l^c} &= O\Big(\frac{1}{L}\Big) \frac 1 {|kl|} \;.
\end{align}
The scalings in $k,l$ on the right-hand side correspond to the leading order behavior in diverging $k$ and $l$ of the $O(L^\alpha)$ in front.
For $k=0$, $\tilde e_0^c = -2/N_0$, and we compute exactly:
\begin{equation}\label{eq:tildee0_int}
    2q \, \mathrm L_\bq^{-1}\tilde e_0^c = -\frac{4}{qN_0}\;,
\end{equation}
which we use to determine scalar products involving $\tilde e_0^c$.
All relevant scalar products have been estimated as $L\to\infty$:
\begin{align}
    2q\argx{e_{01},\mathrm L_\bq^{-1}e_{01}} &= O(1)\\
    2q\argx{\tilde e_k,\mathrm L_\bq^{-1}e_{01}} &= O\Big(\frac{1}{\sqrt{L}}\Big) \frac 1 {|k|}\\
    2q\argx{\tilde e_k,\mathrm L_\bq^{-1}\tilde e_l} &= O\Big(\frac{1}{L}\Big) \frac 1 {|kl|}\\
    2q\argx{\tilde e_0,\mathrm L_\bq^{-1}e_{01}} &= O(\sqrt{L})\\
    2q\argx{\tilde e_0,\mathrm L_\bq^{-1}\tilde e_k} &= O(1) \frac 1 {|k|}\\
    2q\argx{\tilde e_0,\mathrm L_\bq^{-1}\tilde e_0} &= O(L)\;,
\end{align}
where $e_{01}$ stands for either $e_0$ or $e_1$, $k$ and $l$ are both nonzero, and the scalings in $k,l$ indicated on the right-hand side correspond to the leading order behavior in diverging $k$ and $l$ of the $O(L^\alpha)$ in front.

We can now evaluate the various orders in $\mathrm H_\bq \p_z m_c$ as:
\begin{equation}
\begin{split}
    \mathrm H_\bq \p_z m_c = &\frac{A(q)}{2\sigma q^3}\Big(\p_z m_c + O(L^{-2})e_1 + O(L^{-3/2})\tilde e_0 \\
    &+ O(L^{-5/2})\frac{\tilde e_k} {|k|^3}\Big)\;, 
\end{split}\end{equation}
where there is an implicit integral over $k\neq 0$ and we recall that all the $\tilde e_0$, $\tilde e_k$ functions scale as $L^{-1/2}$. Then, by induction,
\begin{equation}\begin{split}
    \mathrm H_\bq^n &\p_z m_c = \argp{\frac{A(q)}{2\sigma q^3}}^n \Big[\p_z m_c + O(L^{-1})\p_z m_c  \\
    &+ O(L^{-2})e_1+O(L^{-3/2})\tilde e_0 + O(L^{-5/2})\frac{\tilde e_k} {|k|^3}\Big]\;.
\end{split}\end{equation}
As in the main text, for $\omega$ of order $O(q^3)$ at most (or, equivalently, $O(L^{-3})$), this allows us to compute:
\begin{equation}
    \begin{split}
        \argp{1 + i\omega\mathrm H_\bq}^{-1} &\p_z m_c = \argp{1+\frac{i\omega A(q)}{2\sigma q^3}}^{-1} \Big[\p_z m_c\\
        &+ O(L^{-1})\p_z m_c + O(L^{-2})e_1 \\
        &+O(L^{-3/2})\tilde e_0 + O(L^{-5/2})\frac{\tilde e_k} {|k|^3}\Big]\;, 
    \end{split}
\end{equation}
which, substituted into the scalar product of Eq.~\eqref{eq:BActionhugly}, leads to
\begin{equation}
\begin{split}
    \langle(\Gamma^\dagger_{\bq,\omega})^{-1}&u_0,\mathrm L_\bq(\Gamma^\dagger_{-\bq,-\omega})^{-1}u_0\rangle = \\
    &\left| i\omega +\frac{2\sigma q^3}{ A(q)}\right|^{-2} \frac{2q}{A(q)}\argc{1 + O(L^{-1})}\;.
\end{split}
\end{equation}
This leads to the action given in Eq.~\eqref{eq:B_Sforh} for the interface height.

\subsection{Intuition for the next-order correction}\label{sec:BShino}
Thanks to the expansion above, we obtain the leading order in $q$ of the dispersion relation as
\begin{equation}
    i\omega = \frac{2\sigma q^3}{A(0)}\;.
\end{equation}
For higher orders in $q$, the eigenvectors other than $\p_z m_c$ start playing a role. We can however provide the reader with some intuition on the behavior of the dispersion relation to the next order. We begin by identifying the key operator $ \mathrm L_\bq (\bq^2 +\Omega_0 )$, for a generic double-well potential $f$. We denote its eigenvalues $\lambda_i(q)$ and the corresponding right and left eigenvectors $R_i(q)$ and $L_i(q)$, respectively. Upon integrating over $\omega$, the leading contribution to $(\Gamma^\dagger_{\bq,\omega})^{-1}$---{\it i.e.}~the slowest mode that controls the interface relaxation---comes from the eigenvalue $\lambda_0(q)$, whose real part is closest to zero. This eigenvalue can be expressed as
\begin{equation}\label{eq:nextorder}
\begin{split}
    \lambda_0(q) &= q^2\frac{\argx{L_0(0),\mathrm L_\bq R_0(q)}}{\argx{L_0(0),R_0(q)}}\\
    &=2q^3 \frac{\int_z e_0(z)\,R_0(q)(z)}{\int_{z,z'} e_0(z)\, \ee^{-q|z-z'|}R_0(q)(z')}\;,
\end{split}
\end{equation}
where $R_0(q)(z)$ is the right eigenvector of the operator $\mathrm L_\bq (\bq^2 +\Omega_0 )$ for the eigenvalue $\lambda_{0}(q)$ and $L_0(0) = \mathrm L_\bq^{-1}e_0$ is the left eigenvector of the operator $\mathrm L_\bq \Omega_0$ for the eigenvalue $0$. The notation  $R_0(q)$ is admittedly a little bit misleading: $R_0(0)$ is the eigenvector of ${\mathrm L}_\bq\Omega_0$, and as such it still features a $\bq$ dependence. Hence the $q$ dependence in $R_0(q)$ labels the added contribution of $q^2$ in ${\mathrm L}_\bq(\bq^2+\Omega_0)$. In general, $R_0(q)$ is unknown, but Shinozaki and Oono~\cite{Shinozaki1993_Dispersion} showed that $\lambda_0(q)$ can be obtained by perturbing the zero eigenvalue of $\mathrm L_\bq\Omega_0$, for which $R_0(0) = e_0$. Therefore in the $q\to 0$ limit, the fraction \eqref{eq:nextorder} has a finite limit: $R_0(q) \to R_0(0) = e_0$ and one recovers the  relation $i\omega = \frac{2\sigma q^3}{A(0)}$. However for non-vanishing $q$, the study of the dispersion relation requires detailed knowledge of $R_0(q)$. Since the dispersion relation is obtained by perturbing $\mathrm L_\bq\Omega_0$ by $q^2 \mathrm L_\bq$, it seems reasonable to assume that the correction to $R_0(0)$ at low $q$ is of order $q^2$. However, the presence of $q$ in the operator $\mathrm L_\bq = q^2 - \p_z^2$ makes proving this statement analytically non-trivial and beyond the scope of this work. To proceed further, Shinozaki and Oono observed numerically that $R_0(q)(z) = e_0(z) + o(q)$. This strongly suggests that the dispersion relation is valid also to the next order:
\begin{equation}
    i\omega = \frac{2\sigma q^3}{A(q)}(1+o(q))
\end{equation}
Numerically, by direct simulation of the partial differential equation \eqref{eq:B}, we not only confirm the theoretical prediction for the leading order behavior, but we also find that the next-to-leading order is correctly given by expanding $A(q)$, in spite of us not having strong analytical support for this (see Fig.~\ref{fig:numerics model B}). 

This means that the expression obtained from the ansatz of Eq.~\eqref{eq:rottenansatz} in \cite{Fausti2021_Capillary} for $\lambda=\zeta=0$, $i\omega = \frac{2\sigma q^3}{A(q)}$ also holds to order $q^4$. Beyond that order, we know from Eq.~\eqref{eq:nextorder} that there are several $O(q^5)$ corrections to the dispersion relation that are not captured by $A(q)$.

\section{Numerical methods}\label{sec:numerics}

In this section we present the numerical scheme we used to obtain the relaxation of interfaces in  models B and AMB. We integrate the stochastic PDEs~\eqref{eq:B}, and \eqref{eq:AMB+} with $\zeta=0$, at zero temperature in a two-dimensional system, using a pseudo-spectral code with periodic boundary conditions and semi-implicit time-stepping. We use the symmetric $\phi^4$ potential 
\begin{equation}\label{eq:f_code}
    f(\phi) = 0.25\argp{-\frac{1}{2}\phi^2 + \frac{1}{4} \phi^4}\;.
\end{equation}

Using existing results on the generalized thermodynamics of motility-induced phase separation~\cite{solon2018generalized2,Solon2018_GeneralizedThermodynamics}, we can numerically solve for the binodals at different values of $\lambda$.
We then use the binodal values to construct the initial condition, with two bands of equal sizes for the dense and dilute phases. We then let this singular profile relax to obtain the steady-state solution. From the resulting profile $m_c$, we compute the quantities that appear in the theoretical prediction for the relaxation time, $\sigma_A$ and $A_A(q)$. (In the passive case, we compute them from the explicit expression of the mean-field profile \eqref{eq:profilmc}).

Starting from the steady state, we then perturb one of the interfaces by a cosine and track its relaxation for different excitation wavelengths.  
For this purpose, we use the local definition for the interface height given by Eq.~\eqref{eq:definterfacevalue}, which is acceptable in the absence of noise. 
We take $\phi_0$ to be the average of the two bulk densities, and identify the interface by linear interpolation. 

We simulated systems of size $L_r=L_z=256$ for model B and $L_r=L_z=512$ for active model B, with a spatial discretization of $1$ in both directions. With these parameters and a $3/2$ anti-aliasing procedure, we have about fifteen points to resolve the interface. Simulations were run for $\lambda=0,1,2$ with a timestep of $0.04$ for a total duration of $\Delta t_{\rm tot}(q_1) = 4.8e+6$ for the mode $q_1=2\pi/256$, adapting the duration such that $q^3 \Delta t_{\rm tot}(q)$ remains constant for the various excited modes.

\begin{figure}
\hspace{-1cm}
    \includegraphics[width=1.0\columnwidth]{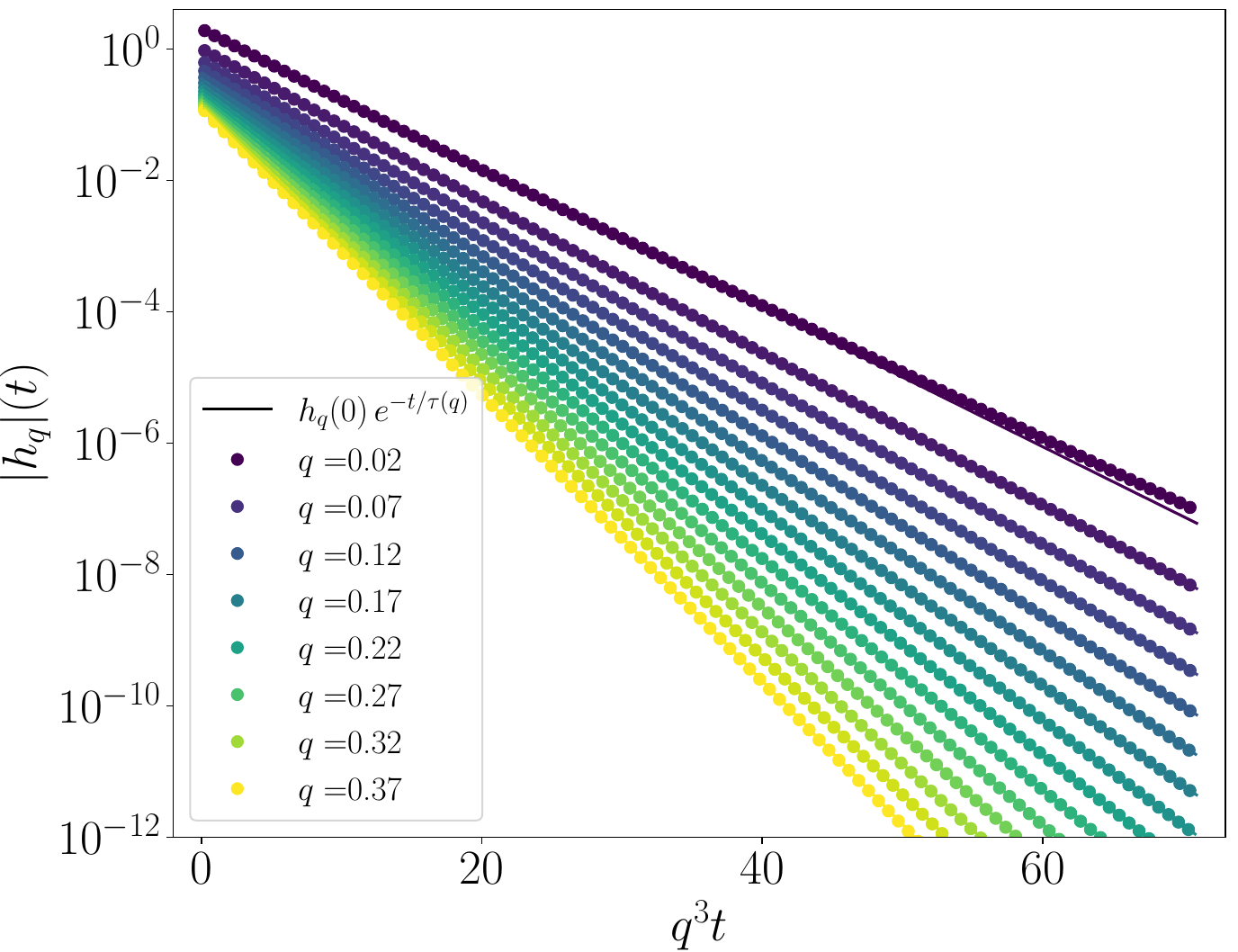}
    \caption{Relaxation of a one-dimensional interface in a two-dimensional model B, starting from an initial condition $\frac{0.1}{q}\cos(qr)$, as a function of $q^3 t$. The dots correspond to the measured Fourier amplitudes of the initially excited modes. The lines, in good agreement with the measurements, are the exponential fits.}\label{fig:relax}
\end{figure}

The theory predicts that, in the linear regime, \textit{i.e.} for small enough initial perturbations, the interface relaxes as $\sim \text{exp}[-t/\tau(q)]$, with a relaxation time $\tau(q)$ depending on the excited mode. In Fig.~\ref{fig:relax}, the amplitudes (the moduli) of the excited Fourier modes of a model B interface are shown  as functions of $q^3 t$ on a logarithmic scale, confirming that the relaxation is exponential.
We extract the numerical value for $\tau(q)$ by fitting the decay by an exponential in a time range where the instantaneous relaxation time is roughly constant.\\ 

\section{Computational details in model H}\label{sec:appendixH}
The operator that controls the behavior of the interface is
\begin{equation}
    \begin{split}
        -\mathrm K_\bq &(\bq^2 + \Omega_0) = \Big[\delta(z-z')(\bq^2 - \p_z'^2)\\
        &+ \p_z m_c(z)T_{zz}(\bq,z-z')\p_{z}m_c(z')\Big]\\
        &\times(\bq^2 + \Omega_0(z'))\;.
    \end{split}
\end{equation}
Following \cite{Jasnow1987_Crossover,Shinozaki1993_Dispersion,shinozaki1993dispersionH}, we introduce $L_0(q)$ and $R_0(q)$, which are,  respectively,  the left and right eigenvectors of this operator for its eigenvalue $\lambda_0(q)$ with real part closest to zero. Then
\begin{align}
    \lambda_0(q) R_0(q)(z) & = \int_{z'} \Big[\delta(z-z')(\bq^2 - \p_z'^2)\notag\\
        &+ \p_z m_c(z)T_{zz}(\bq,z-z')\p_{z}m_c(z')\Big]\notag\\
        &\times(\bq^2 + \Omega_0(z'))R_0(q)(z')\;.
\end{align}
For large viscosities, we compute this eigenvalue by treating the operator $-\mathrm K_\bq (\bq^2 + \Omega_0)$ as a perturbation of $-\mathrm K_\bq \Omega_0$.
The latter admits $R_0(0) = \p_z m_c$ as a right eigenvector associated with the eigenvalue $0$, and the corresponding left eigenvector is $L_0(0) = -\int_{z'}\mathrm K_\bq^{-1}(z,z') \p_z m_c(z')$.
Then, perturbation theory leads to
\begin{equation}\label{eq:H_lambda0}
    \lambda_0(q) =- q^2 \frac{\argx{L_0(0),\int_{z'}\mathrm K_\bq(z,z') R_0(0)(z')}}{\argx{L_0(0),R_0(0)}}\;.
\end{equation}
As given in the main text, the explicit form of the Oseen tensor in Fourier space in $\br$ is
\begin{equation}
    \mathrm T_{zz}(\bq,z-z') = \frac 1{4\eta q}\argp{1 + q|z-z'|}\ee^{-q|z-z'|}\;,
\end{equation}
such that, to leading order in $q$,
\begin{equation}\label{eq:HApp_simplifiedH}
    \mathrm K_\bq(z,z')\simeq-\frac{1}{4\eta q}\p_z m_c (z) \p_z m_c(z')\;.
\end{equation}
Note that this approximation is rather crude, as it is obviously unsuited if applied to functions orthogonal to $\p_z m_c$. 
However, for our purposes, this approximation suffices to conclude that, to lowest order in $q$, $\p_z m_c$ is an eigenvector of $\mathrm K_\bq$, leading to
\begin{equation}\label{eq:H_lambda02}
    \lambda_0(q) \simeq \frac{\sigma q}{4\eta}\;.
\end{equation}

As mentioned in the main text, for small values of the viscosity, the lowest eigenvalue $\lambda_0(q)$ of $-\mathrm K_\bq (\bq^2 + \Omega_0)$ is not obtained from the ground state of the nonperturbed operator. It is instead given by the lower bound of the continuous part of the spectrum, so that the value computed in Eq.~\eqref{eq:H_lambda02} is not the lowest eigenvalue that controls the relaxation. Conversely, in the regime of large viscosities,  the $i\omega \sim q$ dispersion relation can be observed for small but nonvanishing $q$'s \cite{shinozaki1993dispersionH,shinozaki1993spinodal}.

\bibliography{Biblio}
\end{document}